\documentclass[twoside,11pt]{article}

\usepackage{blindtext}

\usepackage[abbrvbib, preprint]{jmlr2e}

\usepackage{amsmath, amsfonts}
\usepackage{enumerate}
\usepackage{enumitem}

\usepackage{xcolor}
\usepackage{algorithm2e}
\usepackage{mathtools}

\usepackage{subcaption}
\usepackage[font=scriptsize,labelfont=bf]{caption} 

\usepackage{tabularx}
\usepackage{array}

\usepackage{booktabs}
\usepackage{longtable}
\usepackage{colortbl}

\usepackage[margin=1in]{geometry}
\usepackage[subtle]{savetrees}

\usepackage{fontawesome}

\usepackage{changepage}

\usepackage{float}

\newcommand{\blind}{1}

\newcommand{\boldf}{\mathbf{f}}

\newcommand{\bx}{\mathbf{x}}
\newcommand{\by}{\mathbf{y}}

\newcommand{\bbeta}{\boldsymbol{\beta}}

\newcommand{\beps}{\boldsymbol{\epsilon}}

\newcommand{\bP}{\mathbf{P}}

\newcommand{\bX}{\mathbf{X}}
\newcommand{\bY}{\mathbf{Y}}

\newcommand{\bSigma}{\boldsymbol{\Sigma}}
\newcommand{\bPsi}{\boldsymbol{\Psi}}

\newcommand{\R}{\mathbb{R}}

\newcommand{\E}{\mathbb{E}}

\newcommand{\Trace}{\textnormal{Trace}}

\DeclarePairedDelimiter{\braces}{\lbrace}{\rbrace}
\DeclarePairedDelimiter{\paren}{(}{)}

\DeclarePairedDelimiter{\abs}{|}{|}
\DeclarePairedDelimiter{\norm}{\|}{\|}

\newcommand{\data}[1][n]{\mathcal{D}_{#1}}

\newcommand{\node}{\mathfrak{t}}

\newcommand{\splits}{\mathcal{S}}

\newcommand{\rsq}{R^2}

\newcommand{\mdi}[1]{\textnormal{MDI}_k\paren{#1}}
\newcommand{\mdip}[1]{\textnormal{MDI}_k^+\paren{#1}}

\newcommand{\plus}{\texttt{+}}
\newcommand{\method}{MDI\plus}
\newcommand{\rfmethod}{RF\plus}

\definecolor{cyan}{cmyk}{1, 0.2, 0, 0}

\definecolor{green}{cmyk}{1,0,1,0}

\definecolor{orange}{cmyk}{0,50,100,0}

\definecolor{amethyst}{rgb}{0.6, 0.4, 0.8}

\SetCommentSty{mycommfont}

\SetKwInput{KwInput}{Input}                
\SetKwInput{KwOutput}{Output}              
\SetKw{KwBy}{by}

\makeatletter
\renewcommand{\paragraph}{%
  \@startsection{paragraph}{4}%
  {\z@}{1.5mm}{-0.6em}%
  {\normalfont\normalsize\bfseries}%
}
\makeatother

\usepackage{lastpage}
\jmlrheading{23}{2025}{1-\pageref{LastPage}}{1/21; Revised 5/22}{9/22}{21-0000}{Agarwal, Kenney, Tan, Tang, and Yu}

\ShortHeadings{\method}{Agarwal, Kenney, Tan, Tang, and Yu}
\firstpageno{1}

\begin{document}

\title{Integrating Random Forests and Generalized Linear Models for Improved Accuracy and Interpretability}

\author{\name Abhineet Agarwal\thanks{Denotes equal contribution} \email aa3797@berkeley.edu \\
       \addr Department of Statistics\\
       University of California, Berkeley\\
       Berkeley, CA 94720, USA
       \AND
       \name Ana M. Kenney$^{*}$ \email akenney1@uci.edu \\
       \addr Department of Statistics\\
       University of California, Irvine\\
       Irvine, CA 92697, USA
       \AND
       \name Yan Shuo Tan$^{*}$ \email yanshuo@nus.edu.sg \\
       \addr Department of Statistics and Data Science\\
       National University of Singapore\\
       Singapore
       \AND
       \name Tiffany M. Tang$^{*}$ \email ttang4@nd.edu \\
       \addr Department of Applied and Computational Mathematics and Statistics\\
       University of Notre Dame\\
       Notre Dame, IN 46556, USA
       \AND
       \name Bin Yu \email binyu@berkeley.edu \\
       \addr Department of Statistics, EECS, CCB\\
       University of California, Berkeley\\
       Berkeley, CA 94720, USA
       }

\editor{My editor}

\maketitle

\begin{abstract}

Random forests (RFs) are among the most popular supervised learning algorithms due to their nonlinear flexibility and ease-of-use.
However, as black box models, they can only be interpreted via algorithmically-defined feature importance methods, such as Mean Decrease in Impurity (MDI), which have been observed to be highly unstable and have ambiguous scientific meaning.
Furthermore, they can perform poorly in the presence of smooth or additive structure.
To address this, we reinterpret decision trees and MDI as linear regression and $R^2$ values, respectively, with respect to engineered features associated with the tree's decision splits.
This allows us to combine the respective strengths of RFs and generalized linear models in a framework called \rfmethod, which also yields an improved feature importance method we call \method.
Through extensive data-inspired simulations and real-world data sets, we show that \rfmethod\;improves prediction accuracy over RFs and that \method\;outperforms popular feature importance measures in identifying signal features, often yielding more than a 10\% improvement over its closest competitor.
In case studies on drug response prediction and breast cancer subtyping, we further show that \method\;extracts well-established genes with significantly greater stability compared to existing feature importance measures.

\end{abstract}

\begin{keywords}
interpretable machine learning, explainable AI, decision trees, ensembles, non-parametrics
\end{keywords}

\section{Introduction}

In many high-impact scientific problems, researchers are often equally or more interested in extracting actionable and reliable interpretations from models of the data than the model predictions themselves. For example, in the medical setting, researchers not only want to develop accurate disease risk prediction models, but also to identify the genes that drive the disease to develop targeted therapies.

Known for their ease of interpretability, generalized linear models (GLMs), including linear regression, logistic regression, and their penalized versions, have long been a cornerstone of statistical analysis in the physical, biological, and social sciences.
In particular, the coefficients of the model provide direct insights into the relationship between predictors and the response and support formal statistical inference, thereby allowing for reliable scientific discovery.
Unfortunately, GLMs are limited by their somewhat rigid structure, which makes it difficult for them to capture interactions, heterogeneity, and other complex structures that exist in real-world datasets and for which we may have little prior information.
Unsurprisingly, their prediction performance on real-world datasets often pales in comparison to that of modern machine learning methods such as random forests (RFs) \citep{breiman2001random}.

RFs are among the most popular supervised learning algorithms because they can adapt to nonlinear relationships, handle high-dimensional data, are robust to low signal-to-noise situations, and can be effectively used out-of-the-box with minimal hyperparameter tuning.
Indeed, they achieve state-of-the-art prediction performance over a wide class of learning problems \citep{olson2018data},
often outperforming deep learning methods on small or moderately-sized tabular datasets \citep{schwartz2022tabularDL}. 
These types of datasets frequently arise in the sciences due to high costs of data collection. 
Given the strong predictive performance of RFs in these settings, practitioners are also keen on using them to extract new scientific insights \citep{basu2018iterative,de2022machine,wang2023epistasis}.
However, RFs are ensembles of hundreds of randomized decision trees, which are not immediately interpretable.
While researchers have attempted to overcome this limitation via feature importance methods, such as Mean Decrease in Impurity (MDI) \citep{breiman2001random}, permutation-based methods (i.e., mean decrease in accuracy (MDA)) \citep{breiman2001random}, and SHAP \citep{lundberg2017unified}, such methods are defined algorithmically, which makes them difficult to interpret scientifically. 
Moreover, these feature importance methods have often been observed to be highly unstable \citep{nicodemus2011stability, kelodjou2024shaping}, leading to the lack of reproducibility.
RFs themselves further suffer from issues relating to their algorithmic definition, such as inductive biases against smooth or additive structure \citep{tan2021cautionary}, as well as an inability to inject useful prior knowledge, such as on the distribution of errors, into the modeling process.

In this paper, we propose \textbf{\rfmethod}, a framework which merges RFs and GLMs in order to simultaneously exploit the strengths of both methods while overcoming their respective weaknesses, thereby helping to bridge the algorithmic- and data-modeling cultures first described by Breiman in 2001 \citep{breiman2001statistical}.
The starting point of \rfmethod~is a seemingly little known connection between decision trees and linear models \citep{klusowski2024large}. 
That is, a decision tree is the best fit linear model on a collection of engineered features associated with splits (i.e., local decision stumps) from the tree.
These engineered features are piecewise constant and have highly localized support, whose mismatch with smooth and additive structure helps to explain the inductive biases of RFs.
Furthermore, the use of ordinary least squares (OLS) reveals a lack of regularization as well as a potential lack of adaptiveness to non-Gaussian error distributions.
To overcome these shortcomings, \rfmethod~gives the analyst the opportunity to (i) expand the engineered feature representation by appending smooth, linear features (e.g., the raw features) and possibly other features (e.g., engineered from prior knowledge), and (ii) replace the implicit assumption of a linear link function and Gaussian error distribution under OLS with other options (e.g., GLMs), while adding explicit regularization.
This generalization of RF to \rfmethod~helps to mitigate the inductive biases of RF while enhancing its flexibility to incorporate prior knowledge and simultaneously making it more stable.

Furthermore, the \rfmethod~framework opens the door to improved interpretations --- a crucial asset given the importance of reliable interpretations for high-impact scientific applications.
More precisely, our framework establishes a new interpretation of MDI, which is the default feature importance method for RFs in \texttt{scikit-learn}.
$\text{MDI}_k$, the score for feature $X_k$, is first computed individually for each tree in an RF ensemble, and is then averaged over all the trees.
We show that for each individual tree, $\text{MDI}_k$ is exactly equal to the training set $R^2$ value of the ``partial OLS model'' obtained by setting all OLS coefficients to be zero apart from those corresponding to local decision stumps that split on $X_k$.
This equivalence allows us to better explain known drawbacks of MDI, such as its instability and biases against correlated and low-entropy features \citep{strobl2007bias, nicodemus2011stability, strobl2008conditional, nicodemus2009predictor, nembrini2019bias}.
Importantly, this leads us to develop \textbf{\method}, which adapts MDI to the \rfmethod~framework by (i) making use of a natural partition of the expanded engineered feature representation, (ii) using partial model predictions from the choice of GLM, (iii) makes use of efficient sample splitting, and moreover (iv) replaces $R^2$ with a user-specified choice of performance metric.
The enhanced flexibility and goodness-of-fit for \rfmethod~allow us to be more confident of \method~values reflecting the inherent structure of the population from which the data is drawn, rather than merely being a description of an algorithmic model.
Meanwhile, the enhanced stability via regularization and sample splitting help to overcome its biases, reduce its variance, leading to better reproducibility.

To aid practitioners with the modeling choices allowed in \rfmethod~and \method, we also provide a data- and stability-driven procedure to guide model selection and aggregation inspired by the Predictability, Computability, and Stability (PCS) framework for veridical data science \citep{yu2020veridical, yu2024veridical}. 
The PCS framework builds upon the three core principles in its name to bridge, unify, and expand on the best ideas from machine learning and statistics for the entire data science life cycle.

We finally demonstrate the effectiveness of \rfmethod\;and \method\;through extensive data-inspired simulations and real-world case studies.
In particular, as better predictive models often enable more reliable interpretations \citep{murdoch2019definitions, yu2020veridical}, we first establish the credibility of using \rfmethod\;as the basis for interpretations by demonstrating its superior prediction accuracy over RF on a wide variety of datasets.
We then conduct extensive simulations, showing that \method\;significantly outperforms other popular feature importance measures in its ability to identify signal features.
These simulations reflect a diverse range of problem structures (e.g., regression/classification, linear/non-linear) and challenges (e.g., low signal-to-noise, outliers, omitted variables, small sample sizes) commonly encountered in the real world. 
We highlight the flexibility of \method\;by showing how tailoring the choice of GLM (e.g., using Huber regression in the presence of outliers) and importance metric to the problem structure at hand can lead to better feature ranking performance. 
We further show that \method\;mitigates the biases of MDI through simulations with highly-correlated features and features with varying levels of entropy. 
Lastly, in two case studies on cancer drug response prediction and breast cancer subtyping, we show that the top-ranked features from \method\;are highly predictive and concur with established domain knowledge with significantly greater stability than other feature importance measures. 
The strong performance of both \rfmethod\;and \method\;demonstrate their utility as powerful and practical tools to extract reliable scientific insights in real-world problems.

\paragraph{Organization.} 
We begin by reviewing related work in Section~\ref{sec:related_work}.
In Section~\ref{sec:new_perspective_on_mdi}, we establish the connection between decision trees and linear regression, which enables us to re-interpret RFs and MDI and to explain their known drawbacks as well as reveal new ones. 
In Section~\ref{sec:mdi_plus}, we build on the linear regression interpretation of RFs and MDI to develop our generalizations, called \rfmethod\;and \method, respectively.
In Section~\ref{sec:prediction_accuracy}, we demonstrate the effectiveness of \rfmethod\;for improving prediction accuracy over RF across a variety of datasets.
Through extensive data-inspired simulations, we demonstrate the effectiveness of \method\;over existing methods in identifying signal features in Section~\ref{sec:sims} and mitigating its known biases in Section~\ref{subsec:mdi_bias_sim}.
In Section~\ref{sec:case_studies}, we apply \rfmethod\;and \method\;to two case studies, investigating the stability and accuracy of its predictions and feature rankings in comparison with existing methods. 
We conclude with a discussion in Section~\ref{sec:discussion}.

\section{Related Work}
\label{sec:related_work}

In this section, we briefly review related extensions of RFs as well as global feature importance measures that are commonly used to interpret RFs in practice.

\subsection{Random Forest Extensions}

Given the strong empirical performance of ordinary RFs, there has unsurprisingly been great interest in extending and improving upon the original RF algorithm \citep{biau2016random}.
While some have tailored RFs to specific problems or tasks such as the online setting \citep{saffari2009line}, quantile regression \citep{meinshausen2006quantile}, survival analysis \citep{ishwaran2008random}, and heterogeneous treatment effect estimation \citep{athey2016recursive, athey2019generalized}, others have proposed sampling or algorithmic techniques to improve the prediction performance of RFs in general \citep[e.g., see][]{bernard2012dynamic, winham2013weighted}.

We highlight a few notable extensions, which are most closely related to the present work.
First, \citet{saha2023random} proposed an extension of RFs, RF-GLS, for spatially-dependent data. 
Interestingly, \citet{saha2023random} exploit the observation that the CART splitting criterion in terms of a linear regression problem and thus extend this to incorporate spatial correlation structures via generalized least squares (GLS).
Though related, this differs from our work, which views the decision tree model, rather than the splitting criterion, via a linear regression perspective.
\citet{friedman2008predictive} take a more similar viewpoint, developing a family of rule-based ensemble models by fitting a LASSO model directly on a collection of features, derived using splits from typically many decision trees. 
However, an equivalence between linear regression and the original decision tree model is not discussed.
\citet{agarwal2022hierarchical} exploit the equivalence between decision trees and linear regression used here (formally introduced in Section~\ref{sec:new_perspective_on_mdi}), but focus mostly on improving the existing RF predictions via a hierarchical shrinkage penalty while relying on existing feature imortance measures for interpretations.

Also related are Bayesian decision trees~\citep{chipman1998bayesian} and ensembles, particularly Bayesian Additive Regression Trees (BART)~\citep{chipman2010bart,hill2020bayesian}.
Such methods put a prior on the space of functions that can be implemented by a decision tree ensemble and combine that with a likelihood for the response data to obtain a posterior that can be sampled from via Markov chain Monte Carlo.
The Bayesian prior regularizes the predictions made by the tree ensemble, similar to \rfmethod.
Furthermore, under this framing, it is natural idea to replace the Gaussian likelihood with other alternatives~\citep{linero2024generalized}, similar to GLM modeling.
However, BART does not consider other feature representations and is more computationally more intensive both in terms of model fitting and extraction of feature importances.

\subsection{Random Forest Feature Importances}


\paragraph{MDI.} 
Our work directly builds upon the Mean Decrease in Impurity (MDI) feature importance measure \citep{breiman2001random}.
Briefly, the MDI of a feature $X_k$ is a weighted average of the impurity decrease from all splits involving $X_k$ in the forest. 
Previous work has theoretically analyzed the population MDI values under some distributional assumptions \citep{louppe2013understanding,scornet2020trees} and provided consistency guarantees \citep{scornet2020trees}.
Notably, \citet{scornet2020trees} showed that the normalized sum of the MDI values of a single tree is the global $\rsq$ (total variance explained) of the model; however, they fell short of connecting the MDI of an individual feature to linear regression $\rsq$ values.
Relatedly, \citet{klusowski2021nonparametric} showed a connection between MDI and linear correlation for decision stump models and proved nonparametric variable selection consistency.
Despite being widely used in practice, MDI is known to be biased towards high entropy features (e.g., continuous features) \citep{strobl2007bias,nicodemus2011stability}.
Many strategies have thus been proposed to improve upon MDI and mitigate its bias, such as by creating artificial uninformative features \citep{sandri2008bias, nembrini2018revival}, using out-of-bag samples \citep{li2019debiased, zhou2021unbiased}, and using a penalized framework for debiasing MDI \citep{loecher2022debiasing, loecher2022unbiased}.
However, unlike \method, they do not address the instability of MDI or the negative inductive biases of RF.

\paragraph{MDA.}
In addition to MDI, \citet{breiman2001random} proposed a permutation-based feature importance measure called Mean Decrease in Accuracy (MDA).
To measure the importance of a feature $X_k$, MDA permutes the values of $X_k$ marginally for out-of-bag (OOB) samples, and calculates the excess prediction loss incurred on these samples. Previous work has studied MDA and variants thereof both empirically \citep{strobl2008conditional,genuer2008random,gromping2009variable,genuer2010variable,hooker2019please} and theoretically \citep{gregorutti2017correlation,ramosaj2019asymptotic,benard2021mda}.
Empirical studies show that MDA performs poorly when features are highly correlated \citep{strobl2008conditional,nicodemus2009predictor,nicodemus2011stability,nembrini2019bias}.
This is because permutations break dependencies between features in the dataset \citep{hooker2019please}. 
Hence, MDA involves evaluating the model out-of-distribution, i.e., on regions of the covariate space with little or no data where its fit to the data is unreliable, making the computed MDA values unstable.
To overcome this, variants of permutation scores have been proposed \citep{strobl2008conditional,hooker2019please} while other works have investigated altering the RF algorithm altogether \citep{hothorn2006unbiased}.
However, they do not fully address the weaknesses of MDA arising from the intrinsic instability and biases of RF.

\paragraph{Other feature importance methods.}
More recently, SHAP (SHapley Additive exPlanations) \citep{lundberg2017unified} was developed as a model-agnostic local feature attribution method based upon approximating Shapley values from economic game theory. These local (i.e., sample-specific) attributions can be summarized into a global feature importance measure by taking a mean over the samples. TreeSHAP \citep{lundberg2020local} is a computationally-efficient implementation of SHAP values for tree-based methods. 
Furthermore, other less well-known feature importance methods have been proposed \citep{ishwaran2007variable,ishwaran2010high,kazemitabar2017variable,saabas,sutera2021global,klusowski2021nonparametric}.
None of these methods account for the intrinsic instability and biases of RF.
Also related are model-agnostic conditional dependence scores \citep{azadkia2021simple,zhang2020floodgate}, leave-one-covariate-out inferences \citep{lei2018distribution, rinaldo2019bootstrapping, gan2022inference}, and importance measures for interactions, such as via iterative random forests that have been widely adopted in the genomics community \citep{basu2018iterative,behr2020learning}. 

\section{A Linear Regression Perspective of Decision Trees}
\label{sec:new_perspective_on_mdi}

The starting point of this work builds upon an interesting connection between decision trees and linear models, which has been used in previous work \citep[e.g., see][]{klusowski2024large, agarwal2022hierarchical} but is not widely known.
To formalize and highlight this connection, let us introduce notation and background.

Assume we are given a dataset $\data = \braces*{\left(\bx_{i},y_{i}\right)}_{i=1}^{n}$, with covariates $\bx_i \in \R^{p}$ and responses $y_i \in \R$.
A decision tree is simply a collection of binary splits that recursively partition the covariate space into disjoint regions, or nodes. 
More formally, a decision tree $\splits$ can be represented by its \textit{tree structure}, or its collection of splits $\splits = \braces*{s_{1}, \ldots, s_{m}}$, where each \emph{split} $s \in \splits$ of a \textit{node} $\node$ partitions it into two children nodes $\node_L = \braces{\bx_0 \in \node \colon x_{0,k} \leq \tau}$ and $\node_R = \braces{\bx_0 \in \node \colon x_{0,k} > \tau}$ for some feature index $k$ and threshold $\tau$.\footnote{Some decision tree algorithms \citep{murthy1994system} allow for multivariate splits that depend on multiple features simultaneously. These algorithms can still be covered under our framing by performing the appropriate feature engineering and augmenting these features that encode the multivariate splits to $\bX$.}
Different tree construction algorithms choose these splits to optimize different criteria.
For example, Classification and Regression Trees (CART) \citep{breiman1984classification}, arguably the most popular decision tree algorithm, grows a tree by selecting the feature and threshold that minimize the \textit{impurity decrease} of $s$, defined as
\begin{equation} \label{eq:impurity_dec}
\hat{\Delta}(s, \data) \coloneqq N\left(\node\right)^{-1}\paren*{\sum_{\bx_i \in \node}\left(y_{i} - \bar{y}_{\node}\right)^{2} - \sum_{\bx_i \in \node_{L}}\left(y_{i}- \bar{y}_{\node_{L}} \right)^{2} - \sum_{\bx_i \in \node_{R}}\left(y_{i}- \bar{y}_{\node_{R}} \right)^{2}},
\end{equation}
where all summations are over samples in $\data$, $N(\node)$ represents the number of samples in node $\node$, and $\bar{y}_{\node}, \bar{y}_{\node_L}, \bar{y}_{\node_R}$ are the mean responses in each node.
An RF is an ensemble of CARTs that are fitted independently of one another on bootstrapped versions $\data^*$ of $\data$.

To measure the importance of each feature in CART as well as other tree-based algorithms (e.g., XGBoost \citep{chen2016xgboost}), the mean decrease in impurity (MDI) is often computed.
Specifically, for each $k = 1, \ldots, p$, the MDI of feature $X_{k}$ is defined as the weighted average of impurity decreases across all splits that split on $X_{k}$:
\begin{equation}\label{eq:MDI_single_tree}
    \mdi{\splits,\data^*} \coloneqq \sum_{s \in \splits^{(k)}} n^{-1}N\left(\node(s)\right)\hat{\Delta}(s, \data^*),
\end{equation}
where $\splits^{(k)} \subseteq \splits$ is the subset of splits that split on the feature $X_{k}$ and $\node(s)$ is the node being split by a split $s$. 
For RFs, the MDI of feature $X_{k}$ is the mean MDI across all CART models in the ensemble. 

\subsection{Decision Trees as a Linear Regression Problem} 
To build the connection between a decision tree $\splits$ and a linear model, we associate to each split $s \in \splits$ the local decision stump function
\begin{equation} \label{eq:local_decision_stump}
    \psi(\bx;s,\data) = \frac{N\left(\node_{R}\right)\mathbf{1}\braces*{\bx \in \node_{L}} - N\left(\node_{L}\right)\mathbf{1}\braces*{\bx \in \node_{R}}}{\sqrt{N\left(\node_{L}\right)N\left(\node_{R}\right)}}.
\end{equation}
Intuitively, $\psi$ is a tri-valued function that indicates whether the sample $\bx$ lies to the left of the threshold, lies to the right of the threshold, or is not contained in node $\node$ at all. 
If $\splits$ has $m$ splits, concatenating these $m$ functions yields the learned feature map $\Psi(\bx;\splits,\data) \coloneqq \left(\psi(\bx;s_{1},\data), \ldots, \psi(\bx;s_{m},\data)\right)$ 
and its corresponding transformed dataset $\Psi\left(\bX;\splits,\data\right) \in \R^{n \times m}$, where $\bX \in \R^{n \times p}$ is the original data matrix. 

Given a decision tree whose model predictions are made by averaging the responses in each leaf node (as in typically the case, e.g., in CART), \citet{klusowski2024large} showed that the decision tree model predictions are equivalent to the predictions from the best fit linear model of the responses $\by$ on the transformed dataset $\Psi\left(\bX; \splits,\data\right)$ (see Lemma 3.1 in a previous version of the work \citep{klusowski2021universal}).
Due to its importance, we make explicit and repeat this statement from the literature in Proposition~\ref{prop:CART_linear}. 
The formal proof is provided in Appendix~\ref{supp:proof_CART_linear}.

\begin{proposition}
    \label{prop:CART_linear}
    Let $\hat{f}$ denote the CART model obtained from splits $\splits$ and data $\data$.
    Let $\hat\bbeta \coloneqq (\hat\beta_1,\ldots,\hat\beta_p)$ be the OLS coefficients obtained when regressing $\by$ on $\Psi\left(\bX;\splits,\data\right)$, and let $\hat\alpha$ be the intercept term.
    Then for any query point $\bx$, we have $\hat f(\bx) = \hat \alpha + \hat\bbeta^T\Psi(\bx;\splits,\data)$.
\end{proposition}

\begin{remark}
    There is a more well-known relationship between decision trees and linear regression models fitted on features indicating the terminal nodes of the tree \citep[e.g., see][]{golea1997generalization, saha2023random}.
We highlight that this is different from Proposition~\ref{prop:CART_linear}, which revolves around a linear regression fit on features engineered from the interior nodes (or splits) of the tree.
The interior node representation used here is more closely related to the work of \citet{friedman2008predictive}, wherein they fit a LASSO model on these interior node features (and possibly other features) to create a new class of rule-based prediction models.
However, \citet{friedman2008predictive} do not explicitly recognize the equivalence between a decision tree model and linear regression, and while Proposition~\ref{prop:CART_linear} fits a linear regression model using only the interior splits of a single tree, \citet{friedman2008predictive} fit a LASSO model using the interior splits from typically many trees (e.g., from an RF).
\end{remark}

\subsection{Mean Decrease in Impurity Importance as an $R^2$}

We further exploit this linear regression perspective to reinterpret the MDI feature importance measure. 
To this end, note that there is a natural partition of decision stumps according to the (original) feature that they split on.
That is, $\splits = \splits^{(1)} \sqcup \cdots \sqcup \splits^{(p)}$, and thus, we can write $\Psi(\bx;\splits,\data) = \paren*{\Psi(\bx;\splits^{(1)},\data), \ldots, \Psi(\bx;\splits^{(p)},\data)}$.
These blocks of $\Psi(\bX;\splits,\data)$ are orthogonal, enabling us to extend Proposition~\ref{prop:CART_linear} and derive the following new connection between MDI and $\rsq$ values.
\begin{theorem}
\label{thm:MDI_r2_equivalence}
Assume we have a tree structure $\splits$ and a dataset $\data = \paren*{\bX, \by}$, with $\bX$ denoting the matrix of covariates and $\by$ denoting the response vector.
For any feature $X_{k}$, we have the following identity:
\begin{equation}
\label{eq:MDI_r2_equivalence}
    \frac{\mdi{\splits,\data}}{n^{-1}\sum_{i=1}^n \paren*{y_i - \bar y}^2} = 1 - \frac{\sum_{i=1}^n \paren*{y_i - \hat y_i^{(k)}}^2}{\sum_{i=1}^n\paren*{y_i - \bar y}^2} \eqqcolon \rsq(\by, \hat \by^{(k)}),
\end{equation}
where $\hat \by^{(k)} = \paren*{\hat y_1^{(k)}, \hat y_2^{(k)}, \ldots, \hat y_n^{(k)}}$ is the vector of fitted response values when regressing $\by \sim \Psi(\bX; \splits^{(k)},\data)$.\footnote{If $\splits^{(k)} = \emptyset$, the fitted model is the constant intercept model, and both the MDI and the $\rsq$ value of this regression are equal to 0.}
\end{theorem}

Here, the fitted values $\hat{\by}^{(k)}$, or \textit{partial model predictions}, are the resulting model predictions using only those decision stumps that split on $X_k$.
Theorem~\ref{thm:MDI_r2_equivalence} thus formalizes the intuition that a more important feature $X_k$ leads to more accurate predictions based solely on $X_k$ and hence a larger MDI.
We also note that the partial model predictions $\hat{\by}^{(k)}$ are precisely the \citet{saabas} local feature importance scores for $X_k$.
As such, Theorem~\ref{thm:MDI_r2_equivalence} is implied by but also helps to clarify Proposition 1 in \citet{li2019debiased}, which equates the MDI of $X_{k}$ to the sample covariance between the Saabas scores and the responses but does not derive this in a linear regression setting.

\subsection{Rewriting RF and MDI via Linear Regression} \label{sec:reinterpreting_mdi}

As a direct consequence of Proposition~\ref{prop:CART_linear} and Theorem~\ref{thm:MDI_r2_equivalence}, we can rewrite the computation of a tree from an RF and its MDI feature importance via the following procedure (Figure~\ref{fig:gmdi}, left).

\vspace{0.5em}
\noindent Given a pre-fitted (fixed) tree, which was trained on a bootstrapped dataset $\data^* = (\bX^*, \by^*)$ and has tree structure $\splits = \splits(\data^*)$:

\vspace{0.5em}
\begin{adjustwidth}{0.75cm}{}
\noindent \textit{Step 1: Obtain transformed dataset on in-bag samples.} Construct the feature map $\Psi(\bx;\splits,\data^*)$ and use it to obtain the transformed in-bag dataset: 
$$\Psi(\bX^*;\splits,\data^*) = [\Psi(\bX^*; \splits^{(1)},\data^*), \ldots, \Psi(\bX^*; \splits^{(p)},\data^*)].$$
For ease of notation, we will henceforth suppress the dependence on $\splits$ and $\data^*$ when describing the MDI procedure, and denote $\Psi^k(-) = \Psi(-; \splits^{(k)},\data^*)$.

\noindent \textit{Step 2: Fit linear model.} Fit an OLS model for $\by$ on $\Psi(\bX^*)$ and obtain the estimated regression coefficients $\hat{\bbeta}$. 
\end{adjustwidth}

\vspace{0.5em}
\noindent By Proposition~\ref{prop:CART_linear}, this fitted OLS model is equivalent to the original tree model. An RF model is thus an ensemble of these linear regression models from Step 2. 

\vspace{0.5em}
\noindent To evaluate the MDI of a feature $X_k$ in a tree from RF, we can continue with the following steps:

\vspace{0.5em}
\begin{adjustwidth}{0.75cm}{}
\noindent \textit{Step 3: Make partial model predictions.} 
For each $k = 1, \ldots, p$, obtain the partial model predictions $\hat{\by}^{(k)}$, as defined in Theorem 2. However, note that due to the orthogonality and centering of $\Psi(\bX^*)$, $\hat{\by}^{(k)}$ as defined in Theorem \ref{thm:MDI_r2_equivalence} is equivalent to imputing the mean value for all stump features in $\Psi(\bX^*)$ that do not split on $X_k$ and then multiplying this modified matrix by $\hat{\bbeta}$. 
Formally, compute $\hat{\by}^{(k)}$ via:
$$
\hat{\by}^{(k)} = [\bar{\Psi}^1(\bX^*), \ldots, \bar{\Psi}^{k-1}(\bX^*), \Psi^k(\bX^*), \bar{\Psi}^{k+1}(\bX^*), \ldots, \bar{\Psi}^p(\bX^*)] \hat{\bbeta},
$$
where $\bar{\Psi}^j(\bX^*) = \frac{1}{n}\mathbf{1}_n\mathbf{1}_n^T\Psi^j(\bX^*)$ and $\mathbf{1}_n$ is an $n \times 1$ vector of all ones.

\noindent \textit{Step 4: Evaluate partial model predictions via $\rsq$.} For each $k = 1, \ldots, p$, the MDI for feature $k$ is precisely the unnormalized $\rsq$ value between the observed responses $\by$ and the partial model predictions $\hat{\by}^{(k)}$, as shown by Theorem~\ref{thm:MDI_r2_equivalence}.
\end{adjustwidth}

\vspace{0.5em}
\noindent For an RF, the MDI of a feature $X_k$ is the average $\rsq$ value across all trees from Step 4.

\subsection{Drawbacks of Decision Trees and MDI}
\label{sec:drawbacks}

Our new linear regression lens allows us to explain known drawbacks of decision trees and MDI as well as reveal new challenges. 
We consolidate these into two main issues --- model mismatch and overfitting --- and discuss ways to overcome them. 
These issues and their potential solutions serve as inspiration for our proposed \rfmethod\;and \method\;frameworks in Section~\ref{sec:mdi_plus}.

\subsubsection{Model Mismatch} 
First, though RF is a nonparametric method that can in principle approximate any functional structure, it comes with a set of inductive biases that allow it to adapt better to some types of structure rather than others.
Our linear reinterpretation of decision trees and MDI highlights two important inductive biases:
\begin{enumerate}[label=(\roman*)]
    \item Trees are trained to learn solely from local decision stump features, which are known to be statistically inefficient at estimating smoothly-varying functions \citep{tsybakov2004introduction} and additive generative models \citep{tan2021cautionary} --- both ubiquitous properties in real-world datasets \citep{hastie1986generalized}.
    \item Furthermore, the fitted tree is inherently a linear regression model while MDI is inherently an $\rsq$ value. However, OLS and $\rsq$ may not be well-suited for all problem structures (e.g., in situations with a categorical response or gross outliers).
\end{enumerate}

\paragraph{How to mitigate model mismatch.} 
Augmenting the local decision stump features with additional (possibly engineered) features can help to alleviate the inductive biases of RFs against additive or smooth structures. 
In particular, augmenting the original $\bX$ data matrix itself can always be done and can help bridge the gap between RFs and linear models.
Moreover, allowing for more flexible prediction models and evaluation metrics in place of OLS and $\rsq$ can help to mitigate the second inductive bias discussed above. 
For example, logistic regression with negative log-loss or Huber regression with negative Huber loss might be more appropriate for classification problems or problems with outliers, respectively.

\subsubsection{Overfitting}\label{subsec:overfitting}

Second, Section~\ref{sec:reinterpreting_mdi} reveals that MDI is implicitly evaluated as a prediction $\rsq$ metric using the same data that was used to train the decision tree model. 
Such ``data snooping'' inevitably leads to overfitting and hence, poor feature importance estimates.
Additional discussion on this overfitting issue and how it is further exacerbated by known biases of the tree-growing process and MDI is provided in Appendix~\ref{app:drawbacks}.

\paragraph{How to mitigate overfitting.} 
In prediction modeling, incorporating regularization in the linear regression model is commonly used to prevent overfitting.
Moreover, sample-splitting approaches (e.g., OOB sampling or leave-one-out (LOO) cross-validation) are crucial to separate the data used for model training and that used for proper evaluation of its generalization error.
Leave-one-out (LOO) cross-validation, in particular, enables the most efficient use of samples, leveraging the entire dataset (in-bag and OOB samples) to both learn the linear model and estimate its out-of-sample error. 
\section{Introducing \rfmethod~and \method}
\label{sec:mdi_plus}

Building upon the ideas and challenges discussed in the previous section, we propose an improved generalization of RFs and MDI, which we call \rfmethod~and \method, respectively.
The proposed approach provides researchers with a highly flexible framework for both prediction and computing feature importances.
In particular, \rfmethod~and \method~allow for a wide range of modeling choices (e.g., the choice of feature augmentation, generalized linear model, and evaluation metric) that can be easily tailored to the data or problem structure at hand.
We first introduce the \rfmethod~and \method~framework given some fixed modeling choices in Section~\ref{subsec:mdi_plus_framework}. 
We then discuss how practitioners can make informed modeling choices using the PCS framework in Section~\ref{subsec:pcs_model_selection}.
Code for \rfmethod~and \method~are provided in a full-fledged python package on Github.%
\footnote{
\if1\blind{\rfmethod~and \method~are integrated into the imodels package \href{https://github.com/csinva/imodels}{\faGithub\,github.com/csinva/imodels}~\citep{singh2021imodels}.}\fi
\if0\blind{Link redacted.}\fi
}

\subsection{The \rfmethod~and \method~Framework}\label{subsec:mdi_plus_framework}
\begin{figure}[h]
    \centering
    \includegraphics[width=.85\textwidth]{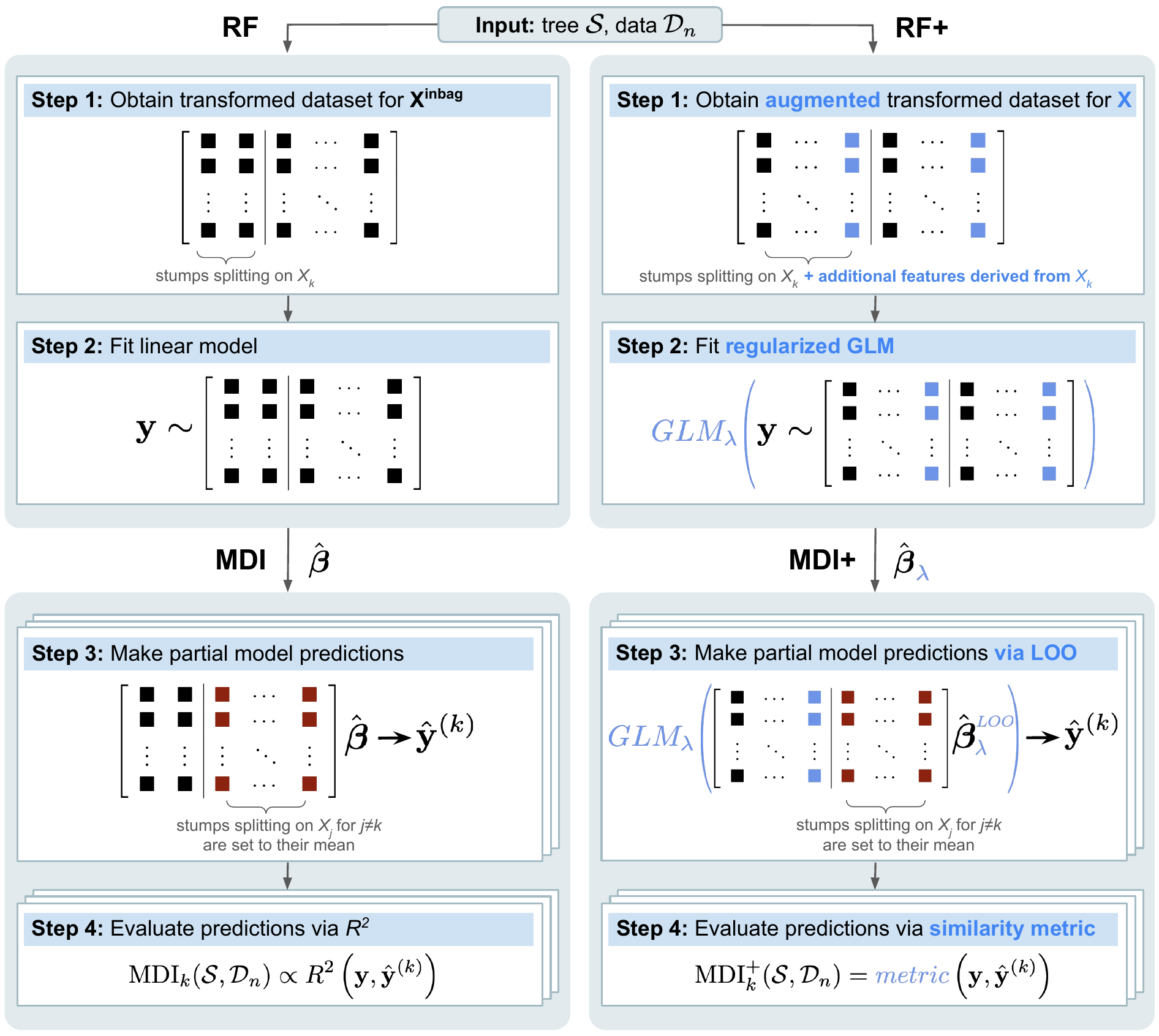}
    \caption{Overview of \method\;for a single tree. For each tree $\mathcal{S}$ in the random forest, \textbf{Step 1:} Obtain the transformed dataset on the in- and out-of-bag samples using stumps from the tree and append the raw and/or
    any additional (possibly engineered) features. \textbf{Step 2:} Fit a regularized GLM. \textbf{Step 3:} Using the fitted GLM, make partial model predictions $\hat{\by}^{(k)}$ for each feature $k = 1, \ldots, p$ (stacked boxes) using a leave-one-out (LOO) data splitting scheme. \textbf{Step 4:} For each $k = 1, \ldots, p$, evaluate partial model predictions via any user-defined similarity metric to obtain the \method\;for feature $k$ in tree $\mathcal{S}$.}
    \label{fig:gmdi} 
\end{figure}

In what follows, we introduce \rfmethod~and \method~for a single tree (Figure~\ref{fig:gmdi}, right) using an analogous scaffolding as RF and MDI (see Section~\ref{sec:reinterpreting_mdi}) but with the main differences highlighted in bold.

\vspace{0.5em}
\noindent Given a pre-fitted (fixed) tree, which was trained on a bootstrapped dataset $\data^*$ and has tree structure $\splits = \splits(\data^*)$, a tree in \rfmethod~is computed as follows:

\begin{adjustwidth}{0.75cm}{}
\noindent \textit{Step 1: Obtain \textbf{augmented} transformed dataset on \textbf{in- and out-of-bag} samples.} For features $k = 1, \ldots, p$, construct an augmented feature map $\tilde\Psi(\bx;\ \splits^{(k)},\data^*)$ by appending the raw feature $X_k$ to the feature map $\Psi(\bx; \splits^{(k)}, \data^*)$.\footnote{We append the raw feature $X_k$ as a default choice. More generally, we can augment the feature map with any engineered feature derived from $X_k$.}
That is, let $\tilde{\Psi}(\bx; \ \mathcal{S}^{(k)}, \data^*) = [\Psi(\bx; \ \mathcal{S}^{(k)}, \data^*), \ x_k]$ if $\splits^{(k)} \neq \emptyset$. 
Denote $\tilde{\Psi}(\bx;  \ \mathcal{S}, \ \data^*) = \left(\tilde{\Psi}(\bx; \ \mathcal{S}^{(1)},\  \data^*), \ldots, \tilde{\Psi}(\bx; \ \mathcal{S}^{(p)}, \ \data^*)  \right)$, and apply $\tilde{\Psi}(\bx; \ \mathcal{S}, \ \data^*)$ to the entire dataset (in- and out-of-bag samples) to obtain $\tilde{\Psi}(\bX; \ \mathcal{S}, \ \data^*)$.
Henceforth, we denote $\tilde{\Psi}(\bX) = \tilde{\Psi}(\bX; \ \mathcal{S}, \ \data^*)$ and $\tilde{\Psi}^{(k)}(\bX) = \tilde{\Psi}(\bX; \ \mathcal{S}^{(k)}; \  \data^*)$.

\noindent \textit{Step 2: Fit \textbf{regularized generalized} linear model (GLM).} Fit a regularized GLM $\mathcal{M}$ with link function $g$ and penalty parameter $\lambda$ for $\by$ using $\tilde{\Psi}(\bX)$ to obtain the estimated regression coefficients $\hat\bbeta_{\lambda}$ and intercept $\hat \alpha_\lambda$. 
We tune $\lambda$ via the approximate leave-one-out (LOO) method described in \citet{rad2020scalable}, which does not require re-fitting the GLM $n$ times. 
Note that we use the full dataset $(\bX, \ \by)$ comprising both in- and out-of-bag samples to perform this as well as subsequent steps. 
\end{adjustwidth}

\vspace{0.5em}
\noindent We define an \rfmethod~model as the ensemble of regularized generalized linear models from Step 2, each trained using a different bootstrapped dataset $\data^*$. 

\vspace{0.5em}
\noindent To evaluate the importance of a feature $X_k$ in a tree from \rfmethod, we continue with the following:

\vspace{0.5em}
\begin{adjustwidth}{0.75cm}{}
\noindent \textit{Step 3: Make partial model prediction via \textbf{LOO}.} For a feature $X_k$, impute the mean value for all features in $\tilde{\Psi}(\bX)$ not derived from $X_k$.
Then, for each feature $k = 1, \ldots, p$, and each sample $i = 1, \ldots, n$, obtain the LOO partial model predictions:
\begin{equation}
\label{eq:partial_loo_prediction}
\hat{y}_i^{k} = g^{-1}\left(
\left[ \bar{\tilde{\Psi}}^1(\bX), \ldots, \bar{\tilde{\Psi}}^{k-1}(\bX), \tilde{\Psi}^k(\bX), \bar{\tilde{\Psi}}^{k+1}(\bX), \ldots, \bar{\tilde{\Psi}}^p(\bX) \right] \hat{\bbeta}_{-i,\lambda} + \alpha_\lambda
\right),
\end{equation}
where $\bar{\tilde{\Psi}}^j(\bX) = \frac{1}{n}\mathbf{1}_n\mathbf{1}_n^T\tilde{\Psi}^j(\bX)$, $\mathbf{1}_n$ is an $n \times 1$ vector of all ones, and $\hat{\bbeta}_{-i,\lambda}$ is the LOO coefficient vector learned without using sample $\bx_i$. 
Again, we use the approximate LOO method of \citet{rad2020scalable} to obtain $\hat{\bbeta}_{-i,\lambda}$ without refitting the GLM.  Denote the LOO partial model predictions for feature $k$ by $\hat{\by}^{(k)} = \left(\hat{y}_1^{k}, \ldots, \  \hat{y}_n^{k}\right)$. 

\noindent \textit{Step 4: Evaluate partial model predictions via \textbf{similarity metric}.} 
Pick a similarity metric $m$ and use it to evaluate the similarity\footnote{The metric should attain larger values on closer arguments.} between the true responses and the partial model predictions. That is, for each $k= 1,\ldots, p$, we define the \method\;for feature $k$ as
\begin{equation}
\label{eq:GMDI_def}
    \mdip{\mathcal{S},\data^*,\tilde\Psi, \mathcal{M}, m} \coloneqq m\left(\by, \hat{\by}^{(k)} \right),
\end{equation}
where $\tilde\Psi$, $\mathcal{M}$ and $m$ represent the choices of augmented feature representation, GLM, and similarity metric, respectively.
\end{adjustwidth}

\vspace{0.5em}
\noindent For an \rfmethod, the \method\;of a feature $k$ is defined as the average of \eqref{eq:GMDI_def} across all trees in the ensemble.\footnote{If a feature is never split on amongst all trees, we set its \method\;to be $-\infty$.}

\vspace{0.5em}
At its core, our proposed \rfmethod~and \method~framework expands upon the linear regression perspective of decision trees and MDI while incorporating specific improvements to remedy the drawbacks described in Section~\ref{sec:drawbacks}.
First, augmenting the raw features $\bX$ (in Step 1) helps to alleviate the inductive bias of RFs and MDI against additive or smooth structures. 
Secondly, using LOO sample-splitting\footnote{OOB sample splitting could have alternatively been used. We however found that LOO sample splitting often outperformed OOB sample splitting in practice.} (in Steps 3 and 4) to evaluate the partial model predictions mitigates overfitting issues that arise from using the same data to both learn the linear model and evaluate its $R^2$ in MDI.
Adding regularization to the GLM (in Step 2) similarly helps to mitigate overfitting and to stabilize the GLM model --- a new and important need given the likely correlated and high-dimensional features in the augmented feature map. 
Lastly, generalizing the linear model and $R^2$ metric to the broader class of GLMs and evaluation metrics (in Steps 2 and 4) provides an invaluable opportunity to tailor the prediction model and the feature importance computation to specific problem structures.

We will later show through extensive simulations that each of these modifications serves a distinctly unique purpose in improving the overall feature importance rankings from \method~(see Section~\ref{sec:sims} and Appendix~\ref{supp:modeling_choices}). 
We will also demonstrate the superior predictive power of \rfmethod~compared to traditional RFs across a variety of real-world datasets in the regression and classification settings (see Section~\ref{sec:prediction_accuracy}).
The strong prediction performance of \rfmethod\;suggests that it better captures the underlying data-generating process observed in reality, thereby providing a powerful new prediction model and giving additional credence to \method\;\citep{murdoch2019definitions, yu2020veridical}.

\subsection{PCS-Informed Model Recommendation}\label{subsec:pcs_model_selection}

A key advantage of the \rfmethod~and \method~framework is its flexibility, enabling the practitioner to tailor the choice of feature augmentation $\tilde\Psi$, regularized GLM $\mathcal{M}$, and similarity metric $m$ to best reflect their prior knowledge and problem structure under study.
For example, using robust regression (e.g., Huber regression) and a robust similarity metric (e.g., negative Huber loss) can improve the performance in the presence of gross measurement errors or outliers (see Section~\ref{sec:sims}).
However, one rarely has sufficient prior knowledge of the problem structure in order to make definitive choices for $\tilde\Psi$, $\mathcal{M}$ and $m$. 
Further, even if there is prior information, a practitioner might be left with multiple choices (e.g., using lasso or ridge regularization in $\mathcal{M}$). 
To address this challenge, we briefly discuss two approaches inspired by the PCS framework \citep{yu2020veridical, yu2024veridical} to help practitioners make these modeling choices in a data-driven manner, and defer details to  Appendix \ref{supp:PCS_model}.

\paragraph{Stability-based selection for  $\tilde\Psi$, $\mathcal{M}$, $m$.} Let $h = \left(\tilde{\Psi}(\bX), \mathcal{M}, m\right)$ denote an \rfmethod/\method\;model defined by the choices of augmented feature representation, GLM, and similarity metric, respectively. 
Accordingly, let $\mathcal{H} = \{h_1, \ldots, h_N\}$ denote the set of possible \rfmethod/\method\;models under consideration.
This first approach selects the \rfmethod/\method\;model that both fits the data well (as measured via prediction performance) and yields the most stable feature importance rankings, described as follows:

\noindent\textit{Step 1: Prediction Screening.} 
For each $h \in \mathcal{H}$, evaluate the prediction performance of $\mathcal{M}$ on a held-out test set, and filter out all $h$ whose prediction performance is worse than that of RF. 
This prediction check ensures that the \rfmethod~model is a reasonably faithful approximation of the underlying data generating process prior to interpreting it \citep{murdoch2019definitions, yu2020veridical}.

\noindent\textit{Step 2: Stability Selection.}
Generate $B$ bootstrapped samples of the fitted trees in the ensemble. 
For each $h$ that passed the prediction screening, evaluate \method\;for each of the $B$ bootstrapped samples of trees, and choose the $h$ which yields the most similar (or stable) feature rankings across the $B$ bootstraps. 
We measure the similarity between different bootstrap samples via Rank-based Overlap \citep{webber2010similarity}. 
While we measure stability via bootstrap sampling as a default, one can also measure stability over different algorithmic perturbations (e.g., random seeds used to train the RF).

\paragraph{Model aggregation via PCS principles.}
As an alternative to the stability-based selection approach which selects a single \rfmethod/\method\;model, it may be sensible to ensemble multiple models that have similar predictive performance (e.g., averaging the predictions and/or feature importance rankings across all $h \in \mathcal{H}$ that passed the prediction screening step described above). 
This model aggregation procedure is motivated by the philosophy that different models can provide competing but equally valid descriptions of reality, and by bringing together these different perspectives, we may be able to obtain more robust and reliable conclusions \citep{murdoch2019definitions, rudin2021interpretable}.

\vspace{0.5em}

\noindent In Appendix \ref{supp:PCS_model}, we conduct a simulation study to establish the efficacy of both approaches. 
There, we show that the GLM and metric selected by the stability-based procedure leads to the best feature ranking performance across different candidate \method\;models. 
While the model aggregation approach does not lead to the best feature ranking performance, it performs competitively, and we present it here since both approaches might be useful in practice, depending on the problem goal or end usage.

\section{\rfmethod\; Prediction Performance}
\label{sec:prediction_accuracy}

\begin{figure}[h!]
    \centering
    \begin{tabularx}{1\textwidth}{@{}l *5{>{\centering\arraybackslash}X}@{}}
    \begin{minipage}{.11\textwidth}
    \phantom{}
    \end{minipage}%
    {\small \textbf{(A) Regression}} & {\small \hspace{10mm} \textbf{(B) Classification}}
    \end{tabularx}\\
    \vspace{1mm}
    \includegraphics[width=.9\textwidth]{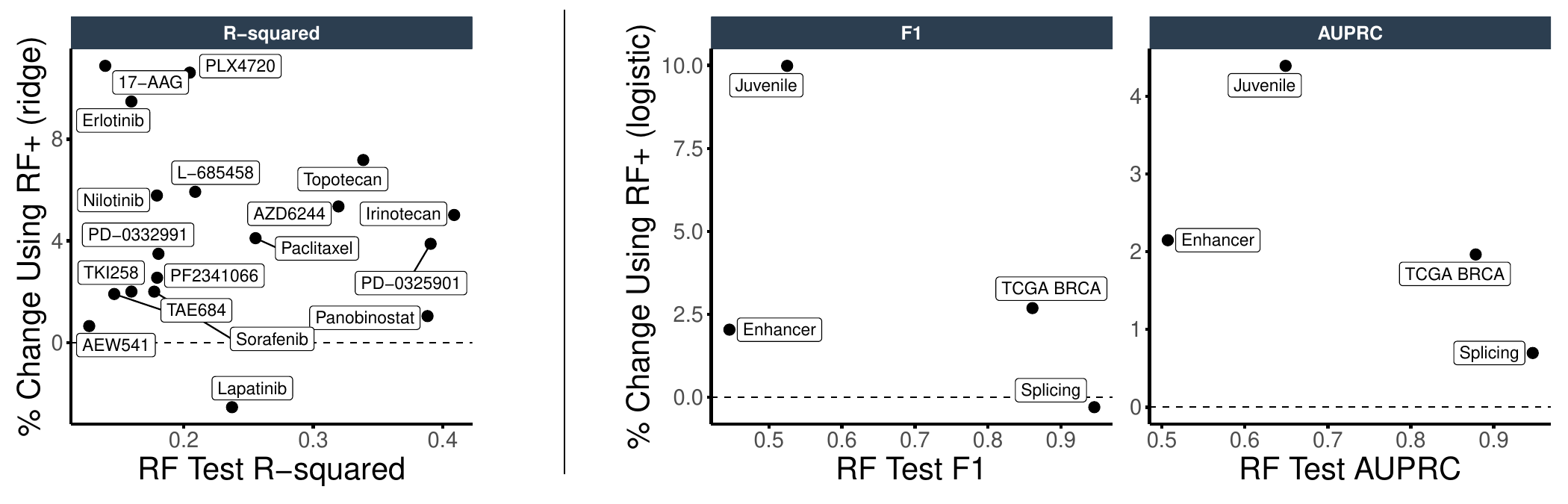}
    \caption{Relative performance of \rfmethod\;(ridge) as compared to RF in both the (A) regression and (B) classification settings. In the regression setting, \rfmethod\;(ridge) increases performance by an average of 4.4\% averaged across all 18 drugs that have test set $\rsq > 0.1$. In the classification setting, \rfmethod\;(logistic) increases performance according to the $F1$ score for three of the four datasets, and on average by $3.6\%$.  \rfmethod\;(logistic) increases AUPRC for all datasets, and on average by $2.3\%$.}
    \label{fig:prediction}
\end{figure}

In this section, we show that \rfmethod\;increases prediction performance over RF for various real-world datasets, spanning regression, binary classification, and multi-class classification tasks (see Table~\ref{tab:datasets} for list of considered datasets). 
Specifically, for regression tasks, we compared the test $\rsq$ performance between RF and \rfmethod~using ridge regression as the GLM and the $\rsq$ metric. 
For classification tasks, we compared the test $F1$ score and area under the precision-recall curve (AUPRC) between RF and \rfmethod~using ridge-regularized logistic regression as the GLM and the log-loss metric.
Figure~\ref{fig:prediction} summarizes the test prediction performance results, averaged across 32 different train-test splits using an 80\%-20\% split.

Of the 24 regression tasks, \rfmethod~yielded a higher average test $\rsq$ than RF for 23 tasks (see Figure~\ref{fig:prediction_supp} in Appendix~\ref{supp:prediction_accuracy}).
In Figure~\ref{fig:prediction}(A), we present only the results for regression tasks where RF yielded an average test set $\rsq > 0.1$, so as to focus on models that have meaningful and better than random guessing predictive power. 
Here, \rfmethod~increased the test $\rsq$ for 17 out of 18 regression tasks with an average percent change increase of $4.4\%$ relative to RF.

Similarly, \rfmethod~improved the $F1$ score for three of the four classification datasets and yielded approximately the same $F1$ score for the splicing dataset, as shown in Figure~\ref{fig:prediction}(B).
This corresponded to an average percent change increase of $3.6\%$ in the $F1$ score.
With respect to the AUPRC score, \rfmethod~increased the AUPRC for all four classification datasets and yielded an average percent change increase of $2.3\%$ relative to RF. 

Many notable works \citep{murdoch2019definitions, yu2020veridical} have advocated that strong prediction performance is a necessary prerequisite for obtaining reliable and trustworthy interpretations.
The superior prediction performance of \rfmethod~compared to RF thus not only demonstrates the utility of \rfmethod~as a powerful prediction model but more importantly, helps to justify and give credance to the use of \rfmethod~as a building block to obtain more reliable feature importances.

\begin{table}[h]
    \centering
    \small
    \renewcommand{\arraystretch}{0.75} 
    \begin{tabular}[t]{lccc}
    \toprule
    \textbf{Dataset} & \textbf{\# Samples} & \textbf{\# Features} & \textbf{Task}\\
    \midrule
    CCLE \citep{barretina2012cancer} & 472 & 5000 & Regression (24 independent tasks)\\
    Juvenile \citep{osofsky1997effects} & 3640 & 277 & Binary Classification\\
    Enhancer \citep{basu2018iterative} & 7809 & 80 & Binary Classification\\
    Splicing \citep{basu2018iterative} & 23823 & 264 & Binary Classification\\
    TCGA Breast Cancer \citep{cancer2012comprehensive} & 1904 & 5000 & Multiclass Classification (5 classes)\\
    \bottomrule
    \end{tabular}
    \caption{Datasets used to assess the prediction performance of \rfmethod~and RF.}
    \label{tab:datasets}
\end{table}

\section{Data-Inspired Feature Ranking Simulations}
\label{sec:sims}
We next highlight the effectiveness of \method\;to identify signal features across simulations, covering three common settings:  regression, classification, and robust regression (i.e., presence of outliers). 
In all settings, we demonstrate the flexibility of \method\;for choosing an appropriate GLM and similarity metric for the problem at hand. 
In addition, for all simulations, we use covariate matrices coming from real-world datasets to capture naturally occurring structures. 
Using these covariate matrices, we simulate responses with analytical functions that depend on a sparse set of features. 
These response functions were chosen to reflect both canonical forms and domain knowledge.
We then measure the ability of \method\;and other feature importance methods to recover these signal features, with the specific metric for recovery discussed below. 
Additional simulations under varying sparsity levels, number of features, and misspecified model settings with omitted variables are provided in Appendix~\ref{supp:additional_feature_ranking}, further supporting the strong empirical performance of \method. 

\subsection{Simulation Setup}
\label{subsec:sim_set_up}
\paragraph{Real-world datasets used.}
For our covariate matrices, we use the following datasets: 
(i) Juvenile dataset ($n = 3640$, $p = 277$) \citep{osofsky1997effects}; 
(ii) a subset of the Cancer Cell Line Encyclopedia (CCLE) RNASeq gene expression dataset ($n = 472$, $p = 1000$) \citep{barretina2012cancer}; 
(iii) Enhancer dataset ($n = 7809$, $p = 41$) \citep{basu2018iterative}; and 
(iv) Splicing dataset ($n = 23823$, $p = 264$) \citep{basu2018iterative}. 
 
\paragraph{Simulated responses. } Using each dataset above as the covariate matrix $\bX$, we consider the following response functions:
\begin{enumerate}[itemsep=0.1em]
    \item Linear model:  $\E[Y ~| ~ X] = \sum_{j = 1}^{5}X_j$; 
    \item Locally-spiky-sparse (LSS) model \citep{behr2021provable}: $\E[Y ~| ~ X] = \sum_{m = 1}^{3} \mathbf{1}(X_{2m-1}>0)\mathbf{1}(X_{2m}>0)$; 
    \item Polynomial interaction model: $\E[Y ~| ~ X] = \sum^3_{m = 1} X_{2m-1} + \sum^3_{m = 1}X_{2m-1}X_{2m}$;
    \item Linear + LSS model: $\E[Y ~| ~ X] = \sum_{m = 1}^{3}X_{2m - 1} + \sum_{m = 1}^{3} \mathbf{1}(X_{2m-1}>0)\mathbf{1}(X_{2m}>0)$.
\end{enumerate}
These regression functions were chosen to reflect several archetypes of real DGPs. (1) and (3) reflect well-studied models in the statistics literature. (2) exhibits the discontinuous interactions observed in biological processes \citep{behr2021provable}.
(4) reflects a combination of linear and discontinuous interactions also thought to be prevalent in genomics. 
For our classification simulations, we pass the mean response function (i.e., $\E[Y ~| ~ X]$) through a logistic link function to generate binary responses.

\paragraph{Feature importance methods under consideration.} We compare \method\;to a number of popular feature importance methods for RFs: MDI \citep{breiman1984classification}, MDI-oob \citep{li2019debiased}, MDA \citep{breiman2001random}, and TreeSHAP \citep{lundberg2017unified}.

\paragraph{RF settings.} For the regression (Section \ref{subsec:regression_results}) and robust regression (Section \ref{subsec:robust_results}) settings, we train an RF regressor using \texttt{scikit-learn} \citep{pedregosa2011scikit} with \textit{n\_estimators}=$100$ (i.e., number of trees), \textit{max\_features}=$p/3$ (i.e., proportion of features subsampled at each node), and \textit{min\_samples\_leaf}=$5$ alongside other default parameters. For classification (Section \ref{subsec:robust_results}),  we use \texttt{scikit-learn}'s RF classifier with \textit{n\_estimators}=$100$, \textit{max\_features}=$\sqrt{p}$, and \textit{min\_samples\_leaf}=$1$ alongside other default parameters.

\paragraph{Feature Ranking Performance Metric.} As in previous work \citep{li2019debiased,yu2020veridical}, we evaluate the performance of each feature importance method by how well it can be used to classify features used in the regression function (signal) vs. those that are not (non-signal). 
Each set of feature importance scores induces a ranking of the features, which can then be evaluated for this classification problem using AUROC.
A high AUROC scores indicates that the signal features are ranked higher (i.e., more important) than the non-signal features. 
Performance results are averaged across 50 simulation replicates. 
In each simulation replicate, we choose the signal features randomly from $\bX$.

\subsection{Regression Simulations}
\label{subsec:regression_results}

We simulate responses as discussed above and introduce additive Gaussian noise: $Y = \E[Y ~|~ X] + \epsilon$, where $\epsilon \sim N(0,\sigma^2)$.
We examine performance across various signal-to-noise ratios as measured by the proportion of variance explained,\footnote{$PVE$ is a monotone transformation of the signal-to-noise ratio such that its range is bounded between 0 and 1 and thus more interpretable. It is also a standard measurement of noise used in many fields (e.g., it is often called \textit{heritability} in genomics). PVE in genomics is estimated to range from between 0.05 to 0.4 \citep{wang2020simple}.} defined as $PVE = \text{Var}\braces*{\E\braces*{Y|X}}/\text{Var}\braces{Y}$.
Specifically, we vary across $PVE$ in $\{0.1,\; 0.2,\; 0.4,\; 0.8\}$ (or equivalently, signal-to-noise ratio in $\{0.11,\; 0.25,\; 0.66,\; 4\}$) and across a range of sample sizes $n$ (i.e., $n \in \{100,\; 250,\; 472\}$ for CCLE and $n \in \{100,\; 250,\; 500,\; 1000,\; 1500\}$ for the other three datasets). 
For \method, we use ridge-regularized regression as the GLM, and $\rsq$ as our similarity metric of choice. 

\begin{figure}[htbp]
\begin{subfigure}{\textwidth}
    \centering
    \begin{tabularx}{1\textwidth}{@{}l *5{>{\centering\arraybackslash}X}@{}}
    \begin{minipage}{.075\textwidth}
    \phantom{}
    \end{minipage}%
    & {\small $PVE=0.1$} & {\small $PVE=0.2$} & {\small $PVE=0.4$} & {\small $PVE=0.8$}
    \end{tabularx}\\
    \vspace{-4pt}
    \hspace{2mm}
    \rotatebox{90}{{\hspace{12.5mm} \centering \small \textbf{LSS}}}
    \includegraphics[width=0.95\textwidth]{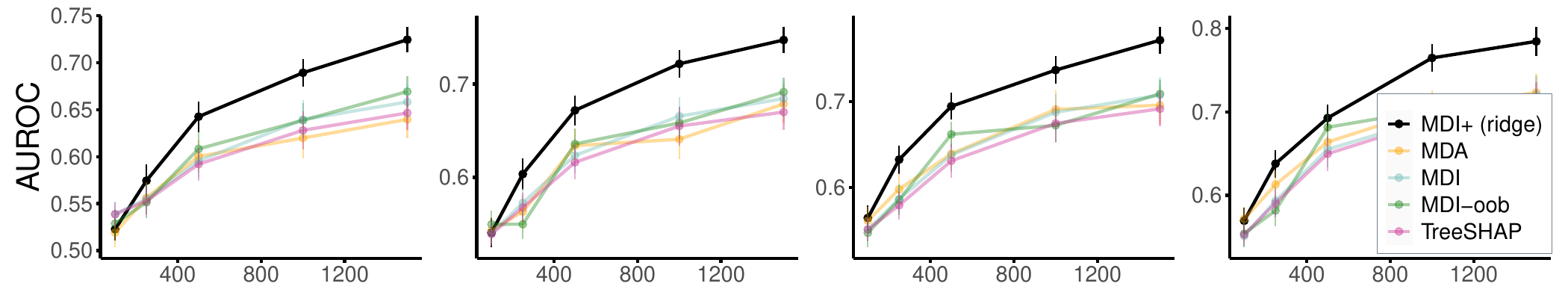}
\end{subfigure}
\begin{subfigure}{\textwidth}
    \centering
    \vspace{-2mm}
    \hspace{2mm}
    \rotatebox{90}{{\hspace{11mm} \centering \small \textbf{Poly. Int.}}}
    \includegraphics[width=0.95\textwidth]{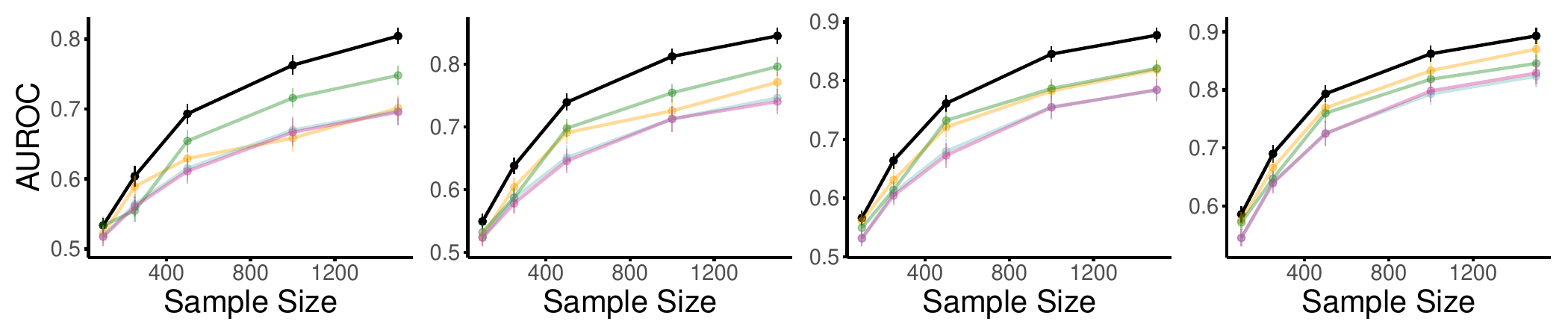}
\end{subfigure}
\caption{\method\;outperforms all other feature importance methods for the data-inspired regression simulations described in Section \ref{subsec:regression_results} using the Splicing dataset. This pattern is evident across various regression functions (specified by row), proportions of variance explained (specified by column), and sample sizes (on the $x$-axis). In all subplots, the AUROC has been averaged across 50 experimental replicates, and error bars represent $\pm$ 1SE.}
\label{fig:regression_sims}
\end{figure}

We display results for the Splicing dataset for the LSS and polynomial interaction model in Figure \ref{fig:regression_sims}. 
Results for other datasets and regression functions are similar and deferred to Appendix \ref{supp:experiments}. 
Across all simulation scenarios, \method\;produces more accurate feature rankings, often enjoying more than a 10\% improvement in AUROC over its closest competitor (typically MDI-oob or MDA).
In particular, \method\;produces the most significant improvement for low PVEs, which is especially important since real-world data in fields such as biology, medicine, and social sciences have low SNRs.
The strong performance of \method\;relies on a number of choices made to mitigate the aforementioned biases such as: using ridge regularization, including the raw feature, and evaluating predictions via LOO.
In Appendix \ref{supp:modeling_choices}, we provide simulations that show how each of these choices individually leads to an increase in the ranking accuracy of \method. 

\subsection{Classification Simulations}
\label{subsec:classification_results}
We simulate responses according to the response functions defined in Section \ref{subsec:sim_set_up}, and introduce noise by randomly flipping a percentage of the binary response labels to the opposite class. 
We measure AUROC as we vary the percentage of corrupted labels in $\{0\%,\; 5\%,\; 15\%,\; 25\% \}$ and the number of samples as before.
To tailor \method\;to the classification setting, we use ridge-regularized logistic regression and negative log-loss as our choice of GLM and similarity metric, respectively.
We compare these choices to those used in the regression setting (i.e., ridge regression and $\rsq$). 
Henceforth, we shall refer to these particular settings as \method\:(logistic) and \method\;(ridge), respectively. 

\begin{figure}[ht]
\begin{subfigure}{\textwidth}
    \centering
    \begin{tabularx}{1\textwidth}{@{}l *5{>{\centering\arraybackslash}X}@{}}
    \begin{minipage}{.075\textwidth}
    \phantom{}
    \end{minipage}%
    & {\small $25\%$ Corrupted} & {\small $15\%$ Corrupted} & {\small $5\%$ Corrupted} & {\small $0\%$ Corrupted}
    \end{tabularx}\\
    \vspace{-2pt}
    \hspace{2mm}
    \rotatebox{90}{{\hspace{9mm} \centering \small \textbf{Logistic}}}
    \includegraphics[width=0.9\textwidth]{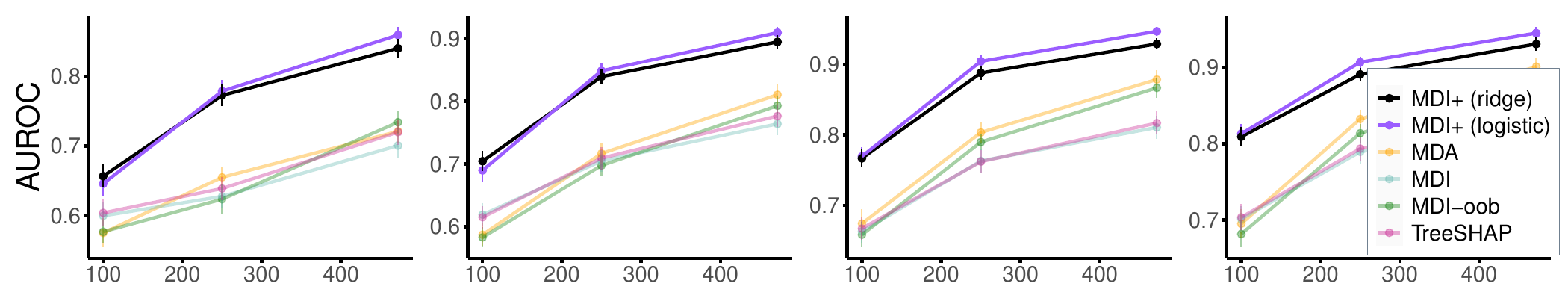}
\end{subfigure}
\begin{subfigure}{\textwidth}
    \centering
    \vspace{-1mm}
    \rotatebox{90}{{\hspace{12mm} \centering \small \textbf{Logistic}}}
    \rotatebox{90}{{\hspace{6mm} \centering \small \textbf{Linear + LSS}}}
    \includegraphics[width=0.9\textwidth]{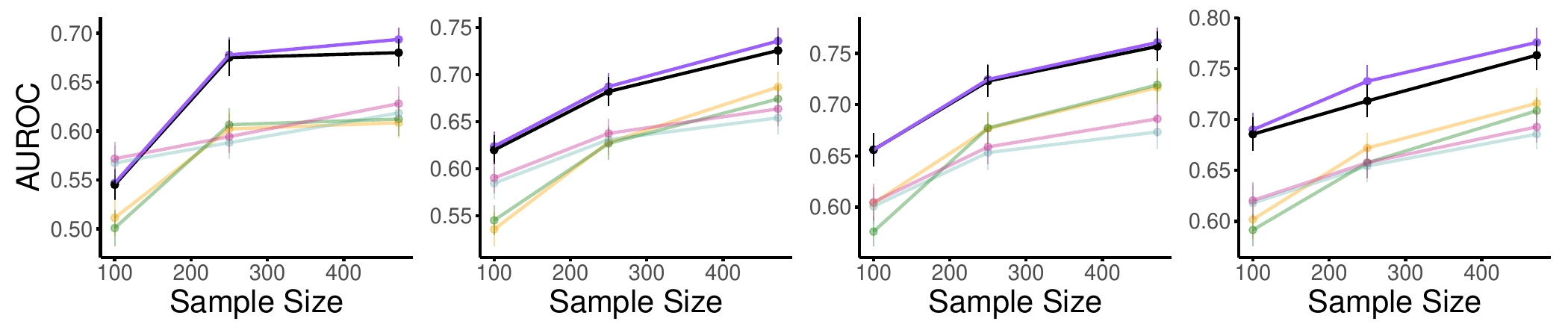}
\end{subfigure}
\caption{Both \method\;(ridge) and \method\;(logistic) outperform all other feature importance methods for the data-inspired classification simulations described in Section~\ref{subsec:classification_results} using the CCLE RNASeq dataset. Furthermore, \method\;(logistic) slightly outperforms \method\;(ridge) in the majority of settings, indicating the benefit of tailoring the choices of \method\;to the data at hand. This pattern is evident across various regression functions (specified by row), proportions of corrupted labels (specified by column), and sample sizes (on the $x$-axis). In all subplots, the AUROC has been averaged across 50 experimental replicates, and error bars represent $\pm$ 1SE.}
\label{fig:classification_sims}
\end{figure}

We display results for the CCLE dataset for the linear and linear + LSS model in Figure \ref{fig:classification_sims}.
Results for other datasets and regression functions are similar and deferred to Appendix \ref{supp:experiments}. 
Both \method\;(logistic) and \method\;(ridge) outperform competitors by over 10\% across most simulation scenarios.
Further, \method\;(logistic) often improves upon \method\;(ridge), indicating the benefit of tailoring the GLMs and metric to the problem at hand. 
Exploring other GLMs and similarity metrics may further improve performance. 

\subsection{Robust Regression Simulations}
\label{subsec:robust_results}
We illustrate another use-case of our framework by showing how \method\;can be tailored to account for the presence of outliers. 
We simulate responses as done in the regression setting, and introduce outliers as follows. 
We first select samples in the top and bottom $q/2\%$ quantiles for a randomly chosen \emph{non-signal feature}. 
For these selected samples, we corrupt their responses by drawing them from $N(\mu_{corrupt}, 1)$ and $N(-\mu_{corrupt}, 1)$ for the bottom and top quantile samples, respectively.
In our simulations, we vary across $q$ in $\{0,\; 1,\; 2.5,\; 5\}$ and $\mu_{corrupt}$ in $\{10,\; 25\}$ for a variety of sample sizes $n$.
We tailor \method\;to this setting by using a robust version of ridge regression \citep{owen2007robust} as our choice of regularized GLM, and negative Huber loss as our similarity metric. 
We compare these choices to those used in the regression setting (ridge regression and $\rsq$).
We shall refer to these particular settings as \method\;(Huber) and \method\;(ridge).

Simulation results for the Enhancer dataset with responses simulated via the LSS function are shown in Figure \ref{fig:robust_sims}.
Results for other datasets and regression functions are similar and provided in Appendix~\ref{supp:experiments}. 
We observe that \method\;(Huber)'s performance is more robust than competing methods including \method\;(ridge) and often remains accurate as both $\mu_{corrupt}$ and the percentage of outliers increase.

\begin{figure}[htp]

\begin{subfigure}{\textwidth}
    \centering
    \begin{tabularx}{1\textwidth}{@{}l *5{>{\centering\arraybackslash}X}@{}}
    \begin{minipage}{.092\textwidth}
    \phantom{}
    \end{minipage}%
    & {\small  $5\%$ Outliers} & {\small $2.5\%$ Outliers} & {\small $1\%$ Outliers} & {\small $0\%$ Outliers}
    \end{tabularx}\\
    \vspace{-2pt}
    \rotatebox{90}{{\hspace{-14mm} \centering \small \textbf{LSS with Outliers}}}
    \hspace{1.25mm}
    \rotatebox{90}{{\hspace{6.5mm} \centering \small $\mu_{corrupt}= 10$}}  \includegraphics[width=0.92\textwidth]{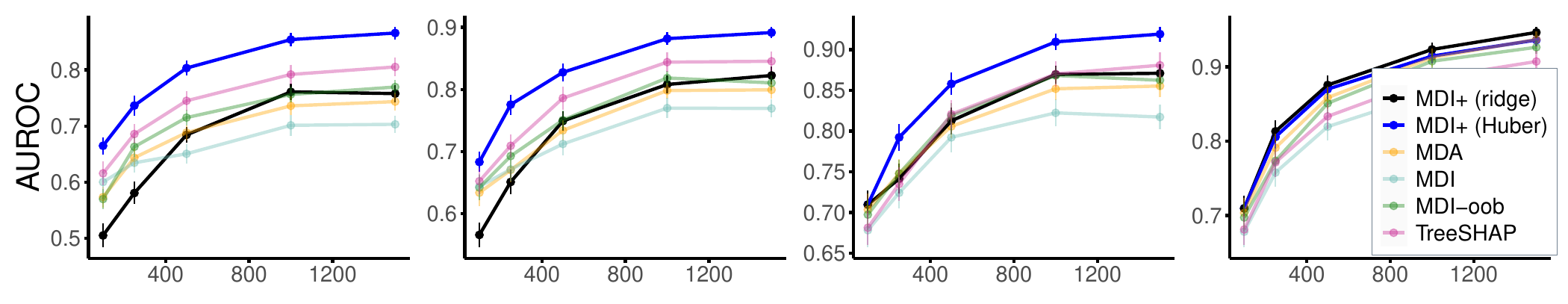}
\end{subfigure}
\begin{subfigure}{\textwidth}
    \centering
    \rotatebox{90}{{\hspace{10mm} \centering \small \textbf{}}}
    \hspace{3.25mm}
    \rotatebox{90}{{\hspace{9.5mm} \centering \small $\mu_{corrupt}= 25$}}
    \includegraphics[width=0.92\textwidth]{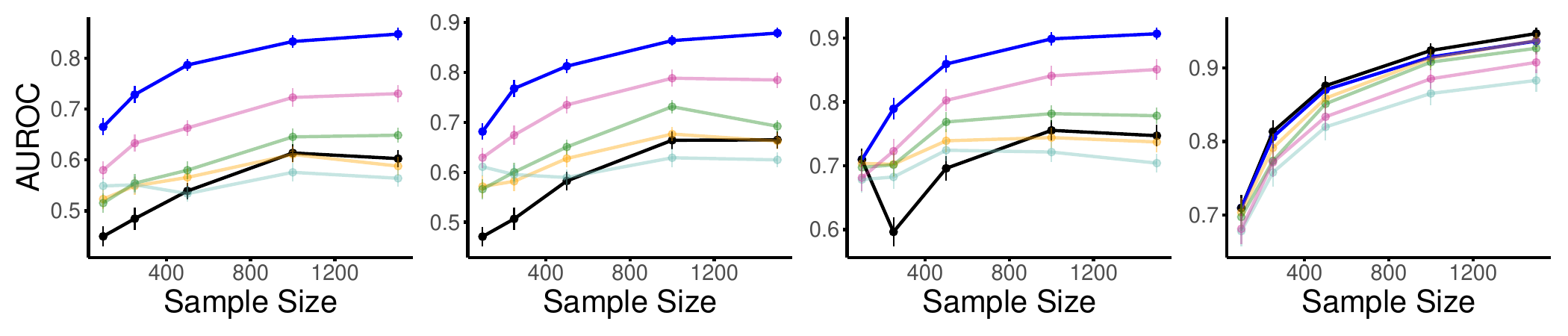}
\end{subfigure}
\caption{Under the LSS with outliers regression setting using the Enhancer dataset (described in Section~\ref{subsec:robust_results}), \method\;(Huber)'s performance remains suffers far less than other methods including \method\;(Ridge) as the level of corruption $\mu_{corrupt}$ (specified by row) and the proportion of outliers (specified by column) grow. This pattern also holds across sample sizes (on the $x$-axis). In all subplots, the AUROC has been averaged across 50 experimental replicates, and error bars represent $\pm$ 1SE.}
\label{fig:robust_sims}
\end{figure}

\section{\method\;Overcomes Biases of MDI}
\label{subsec:mdi_bias_sim}
In this section, we conduct simulations to show that \method\;overcomes the known biases of MDI against highly-correlated, and low-entropy features. 

\subsection{Correlated Feature Bias}
\label{sec:correlation-bias}

\paragraph{Experimental details.} We draw $\bx_i \sim N(\mathbf{0}, \bSigma)$, where $\bx_i \in \R^{100}$ and $\bSigma \in \R^{100 \times 100}$ has the following block-covariance structure: 
Features  $X_1, \ldots, X_{50}$ have pairwise correlation $\rho$, and features  $X_{51}, \ldots, X_{100}$ are independent of all other features. 
We then simulate responses from the linear+LSS model (see Section~\ref{subsec:sim_set_up}) with $PVE$ in $\braces{0.1,\; 0.4}$. 
Denote the group of signal features (i.e., $X_1, \ldots, X_6$) as Sig, the non-signal features that have non-zero correlation with the signal features (i.e., $X_7, \ldots, X_{50}$) as C-NSig, and the non-signal uncorrelated features (i.e., $X_{51}, \ldots, X_{100}$) as NSig. 
We generate $n= 250$ samples, and vary the correlation across $\rho$ in $\braces{0.5,\; 0.6,\; 0.7,\; 0.8,\; 0.9,\; 0.99}$. For all feature importance methods used in our simulation study in Section~\ref{sec:sims}, we compute the average rank of features within each group (Sig, C-NSig, and NSig). 

\begin{figure}[htbp]
    \centering
    \begin{subfigure}{\textwidth}
    \centering
    \rotatebox{90}{{\hspace{9.5mm} \centering \small \textbf{$PVE = 0.1$}}}
    \hspace{.25mm}
    \includegraphics[width=0.96\textwidth]{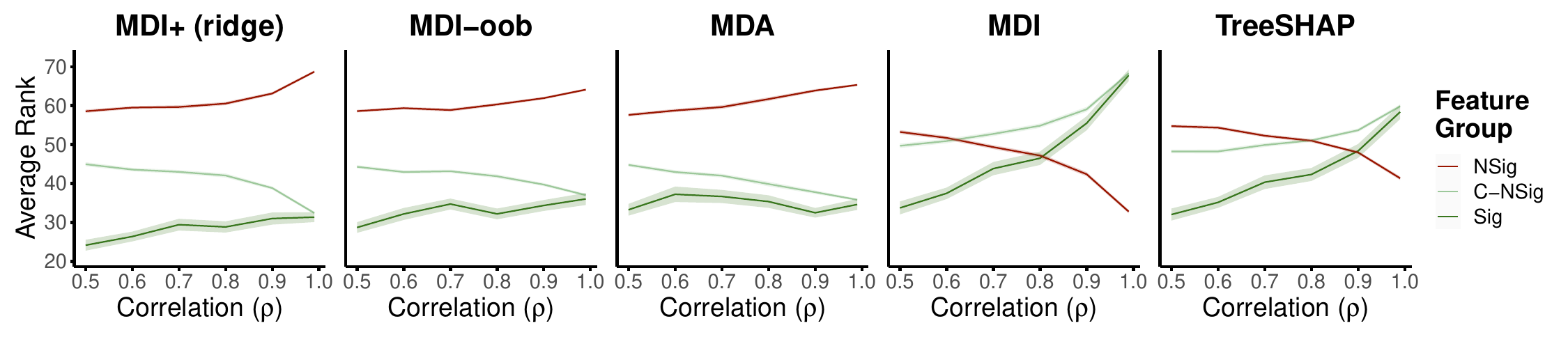}
    \end{subfigure}
    
    \begin{subfigure}{\textwidth}
    \centering
    \rotatebox{90}{{\hspace{9.5mm} \centering \small \textbf{$PVE = 0.4$}}}
    \hspace{.25mm}
    \includegraphics[width=0.96\textwidth]{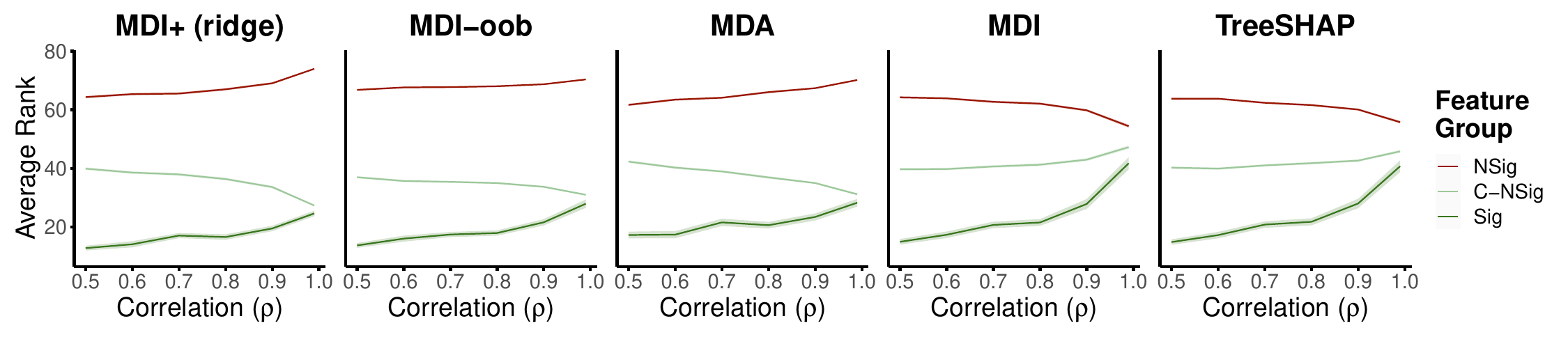}
    \end{subfigure}

    \caption{\method\;improves upon the bias that MDI has against selecting features that are highly correlated. We show the average ranks ($\pm 1 SE$) within feature groups (Sig, C-NSig, NSig) for various correlation ($\rho$) levels over 50 replicates. When the signal is moderate-to-high ($PVE = 0.4$, bottom), all methods rank the true signal features (Sig, dark green) as more important than the non-signal features (NSig, red) for all $\rho$; however, the gap is small for MDI and TreeSHAP. When the signal is low ($PVE = 0.1$, top), \method\;, MDI-oob, and MDA are still able to identify the true signal features (Sig, dark green) as most important. In contrast, MDI and TreeSHAP rank the non-signal features (NSig, red) as more important than the true signal (Sig, dark green) and the correlated, non-signal (C-NSig, light green) feature groups when the $\rho$ is large. 
    }
    \label{fig:correlation-sim}
\end{figure}

\paragraph{Results.} The results for the average ranks of features per group are shown in Figure~\ref{fig:correlation-sim}. 
Across all levels of correlation and $PVE$s, \method, MDI-oob, and MDA rank the true signal features (Sig, dark green) as the most important feature group and the uncorrelated non-signal group (NSig, red) as the least important feature group by a sizeable margin. 
MDI's behavior can be explained by Proposition \ref{prop:overfitting} and Figure \ref{fig:entropy_splits}, which displays the average percentage of RF splits per feature in each group. 
As $\rho$ increases, Figure \ref{fig:entropy_splits} shows the percentage of splits on NSig features increases, while that of Sig and C-NSig decreases. 
Intuitively, this occurs because Sig and C-NSig features, being correlated with each other, result in similar decision stump functions and must compete over splits.
Since MDI grows with the number of splits as established in Proposition \ref{prop:overfitting}, this leads to an overestimate of MDI for the NSig features. 
Moreover, MDI and TreeSHAP are measured on in-bag (training) samples, thereby amplifying biases that are learned during the tree construction. 
In contrast, \method, MDI-oob, and MDA use sample-splitting, which helps to mitigate this overfitting issue. 
A direct comparison between \method\;with and without LOO sample splitting helps to confirm this intuition (see Figure~\ref{fig:correlation-sim-gmdi-loo}).

\subsection{Entropy Bias}
\label{sec:entropy-bias}

\paragraph{Experimental details.} Following the simulation setup proposed in \citet{li2019debiased}, we sample 5 features: $X_{1}$ from a Bernoulli distribution with $p = 0.5$, $X_{2}$ from a standard Gaussian distribution, and $X_3$, $X_4$, and $X_5$ from a uniform discrete distribution with 4, 10, and 20 categories, respectively. 
To investigate entropy bias, we simulate the response as a function of only the lowest entropy feature, $X_1$, under (1) the regression setting via $Y = X_1 + N(0, \sigma^2)$ where $\sigma^2$ is chosen to achieve $PVE=0.1$ and (2) the binary classification setting via $\mathbb{P}(Y = 1 \mid X) = \frac{1 + X_{1}}{3}$.
We vary the number of samples $n \in \braces*{50,\; 100,\; 250,\; 500,\; 1000}$ and measure the rank for each feature across 50 replicates.

\begin{figure}[htbp]
    \centering
    \includegraphics[width=.8\textwidth]{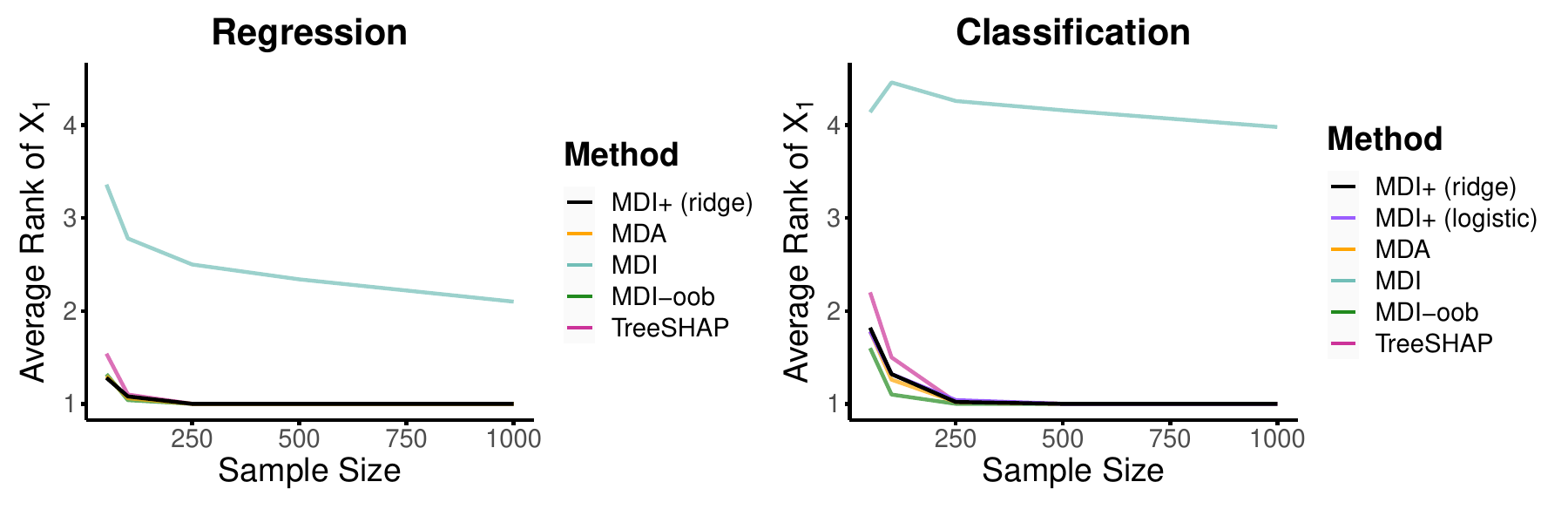}
    \caption{\method\;improves upon the bias that MDI has against selecting features with lower entropy in both the regression (left) and classification (right) simulation settings described in Section~\ref{sec:entropy-bias}. The feature ranking of the solo signal feature, $X_1$, averaged across 50 replicates, is shown on the y-axis. Here, a lower value indicates greater importance.}
    \label{fig:entropy_sim}
\end{figure}

\paragraph{Results.} Figure~\ref{fig:entropy_sim} displays the average rank of $X_1$, the solo signal feature, for each feature importance method.
Here, only MDI is unable to identify $X_1$ as the signal feature.  
This occurs because $X_1$ has the lowest entropy relative to other features, and is hence split upon the least (see Figure~\ref{fig:entropy_splits}). 
As a result of Proposition \ref{prop:overfitting}, this leads to an overestimate of MDI for high-entropy features $X_2, \ldots, X_5$ that are split upon more often. 
To combat this, \method\;employs regularization to control the effective degrees of freedom and sample-splitting to mitigate biases learned during the tree construction.
A direct comparison of \method\;with and without ridge regularization and LOO sample-splitting in Figure \ref{fig:entropy_sim-gmdi-loo} helps confirm this. 

\section{Case Studies}
\label{sec:case_studies}

In this section, we investigate two case studies using \method\;as well as other feature importance measures to identify important features in predicting drug responses and breast cancer subtypes.
In both case studies, we show that \rfmethod\;increases the prediction accuracy over RF and that the feature rankings from \method\;agree with established domain knowledge with significantly greater stability  than other feature importance measures across different perturbations (i.e., across different train-test splits and random seeds used to train the RF).
Given that both predictability and stability are important prerequisites for interpretability \citep{murdoch2019definitions, yu2020veridical}, these findings showcase the effectiveness of \method\;for extracting reliable interpretations in real-world scientific problems.

\subsection{Case Study I: Drug Response Prediction} \label{sec:case_study_ccle}

Accurately predicting a cancer drug's efficacy for a patient before prescribing it can tremendously improve both the patient's health and financial well-being, given the exorbitant costs of many cancer drugs. 
Moreover, identifying the important genetic predictors that are influential of the drug response can provide valuable insights into potential targets and novel candidates for future preclinical research.
Towards this end, we leverage data from the Cancer Cell Line Encyclopedia (CCLE) \citep{barretina2012cancer} to build accurate drug response models and identify genes whose expression levels are highly predictive of a drug's response.

\paragraph{Data and Methods.} We use gene expression data $\bX \in \R^{472 \times 5000}$, measured via RNASeq, from $n = 472$ cell lines and $p = 5000$ genes after filtering (details in Appendix~\ref{supp:ccle_case_study}). For each cell line, the response of 24 different drugs was measured, yielding a multivariate response matrix $\bY \in \R^{472 \times 24}$. We split the data into 80\% training and 20\% test. Using the training data, we fit 24 separate RFs, one to predict the response for each drug using the gene expression data $\bX$, and investigate the gene importance rankings for each drug. We repeat this for 32 different train-test splits. Here, the RF settings and feature importance methods used are the same as those in Section~\ref{subsec:sim_set_up} for regression.

\paragraph{Prediction accuracy.} For 23 out of the 24 drugs, the test $\rsq$, averaged across the 32 train-test splits, was higher for \rfmethod\;than RF.
Furthermore, \rfmethod\;improved the test $\rsq$ by an average of 4\% across the 18 drugs with RF test $\rsq>0.1$. We use this threshold of $\rsq>0.1$ to focus on models that have non-trivial predictive power; however, the improvement from \rfmethod\;is even higher without this filter (see Appendix~\ref{supp:prediction_accuracy}).

\paragraph{Accuracy of gene importance rankings.} \method\;identified several well-known genes that have been previously identified as drug targets in the literature. In particular, every known gene expression predictor of drug response identified in the original work on CCLE \citep{barretina2012cancer}, except for one, was ranked in the top $5$ by \method\;for their respective drugs. The one gene outside the top $5$ was \textit{MDM2}, a known predictor for the drug Nutlin-3 \citep{barretina2012cancer}. Still, \method\;ranked \textit{MDM2} 17\textsuperscript{th}, which is higher than the rankings given by its competitors (MDI-oob: 22\textsuperscript{nd}, TreeSHAP: 30\textsuperscript{th}, MDI: 35\textsuperscript{th}, MDA: 53\textsuperscript{rd}). A full list of the top 5 genes identified for each drug is provided in Appendix~\ref{supp:ccle_case_study}.

To evaluate the gene importances beyond evidence from the scientific literature, we also assessed the predictive power of the top 10 genes from each feature importance method. We found that the top 10 genes from \method\;were often more predictive than that from other feature importance methods. For brevity, we defer details and additional discussion regarding this prediction analysis to Appendix~\ref{supp:ccle_case_study}. 

\begin{figure}[htp]
    \centering
     \begin{subfigure}[b]{.85\textwidth}
         \centering
         \captionsetup{font=normal,labelfont=normal}
         \caption{Stability of Features Ranked in Top 10}
         \includegraphics[width=1\textwidth]{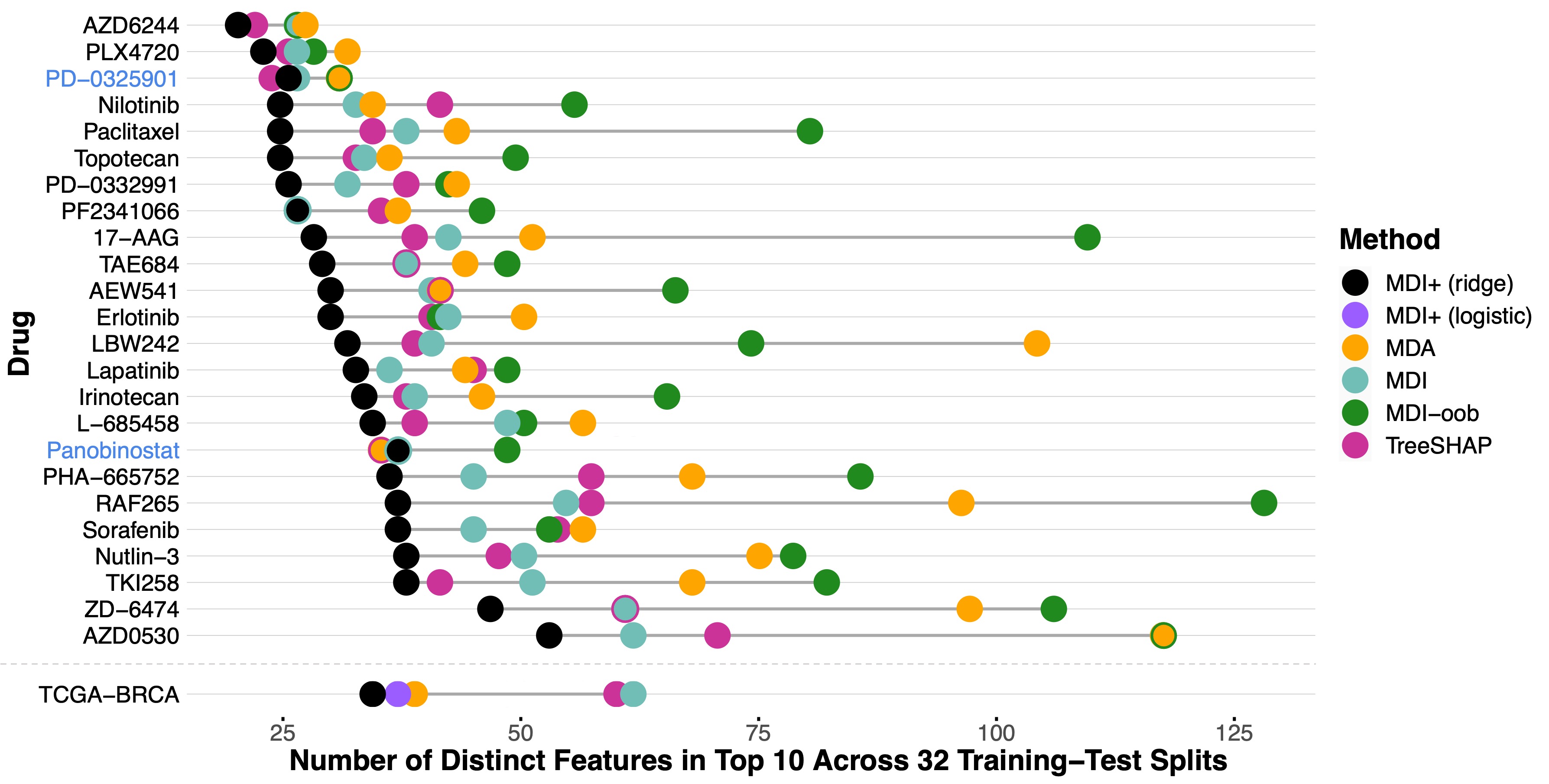}
         \label{fig:casestudy-stability-top10}
         \vspace{-4mm}
     \end{subfigure}

     \begin{subfigure}[b]{0.97\textwidth}
        \centering
        \captionsetup{font=normal,labelfont=normal}
        \caption{Stability of Top 5 Ranked Features}
         \includegraphics[width=1\textwidth]{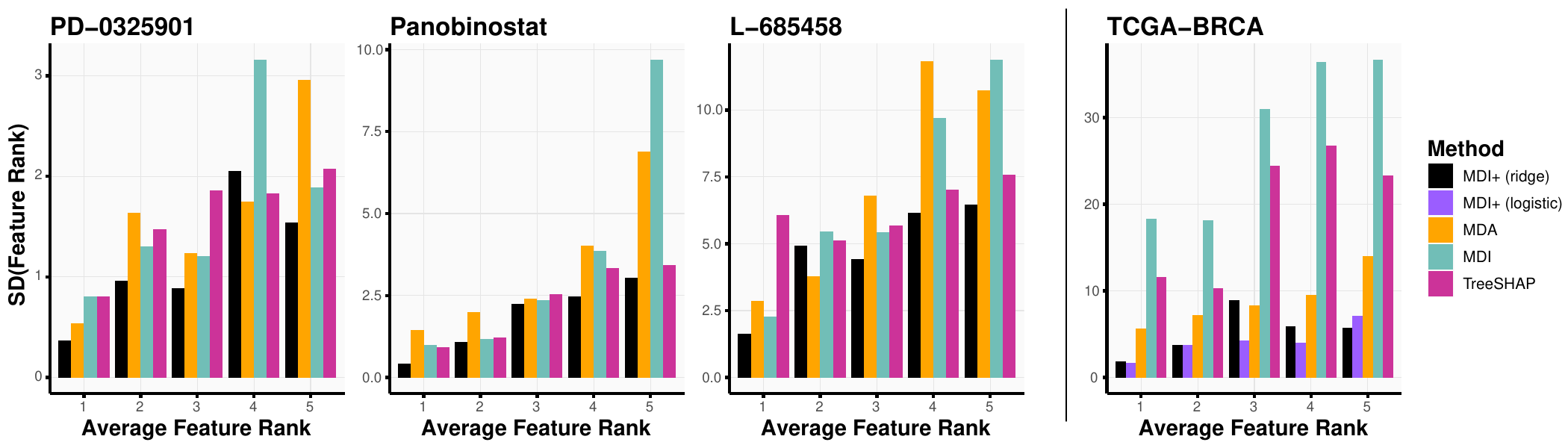}
         \label{fig:casestudy-stability-dist}
     \end{subfigure}
    \caption{
    In the CCLE drug response and TCGA breast cancer (TCGA-BRCA) subtype case studies, \method\;provided the most stable feature importance rankings across 32 train-test splits. Specifically,
    (a) for 22 out of the 24 drugs in the CCLE case study (the exceptions being PD-0325901 and Panobinostat, highlighted in blue) and in the TCGA-BRCA case study, \method\;had the fewest number of distinct features that were ranked in the top 10 across the 32 train-test splits. Note that the outlined points denote ties and that \method\;(logistic) is only used in the classification setting (i.e., for TCGA-BRCA).
    Moreover, as shown in (b), the standard deviation of the top 5 features is smaller for \method\;as compared to other methods. We provide the results for three CCLE drugs and the TCGA-BRCA subtypes here. MDI-oob is also omitted due to its high instability. Full results can be found in Appendix~\ref{supp:case_study}.
    }
    \label{fig:ccle-stability}
\end{figure}

\paragraph{Stability of gene importance rankings.} Lastly, for each feature importance method, we investigated the stability of gene importance rankings across the 32 train-test splits. Methods that exhibit greater stability are highly advantageous in practice, since it is undesirable for interpretations to change due to arbitrary choices like train-test splits and random seeds. In Figure~\ref{fig:casestudy-stability-top10}, we display one measure of stability, namely, the number of distinct genes ranked in the top 10 across the 32 train-test splits. For 22 out of the 24 drugs (exceptions being PD-0325901 and Panobinostat), \method\;had the fewest number of distinct genes that were ranked in the top 10 across the 32 train-test splits. Moreover, for all 24 drugs, \method\;had the highest number of genes that were always ranked in the top 10 across all train-test splits. Interestingly, sample-splitting techniques such as MDA and MDI-oob were significantly less stable than other methods, highlighting the downside of using fewer samples to measure feature importances. On the other hand, \method, which leverages LOO and regularization, overcomes this drawback. 

To further establish the stability of \method, we examined the distribution of feature rankings of each method for the five most important genes across the 32 splits. As seen in  Figure~\ref{fig:casestudy-stability-dist} (results for all drugs in Figure~\ref{fig:ccle-stability-dist-all-zoom}), feature rankings of \method\;tend to have smaller variance as compared to other methods. This remains true for PD-0325901 and Panobinostat, for which \method\;did not perform optimally according to the aforementioned top 10 stability metric. A similar improvement in stability is also seen when fixing the training data and only altering the random seed used for fitting the RF (see Appendix~\ref{supp:ccle_case_study}).
This improved stability of \method, in addition to the increase in prediction performance from \rfmethod, demonstrate the practical advantages of \method\;in this real-data case study on drug response prediction.

\subsection{Case Study II: Breast Cancer Subtype Prediction} \label{sec:case_study_tcga}

Breast cancer is known to be a heterogeneous disease \citep{cancer2012comprehensive} and is thus often classified into various molecular subtypes. PAM50 is one prominent breast cancer subtyping method, which is frequently used to inform treatment plans \citep{parker2009supervised}. PAM50 divides breast cancers into five intrinsic subtypes: Luminal A, Luminal B, human epidermal growth factor receptor 2 (HER2)-enriched, Basal-like, and Normal-like.
Each subtype differs based upon their biological properties and generally corresponds to a different prognosis. Luminal A typically has the best prognosis, while HER2-enriched and Basal-like are more aggressive.
In this study, we aim to predict the PAM50-defined breast cancer subtypes using gene expression data from The Cancer Genome Atlas (TCGA) and more importantly, identify the genes that heavily determine the breast cancer subtype.

\paragraph{Data and Methods.} Using data from the TCGA, we have RNASeq gene expression data from $n = 1083$ individuals and $p = 5000$ genes after filtering (details in Appendix~\ref{supp:tcga_case_study}). We also have the PAM50-defined breast cancer subtype for each of these individuals, giving rise to a multi-class classification problem with five classes. As in Section~\ref{sec:case_study_ccle}, we split the data into 80\% training and 20\% test and evaluate results across 32 train-test splits. The RF settings and feature importance methods used are the same as those in Section~\ref{subsec:sim_set_up} for classification, but excluding MDI-oob which has not been implemented for multi-class classification.

\paragraph{Prediction accuracy.} Comparing RF and \rfmethod\;across the 32 train-test splits, \rfmethod\;(logistic) yielded the best prediction performance over a variety of metrics (see Table~\ref{tab:tcga-prediction}). In particular, \rfmethod\;(logistic) yielded an average classification accuracy of 88.4\%, compared to 87.3\% from \rfmethod\;(ridge) and 86.1\% from RF. The improvement of \rfmethod\;(logistic) over \rfmethod\;(ridge) illustrates the benefit of tailoring the GLM to the data. 

\paragraph{Accuracy of gene importance rankings.}  Both \method\;(ridge) and \method\;(logistic) produced the same set of top 10 genes, albeit in a different order (see Appendix~\ref{supp:tcga_case_study}). For each of these 10 genes, there is significant literature, supporting the gene's involvement in the development and progression of breast cancer. For example, \textit{ESR1}, the top-ranked gene across all feature importance methods, has been widely studied over the last decade due to its prominent role in the pathogenesis of breast cancer \citep{toy2013esr1} as well as in novel promising therapies \citep{brett2021esr1}. \textit{ESR1} encodes the estrogen receptor-$\alpha$ protein and is known to cause increased resistance to standard-of-care endocrine therapy \citep{brett2021esr1}. Several other genes in the top 10, namely, \textit{FOXA1}, \textit{FOXM1}, and \textit{MLPH}, are also associated with estrogen receptor-positive breast cancer and increased resistance to endocrine therapy \citep{fu2019foxa1,bergamaschi2014forkhead,thakkar2010identification}. Beyond this, \textit{GATA3} (ranked 2\textsuperscript{nd} in \method\;(logistic) and 4\textsuperscript{th} in \method\;(ridge)) was shown to control luminal breast cancer predominantly through differential regulation of other genes including \textit{THSD4} \citep{cohen2014shift} while \textit{FOXC1} can counteract \textit{GATA3} binding and impact endocrine resistance \citep{yu2016foxc1}. 

We further show in Figure~\ref{fig:tcga-prediction} that the top 25 genes from \method\;(logistic) and \method\;(ridge) are more predictive of the PAM50 breast cancer subtypes than the top 25 genes from other feature importance measures (details in Appendix~\ref{supp:case_study}). This top 25 from \method\;includes genes such as \textit{RRM2} and \textit{SFRP1} that are not in the top 25 from any of the competing feature importance methods. However, these genes are known to be associated with breast cancer. For instance, up-regulation of \textit{RRM2} was shown to enhance breast cancer cell proliferation \citep{liang2019dscam}, while loss of \textit{SFRP1} expression is associated with breast tumor stage and poor prognosis when present in early stage breast tumors \citep{klopocki2004loss}. 

\paragraph{Stability of gene importance rankings.} As in the CCLE case study, we evaluated the stability of the gene importance rankings across 32 train-test splits and found that \method\;gave the most stable rankings. In Figure~\ref{fig:casestudy-stability-top10}, \method\;(ridge), followed by \method\;(logistic), yielded the smallest number of distinct features in the top 10 across the 32 train-test splits. Moreover, for the top five features, the variability of the feature importance rankings across the 32 train-test splits is much smaller for \method\;(ridge) and \method\;(logistic) compared to its competitors (Figure~\ref{fig:casestudy-stability-dist}). 

\section{Discussion}
\label{sec:discussion}

By reinterpreting decision trees and MDI via a linear regression perspective, we provide a novel framework for expanding the prediction and interpretation capabilities of RFs using \rfmethod\;and \method.
Specifically, we exploit a seemingly little known interpretation of decision trees as a representation learning algorithm, connecting decision trees to a linear regression fit and MDI to an $\rsq$ value from the linear regression on this learned representation.
Our \rfmethod\;and \method\;framework provides researchers with the opportunity to improve upon the already powerful RF predictions and MDI interpretations by (1)~replacing the implicit OLS model and/or $\rsq$ metric with other models and/or metrics that are tailored to a given data structure, (2)~incorporating additional features or knowledge to mitigate the known biases of decision tree-based techniques, and (3)~utilizing regularization and efficient sample splitting to avoid overfitting issues.

Importantly, the linear regression interpretation of decision trees, highlighted here, builds a valuable bridge between GLMs in the traditional statistics literature and RFs in the modern machine learning literature.
Our work only scratches the surface of possible future research directions that combine the flexible power of decision tree-based modeling and classical, theoretically-grounded statistical tools.
For example, inferential tools that have already been developed and well-studied for GLMs could prove useful to develop a measure of statistical significance for RF-based feature importances.
Such an advancement would be hugely impactful, given the wide use of RFs and its feature importances in scientific problems and other high-stakes applications.

Relatedly, despite the strong empirical performance of \method\;demonstrated here, we recognize that our data-inspired simulations and real-world case studies only represent a first step in establishing the trustworthiness of \method, or any feature importance measure.
Given the potentially high-stakes and downstream impacts of using feature importance measures in real-world scientific problems, developing a formal protocol and benchmarks to evaluate the effectiveness of feature importance measures is needed and will have a significant impact on the applicability of these methods in practice. 
One possible pathway to constructing this protocol is via the PCS framework \citep{yu2020veridical, yu2024veridical}. 
In particular, the PCS framework advocates for holistic evaluation of feature importances including its predictability, and stability, as we have begun to explore in our case studies. 
We leave it as important future work to formalize and expand these ideas.

Beyond this, our work opens the door to many other natural directions for future work.
First, \method\;need not be limited to RFs and can be defined for any tree ensemble, and in particular, XGBoost \citep{chen2016xgboost} and FIGS \citep{tan2022fast}.
Second, as we illustrated in Sections~\ref{sec:prediction_accuracy} and \ref{sec:sims}, exploiting the flexibility of \rfmethod\;and \method\;can substantially improve performance for structured data. Though we demonstrated this in the regression, classification, and robust regression settings, there are many other contexts to consider such as multitarget prediction, survival analysis, longitudinal data analysis, etc. 
In addition, while we have provided practitioners with a PCS-based approach to assist practitioners with modeling choices in the \rfmethod\;and \method\;framework, further investigation is needed (e.g., more sophisticated approaches to combining scores from different \method\;models). 
Finally, the flexibility of \rfmethod\;as a general prediction method may be of significant independent interest. Beyond fitting GLMs on the augmented transformed dataset, \rfmethod\;allows practitioners to fit any ML algorithm on the augmented transformed dataset. This provides a novel avenue to incorporate the advantages of tree-based methodology and other ML tools for prediction.

\if1\blind{
    \section{Acknowledgements}
\label{sec:acknowledgements}
We gratefully acknowledge partial support from 
NSF TRIPODS Grant 1740855, DMS-2209975, 1613002, 1953191, 2015341, IIS 1741340, ONR grant N00014-17-1-2176, the Center for Science of Information (CSoI), an NSF Science and Technology Center, under grant
agreement CCF-0939370, NSF grant 2023505 on
Collaborative Research: Foundations of Data Science Institute (FODSI),
the NSF and the Simons Foundation for the Collaboration on the Theoretical Foundations of Deep Learning through awards DMS-2031883 and 814639, a grant from the Weill Neurohub, a Chan Zuckerberg Biohub Intercampus Research Award, and NIH R01GM152718. TMT acknowledges support from the NSF Graduate Research Fellowship Program DGE-2146752. 
YT was partially supported by NUS Start-up Grant A-8000448-00-00 and MOE AcRF Tier 1 Grant A-8002498-00-00.

} \fi

\appendix

\appendix 

\section{Main Proofs}

We begin by reminding the reader of some definitions made in Section \ref{sec:new_perspective_on_mdi}.
A split $s$ of a node $\node$ partitions it into two children nodes $\node_L = \braces{\bx_0 \in \node \colon x_{0,k} \leq \tau}$ and $\node_R = \braces{\bx_0 \in \node \colon x_{0,k} > \tau}$ for some feature index $k$ and threshold $\tau$.
Given a dataset $\data = (\bX,\by)$, we associate to $s$ the local decision stump function
\begin{equation} \label{eq:local_decision_stump_appendix}
    \psi(\bx;s,\data) = \frac{N\left(\node_{R}\right)\mathbf{1}\braces*{\bx \in \node_{L}} - N\left(\node_{L}\right)\mathbf{1}\braces*{\bx \in \node_{R}}}{\sqrt{N\left(\node_{L}\right)N\left(\node_{R}\right)}},
\end{equation}
where $N(\node)$ denotes the number of sample points in $\data$ that lie in $\node$.
A CART model comprises a collection of recursive splits $\splits = \braces*{s_{1}, \ldots, s_{m}}$ that defines its tree structure, as well as values for the predictions on each leaf on the tree.
Given $\data$, these values are set to be the mean responses over each leaf.

If $\splits$ has $m$ splits, concatenating these $m$ functions yields the learned feature map $\Psi(\bx;\splits,\data) \coloneqq \left(\psi(\bx;s_{1},\data), \ldots, \psi(\bx;s_{m},\data)\right)$ 
and its corresponding transformed dataset $\Psi\left(\bX;\splits,\data\right) \in \R^{n \times m}$.

\subsection{Proof of Proposition~\ref{prop:CART_linear}}\label{supp:proof_CART_linear}

Fix $\bx$ and without loss of generality, re-index the nodes in the tree so that $\node_1 \supset \node_2 \supset \cdots \supset \node_l$ corresponds to its root-to-leaf path in the CART tree.
Assume further without loss of generality that split $s_j$ corresponds to node $j$ and $\node_{j+1}$ is the left child of $\node_{j}$ for $j=1,\ldots,l-1$.
Also assume that $\hat\alpha = \bar y = 0$.
For ease of notation, denote $\bPsi = \Psi(\bX;\splits,\data)$.
The OLS solution can be written as $\hat\bbeta = \paren*{\bPsi^T\bPsi}^{-1}\bPsi^T\by$. Because the columns of $\bPsi$ are orthogonal, the Gram matrix $\paren*{\bPsi^T\bPsi}^{-1}$ is diagonal with $j$-th entry $\paren*{\bPsi^T\bPsi}^{-1}_{jj} = N(\node_j)^{-1}$.

With a view to computing a formula for $\hat \beta_j$, we first let consider an arbitrary split $s$ of a node $\node$ into $\node_L$ and $\node_R$, let $\psi(s)$ denote the corresponding column vector, and use \eqref{eq:local_decision_stump_appendix} to write

\begin{equation*}
    \begin{split}
        \psi(s)^T\by & = \frac{N(\node_R)\sum_{\bx_i \in \node_L}y_i - N(\node_L)\sum_{\bx_i \in \node_R} y_i}{\sqrt{N(\node_L)N(\node_R)}} \\
        & = \sqrt{N(\node_L)N(\node_R)} \paren*{\bar{y}_{\node_{L}} -  \bar{y}_{\node_{R}}} \\
        & = \sqrt{N(\node_L)N(\node_R)} \paren*{\bar{y}_{\node_{L}} -  \frac{N(\node)}{N(\node_R)}\bar{y}_{\node} - \frac{N(\node_L)}{N(\node_R)}\bar{y}_{\node_{L}}} \\
        & = \sqrt{\frac{N(\node_L)N(\node)^2}{N(\node) - N(\node_L)}} \paren*{\bar y_{\node_L} - \bar y_{\node}}.
    \end{split}
\end{equation*}
Choosing $\node = \node_j$, we get
\begin{equation*}
    \begin{split}
        \hat\beta_j & = \paren*{\bPsi^T\bPsi}^{-1}_{jj} \psi(\node_j)^T\by = \sqrt{\frac{N(\node_{j+1})}{N(\node_j) - N(\node_{j+1})}} \paren*{\bar y_{\node_{j+1}} - \bar y_{\node_j}}.
    \end{split}
\end{equation*}
Using \eqref{eq:local_decision_stump_appendix} again, the $j$-th entry of the embedding $\Psi(\bx;\splits,\data)$ satisfies
\begin{equation*}
    \Psi(\bx;\splits,\data)_j = \frac{N(\node_j) - N(\node_{j+1})}{\sqrt{N(\node_{j+1})\paren*{N(\node_j) - N(\node_{j+1}})}}.
\end{equation*}
Multiplying these two expressions gives
\begin{equation*}
    \hat\beta_j\Psi(\bx;\splits,\data)_j = \bar y_{\node_{j+1}} - \bar y_{\node_j}.
\end{equation*}
Since the only nonzero entries of $\Psi(\bx;\splits,\data)$ are those in indices $j=1,2,\ldots,l-1$ (i.e. along the root to leaf path), 
adding up over all these values and using the telescoping property gives us
\begin{equation*}
    \hat\bbeta^T\Psi(\bx;\splits,\data) = \bar y_{\node_l} = \hat f(\bx).
\end{equation*}

\subsection{Proof of Theorem~\ref{thm:MDI_r2_equivalence}}
\label{supp:proof_prop}

Using the law of total variance, we may rewrite the formula for impurity decrease for a node $\node$ via
\begin{align*}
    \hat{\Delta}(s, \data^*) = N\left(\node\right)^{-1}\paren*{\sum_{\bx_i \in \node}\left(y_{i} - \bar{y}_{\node}\right)^{2} - \sum_{\bx_i \in \node_{L}}\left(y_{i}- \bar{y}_{\node_{L}} \right)^{2} - \sum_{\bx_i \in \node_{R}}\left(y_{i}- \bar{y}_{\node_{R}} \right)^{2}}
    = \frac{N\left(\node_{L}\right)N\left(\node_{R}\right)}{N\left(\node\right)^2}\left(\bar{y}_{\node_{L}} -  \bar{y}_{\node_{R}}\right)^{2}.
\end{align*}
Note that we use a binary indicator of a sample lying in the right child node as the conditioning variable.
Next, using \eqref{eq:local_decision_stump_appendix} and using $\psi(s)$ to denote the resulting feature vector on the training set, we compute
\begin{align*}
    \psi(s)^T\by = \frac{N(\node_R)\sum_{\bx_i \in \node_L}y_i - N(\node_L)\sum_{\bx_i \in \node_R} y_i}{\sqrt{N(\node_L)N(\node_R)}} = \sqrt{N(\node_L)N(\node_R)} \paren*{\bar{y}_{\node_{L}} -  \bar{y}_{\node_{R}}}.
\end{align*}
By combining the above two equations, we obtain the formula
\begin{equation} \label{eq:impurity_dec_equivalence}
    \hat{\Delta}(s, \data^*) = \frac{\paren*{\psi(s)^T\by}^2}{N(\node)^2}.
\end{equation} 
Now, let $s_1,\ldots,s_m$ denote the splits on feature $X_k$, which means that we have $\Psi^k = \left[\psi(s_1),\cdots,\psi(s_m)\right]$.
Then,

\begin{align*}
    \norm*{\Psi^{k}(\bX)\paren*{\Psi^{k}(\bX)^T\Psi^{k}(\bX)}^{-1}\Psi^{k}(\bX)^T\by}_{2}^{2} 
    & = \by^T \Psi^{k}(\bX) \paren*{\Psi^{k}(\bX)^T\Psi^{k}(\bX)}^{-1}\Psi^{k}(\bX)^T\by \\
    & = \sum_{i=1}^m \frac{\paren*{\psi(s_i)^T\by}^2}{N(\node_i)} \\
    & = \sum_{i=1}^m N(\node_i)\hat{\Delta}(s_i, \data^*),
\end{align*}
where the second equality comes from recognizing that $\paren*{\Psi^{k}(\bX)^T\Psi^{k}(\bX)}^{-1}$ is a diagonal matrix with $i$-th diagonal entry equal to $N(\node_i)^{-1}$, and the third equality comes from plugging in \eqref{eq:impurity_dec_equivalence}.
Recognizing that the right-hand side is simply $n \textnormal{MDI}_k(\mathcal{S}, \mathcal{D}_n^*)$ completes the proof of the left equality in \eqref{eq:MDI_r2_equivalence}.
The right equality follows from the definition of training $\rsq$.

\section{Additional Discussion of MDI Overfitting Bias}
\label{app:drawbacks}

While MDI generally suffers from overfitting as discussed in Section~\ref{subsec:overfitting}, this overfitting can be more precisely explained as a ``differential optimism bias.''
For a fixed design with homoskedastic noise, it is known that the average difference between training and held-out test mean-squared error (i.e., the \textit{optimism bias}) for linear models is twice the number of degrees of freedom scaled by the noise variance~\citep{hastie2009elements}. 
Combining Theorem~\ref{thm:MDI_r2_equivalence} with a similar set of calculations leads to the following optimism bias of MDI.

\begin{proposition}
    \label{prop:overfitting}
    Suppose $\data$ is generated according to a fixed design with fixed covariate vectors $\bx_1, \ldots, \bx_n$ and responses $y_i = f(\bx_i) + \epsilon_i$ with $\epsilon_1,\ldots,\epsilon_n$ independent and satisfying $Var(\epsilon_i) = \sigma^2$ for $i=1, \ldots, n$.
    Let $\data^0 = \braces*{(\bx_i, f(\bx_i))}$ denote the noiseless dataset.
    Then for any fixed tree structure $\splits$, we have 
    \begin{align*}
        \E\braces*{\mdi{\splits,\data}} = \mdi{\splits,\data^0} + \frac{\sigma^2  \abs*{\splits^{(k)}}}{n}.
    \end{align*}

\end{proposition}

\begin{proof}
Denote $\bP \coloneqq \Psi^{k}(\bX)\paren*{\Psi^{k}(\bX)^T\Psi^{k}(\bX)}^{-1}\Psi^{k}(\bX)^T$.
We have shown in the proof of Theorem \ref{thm:MDI_r2_equivalence} that
\begin{equation*}
    \textnormal{MDI}_k(\splits,\data) = \frac{1}{n}\norm*{\bP\by}_2^2.
\end{equation*}
Now write $\by = \boldf + \beps$ where $\boldf = (f(\bx_1),\ldots,f(\bx_n))^T$ and $\beps = (\epsilon_1,\ldots,\epsilon_n)^T$.
Under a fixed design, $\boldf$ is deterministic while $\beps$ is random.
Taking expectations therefore gives
\begin{align*}
    \E\braces*{\textnormal{MDI}_k(\splits,\data)} & = \E\braces*{\frac{1}{n}\norm*{\bP(\boldf + \beps)}_2^2} \nonumber \\
    & = \frac{1}{n}\norm*{\bP\boldf}_2^2 + \frac{1}{n}\E\braces*{\norm*{\bP\beps}_2^2} \nonumber \\
    & = \frac{1}{n}\norm*{\bP\boldf}_2^2 + \frac{\sigma^2\text{Trace}(\bP)}{n}.
\end{align*}
By orthogonality of local decision stumps, $\Psi^k(\bX)$ has full rank, and $\Trace(\bP)$ is just the number of columns, which is equal to $\abs*{\mathcal{S}^{(k)}}$.
Finally, notice that
\begin{equation*}
    \textnormal{MDI}_k(\splits,\data^0) = \frac{1}{n}\norm*{\bP\boldf}_2^2.
\end{equation*}%
\end{proof}

%
Proposition \ref{prop:overfitting} reveals that (i) $\textnormal{MDI}_k(\splits,\data)$ is optimistic for $\textnormal{MDI}_k(\splits,\data^0)$ (i.e., the noiseless MDI value, or equivalently, the proportion of variance in $f(\bx)$ explained by $\Psi(\bx; \splits^{(k)},\data)$) and (ii) the optimism bias is larger for features $X_k$ that have more splits, leading to a \textit{differential optimism bias}. 
This differential optimism is problematic because the split frequency of a feature $X_k$ reflects not only the feature's inherent influence on the response, but also some biases due to the tree-growing process of CART. 
In particular, CART's splitting rule results in highly-correlated and low-entropy features receiving fewer splits (see Section \ref{subsec:mdi_bias_sim}).
As a result, signal features with these characteristics often have lower MDI than non-signal features without these characteristics.

Though Proposition~\ref{prop:overfitting} requires a fixed tree structure, when $\splits$ is random, the result above can be applied conditionally on $\splits$, as long as it is independent of $\data$. 
This independence can be achieved if OOB samples are used to compute MDI, as is sometimes done in practice.
However, Proposition~\ref{prop:overfitting} shows that this does not resolve the differential optimism bias.
More commonly, in-bag samples are used to both generate $\splits$ and compute MDI. In this case, Proposition~\ref{prop:overfitting} does not apply directly, but because the splits $\splits$ tend to be correlated with the noise, the amount of overfitting (i.e., differential optimism bias) should be worse.
Instead, in the \rfmethod\;and \method\;framework, to more effectively mitigate the overfitting issue, we incorporate (i) regularization, which is known to reduce the effective degrees of freedom (i.e., the number of splits), and (ii) sample splitting to separate the data used to learn the prediction model and that used to evaluate its generalization error.

\section{Feature Ranking Performance Simulations}
\label{supp:experiments}

In this section, we include additional simulation results to supplement Section~\ref{sec:sims}. Unless specified otherwise, we follow the simulation protocol described in Section~\ref{sec:sims}, and all plots show the mean evaluation metric, averaged across 50 experimental replicates, with error bars denoting $\pm$ 1SE. 

\paragraph{Raw Data Availability.} The Juvenile dataset can be downloaded using the \texttt{imodels} python package \citep{singh2021imodels}. The Enhancer and Splicing datasets were taken from \citep{basu2018iterative}. The CCLE dataset can be downloaded from DepMap Public 18Q3 (\url{https://depmap.org/portal/download/}). 

\paragraph{Data Preprocessing.} We performed basic data cleaning on these datasets before using them as covariate matrices in the simulation study. Specifically, for all four datasets, we removed constant and duplicated columns. We also applied a $\log(x+1)$ transformation to all values in the Enhancer and CCLE datasets as these are count-valued and highly right-skewed. For the Enhancer dataset, the raw data contained repeated measurements (i.e., multiple feature columns) for the same transcription factor; we thus removed all but one measurement (i.e., column) for each transcription factor. Note finally that the CCLE dataset contains 50114 features, but we chose a random 
subset of 1000 features to use in the covariate matrix in each simulation replicate. 

\paragraph{Preprocessed Data and Code Availability.} The preprocessed datasets can be found on Zenodo at
\if1\blind{\url{https://zenodo.org/record/8111870}}\fi
\if0\blind{\url{redacted}}\fi. 
Code to run all simulations can be found on GitHub at 
\if1\blind{\url{https://github.com/Yu-Group/imodels-experiments}}\fi
\if0\blind{\url{redacted}}\fi.

\begin{figure}[h!]
\begin{subfigure}{\textwidth}
    \centering
    \vspace{-5mm}
    \begin{tabularx}{0.95\textwidth}{@{}l *5{>{\centering\arraybackslash}X}@{}}
    \begin{minipage}{.12\textwidth}
    \phantom{}
    \end{minipage}%
    & {\small $PVE=0.1$} & {\small $PVE=0.2$} & {\small $PVE=0.4$} & {\small $PVE=0.8$}
    \end{tabularx}\\
    \vspace{-1mm}
    \hspace{2mm}
    \rotatebox{90}{{\hspace{-44mm} \centering \small \textbf{(A) Regression: Linear}}}
    \hspace{1mm}
    \rotatebox{90}{{\hspace{6mm} \centering \small \textsc{Enhancer}}}
    \includegraphics[width=0.85\textwidth]{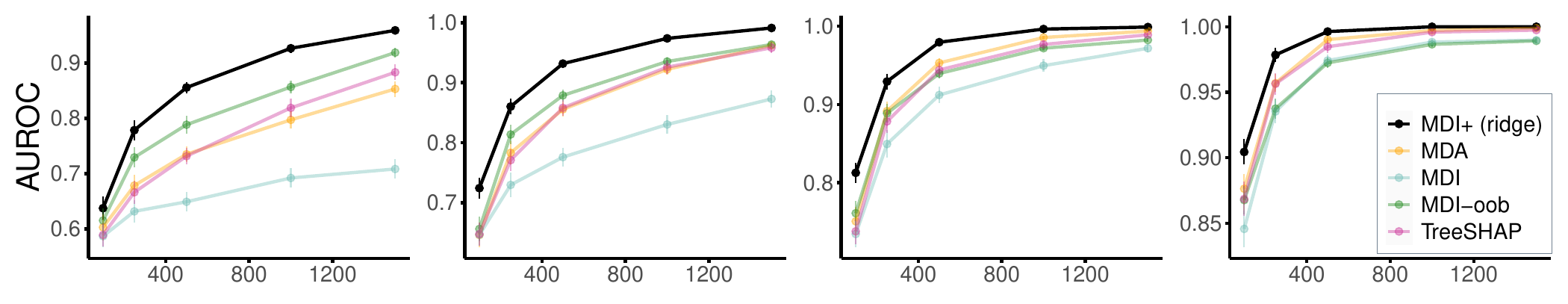}
\end{subfigure}
\begin{subfigure}{\textwidth}
    \centering
    \vspace{-2mm}
    \hspace{2mm}
    \rotatebox{90}{{\centering \small \phantom{\textbf{(A) LRg}}}}
    \hspace{1mm}
    \rotatebox{90}{{\hspace{8.5mm} \centering \small \textsc{CCLE}}}
    \includegraphics[width=0.85\textwidth]{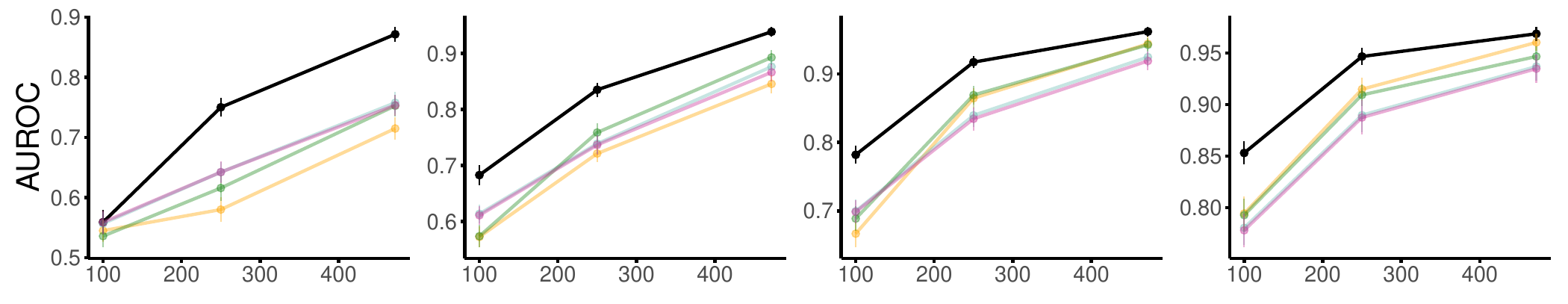}
\end{subfigure}
\begin{subfigure}{\textwidth}
    \centering
    \vspace{-2mm}
    \hspace{2mm}
    \rotatebox{90}{{\centering \small \phantom{\textbf{(A) LRg}}}}
    \hspace{1mm}
    \rotatebox{90}{{\hspace{6.25mm} \centering \small \textsc{Juvenile}}}
    \includegraphics[width=0.85\textwidth]{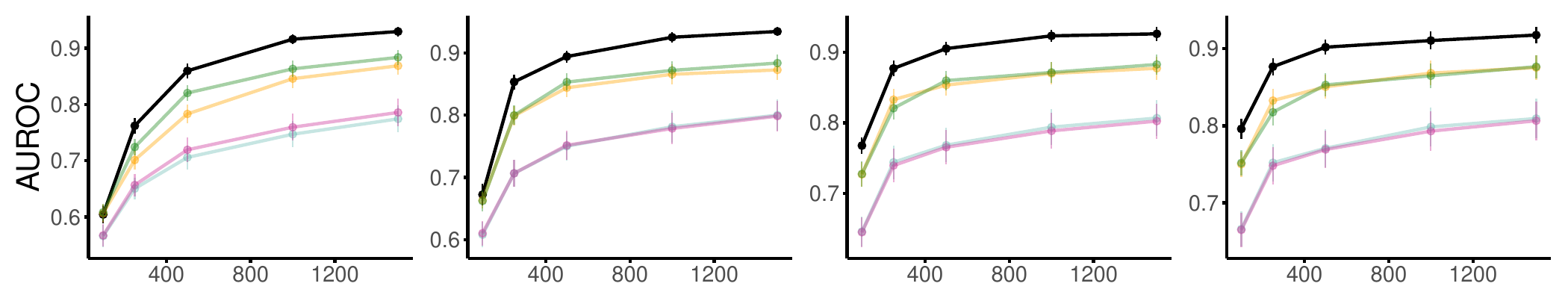}
\end{subfigure}
\begin{subfigure}{\textwidth}
    \centering
    \vspace{-2mm}
    \hspace{2mm}
    \rotatebox{90}{{\centering \small \phantom{\textbf{(A) LRg}}}}
    \hspace{1mm}
    \rotatebox{90}{{\hspace{10mm} \centering \small \textsc{Splicing}}}
    \includegraphics[width=0.85\textwidth]{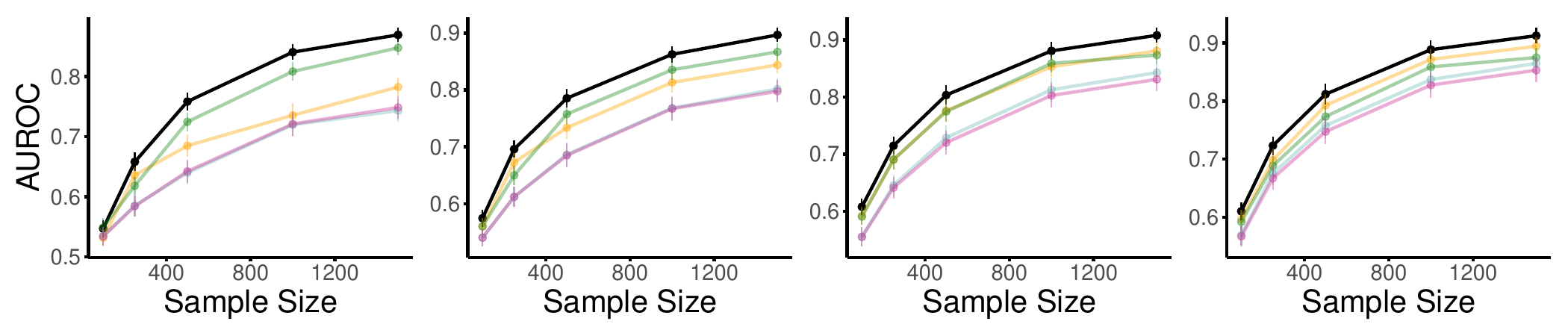}
    \vspace{-5mm}
\end{subfigure}
\noindent\makebox[\textwidth]{\hfill\rule{0.95\textwidth}{0.4pt}\hfill}
\begin{subfigure}{\textwidth}
    \centering
    \begin{tabularx}{0.91\textwidth}{@{}l *5{>{\centering\arraybackslash}X}@{}}
    \begin{minipage}{.07\textwidth}
    \phantom{}
    \end{minipage}%
    \end{tabularx}\\
    \vspace{-5mm}
    \hspace{2mm}
    \rotatebox{90}{{\hspace{-42mm} \centering \small \textbf{(B) Regression: LSS}}}
    \hspace{1mm}
    \rotatebox{90}{{\hspace{6mm} \centering \small \textsc{Enhancer}}}
    \includegraphics[width=0.85\textwidth]{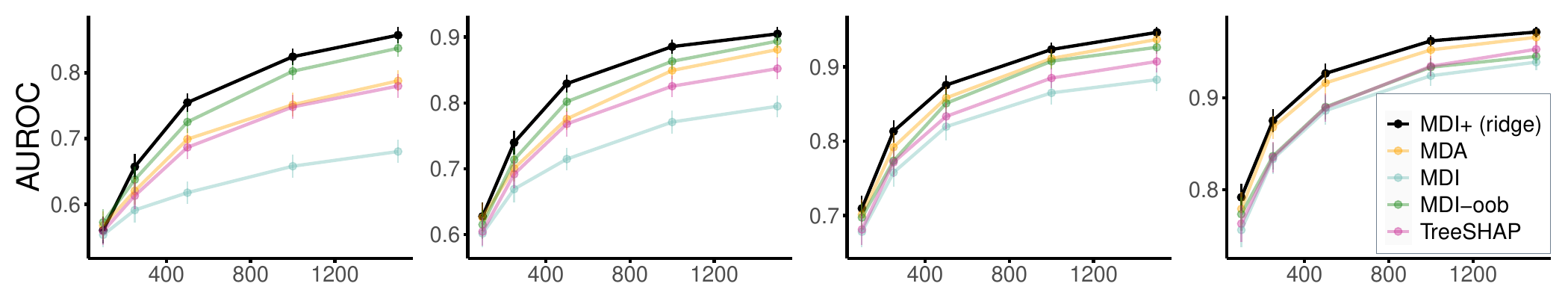}
\end{subfigure}
\begin{subfigure}{\textwidth}
    \centering
    \vspace{-2mm}
    \hspace{2mm}
    \rotatebox{90}{{\centering \small \phantom{\textbf{(A) LRg}}}}
    \hspace{1mm}
    \rotatebox{90}{{\hspace{8.5mm} \centering \small \textsc{CCLE}}}
    \includegraphics[width=0.85\textwidth]{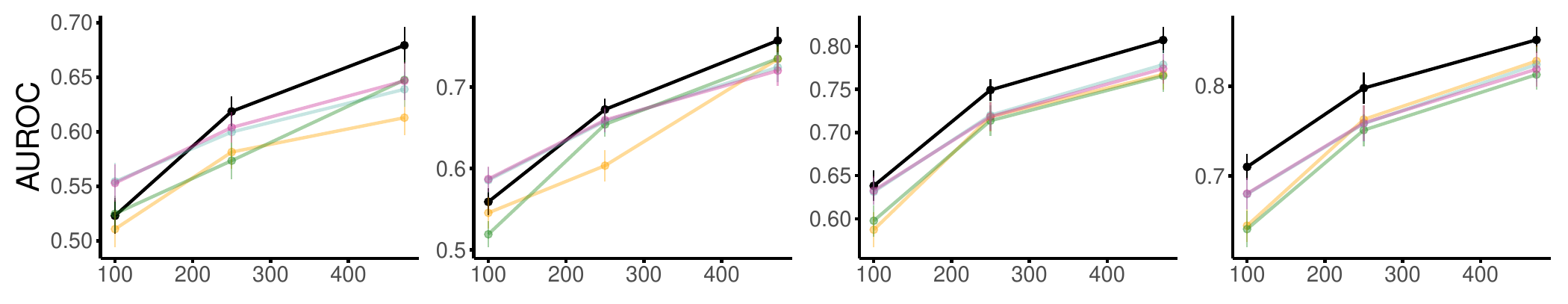}
\end{subfigure}
\begin{subfigure}{\textwidth}
    \centering
    \vspace{-2mm}
    \hspace{2mm}
    \rotatebox{90}{{\centering \small \phantom{\textbf{(A) LRg}}}}
    \hspace{1mm}
    \rotatebox{90}{{\hspace{6.25mm} \centering \small \textsc{Juvenile}}}
    \includegraphics[width=0.85\textwidth]{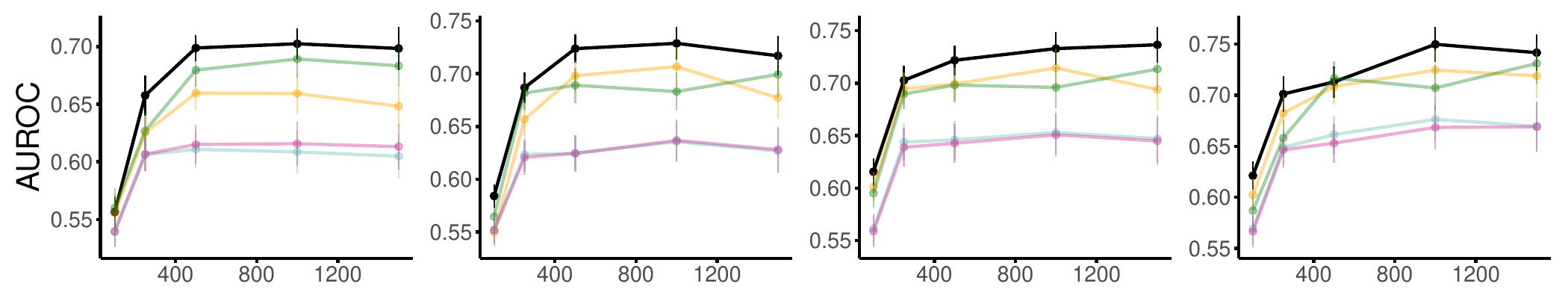}
\end{subfigure}
\begin{subfigure}{\textwidth}
    \centering
    \vspace{-2mm}
    \hspace{2mm}
    \rotatebox{90}{{\centering \small \phantom{\textbf{(A) LRg}}}}
    \hspace{1mm}
    \rotatebox{90}{{\hspace{10mm} \centering \small \textsc{Splicing}}}
    \includegraphics[width=0.85\textwidth]{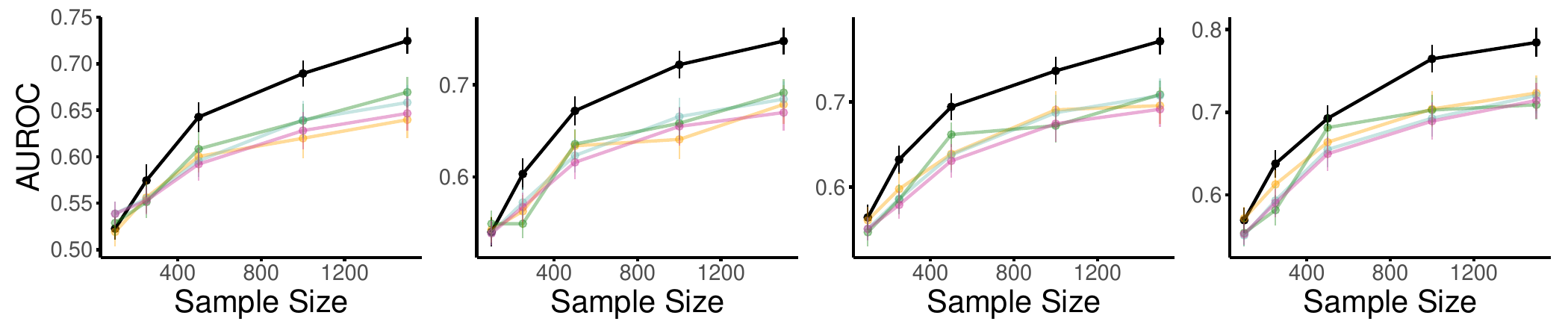}
\end{subfigure}
\caption{\method\;(ridge) outperforms other feature importance methods under the (A) linear regression and (B) LSS regression settings described in Section~\ref{subsec:sim_set_up}. This pattern is evident across various datasets (specified by row), proportions of variance explained (specified by column), and sample sizes (on the $x$-axis). In all subplots, the AUROC has been averaged across 50 experimental replicates, and error bars represent $\pm$ 1SE.}
\label{fig:reg_linear_appendix}
\end{figure}

\begin{figure}[h!]
\begin{subfigure}{\textwidth}
    \centering
    \vspace{-5mm}
    \begin{tabularx}{0.95\textwidth}{@{}l *5{>{\centering\arraybackslash}X}@{}}
    \begin{minipage}{.12\textwidth}
    \phantom{}
    \end{minipage}%
    & {\small $PVE=0.1$} & {\small $PVE=0.2$} & {\small $PVE=0.4$} & {\small $PVE=0.8$}
    \end{tabularx}\\
    \vspace{-1mm}
    \hspace{2mm}
    \rotatebox{90}{{\hspace{-54mm} \centering \small \textbf{(A) Regression: Polynomial Interaction}}}
    \hspace{1mm}
    \rotatebox{90}{{\hspace{6mm} \centering \small \textsc{Enhancer}}}
    \includegraphics[width=0.85\textwidth]{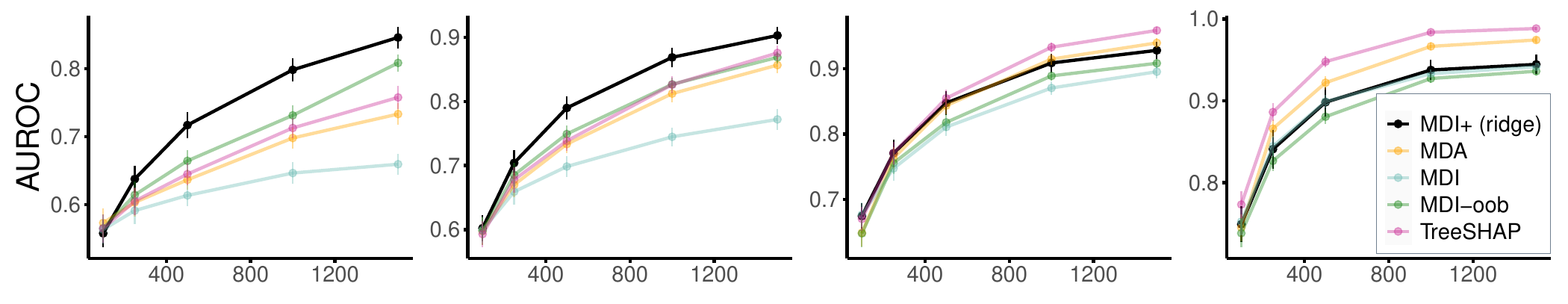}
\end{subfigure}
\begin{subfigure}{\textwidth}
    \centering
    \vspace{-2mm}
    \hspace{2mm}
    \rotatebox{90}{{\centering \small \phantom{\textbf{(A) LRg}}}}
    \hspace{1mm}
    \rotatebox{90}{{\hspace{8.5mm} \centering \small \textsc{CCLE}}}
    \includegraphics[width=0.85\textwidth]{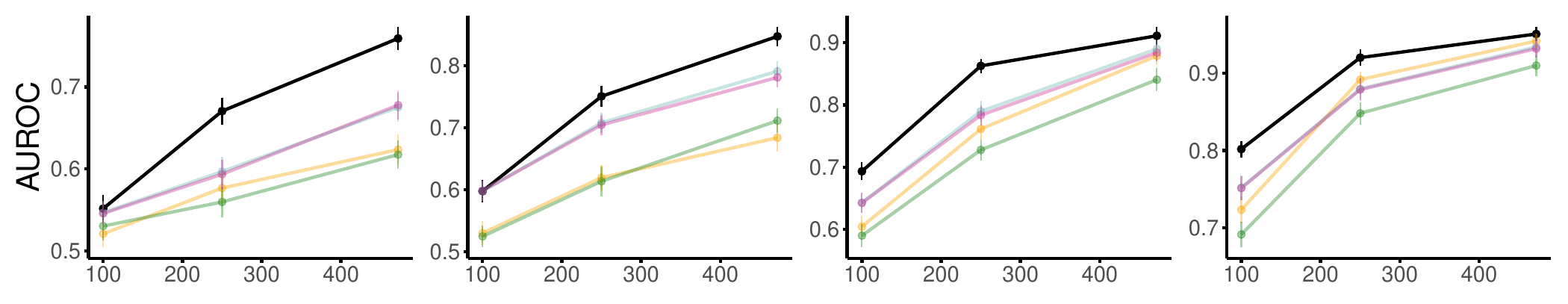}
\end{subfigure}
\begin{subfigure}{\textwidth}
    \centering
    \vspace{-2mm}
    \hspace{2mm}
    \rotatebox{90}{{\centering \small \phantom{\textbf{(A) LRg}}}}
    \hspace{1mm}
    \rotatebox{90}{{\hspace{6.25mm} \centering \small \textsc{Juvenile}}}
    \includegraphics[width=0.85\textwidth]{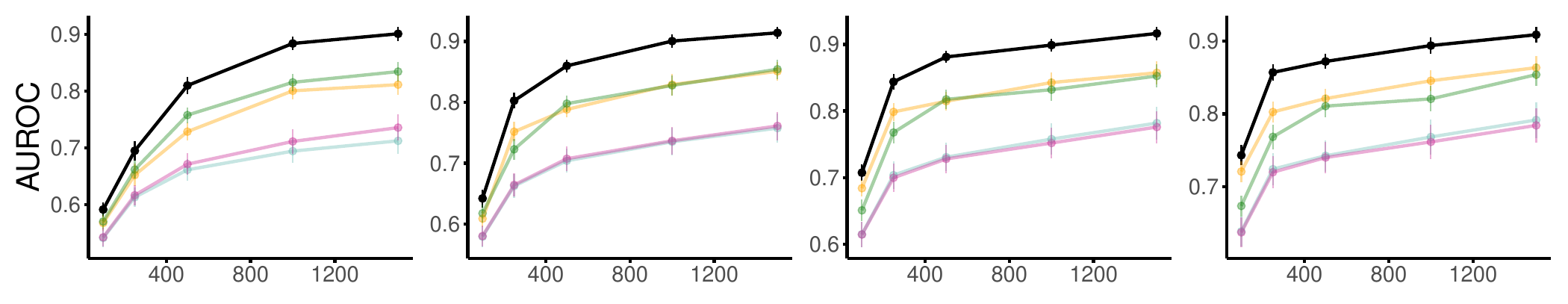}
\end{subfigure}
\begin{subfigure}{\textwidth}
    \centering
    \vspace{-2mm}
    \hspace{2mm}
    \rotatebox{90}{{\centering \small \phantom{\textbf{(A) LRg}}}}
    \hspace{1mm}
    \rotatebox{90}{{\hspace{10mm} \centering \small \textsc{Splicing}}}
    \includegraphics[width=0.85\textwidth]{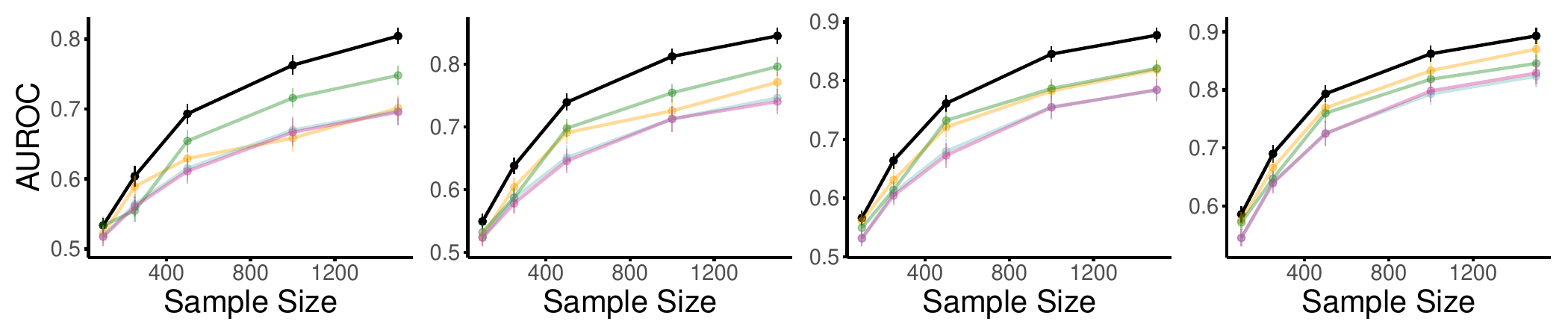}
    \vspace{-5mm}
\end{subfigure}
\noindent\makebox[\textwidth]{\hfill\rule{0.95\textwidth}{0.4pt}\hfill}
\begin{subfigure}{\textwidth}
    \centering
    \begin{tabularx}{1\textwidth}{@{}l *5{>{\centering\arraybackslash}X}@{}}
    \begin{minipage}{.075\textwidth}
    \phantom{}
    \end{minipage}%
    \end{tabularx}\\
    \vspace{-5mm}
    \hspace{2mm}
    \rotatebox{90}{{\hspace{-50mm} \centering \small \textbf{(B) Regression: Linear + LSS}}}
    \hspace{1mm}
    \rotatebox{90}{{\hspace{6mm} \centering \small \textsc{Enhancer}}}
    \includegraphics[width=0.85\textwidth]{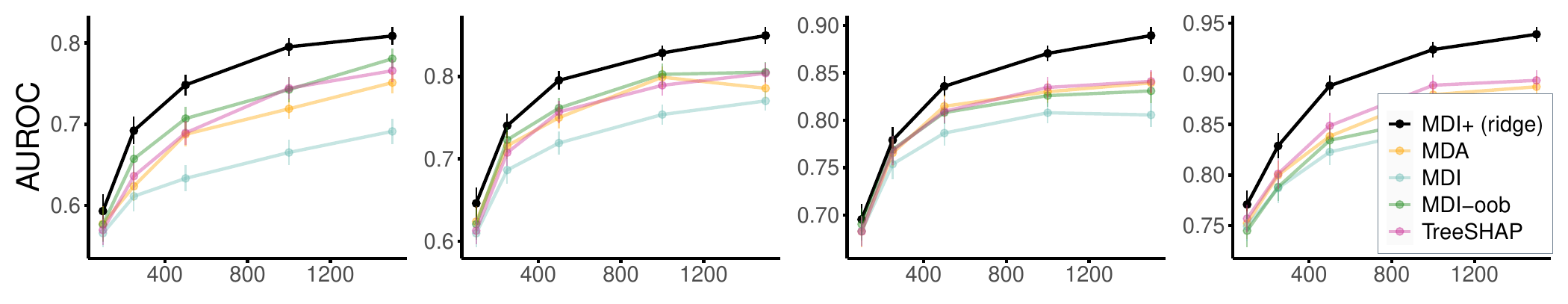}
\end{subfigure}
\begin{subfigure}{\textwidth}
    \centering
    \vspace{-2mm}
    \hspace{2mm}
    \rotatebox{90}{{\centering \small \phantom{\textbf{(A) LRg}}}}
    \hspace{1mm}
    \rotatebox{90}{{\hspace{8.5mm} \centering \small \textsc{CCLE}}}
    \includegraphics[width=0.85\textwidth]{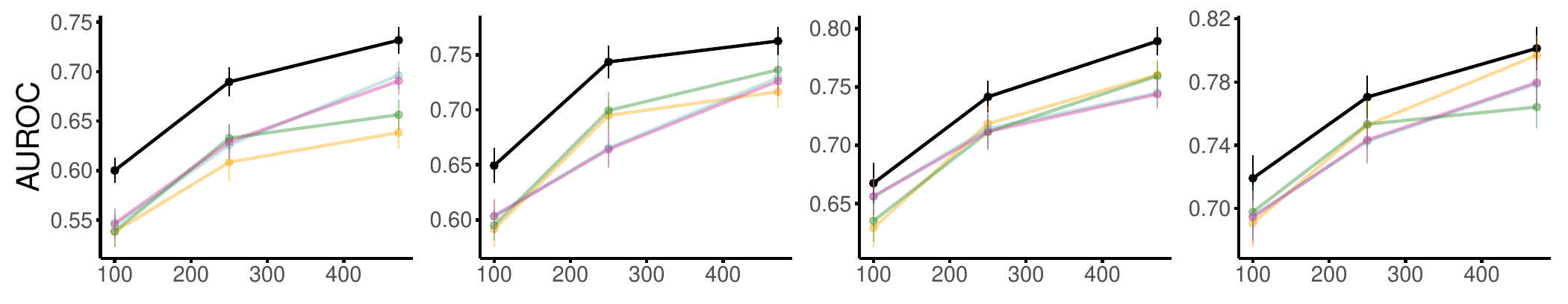}
\end{subfigure}
\begin{subfigure}{\textwidth}
    \centering
    \vspace{-2mm}
    \hspace{2mm}
    \rotatebox{90}{{\centering \small \phantom{\textbf{(A) LRg}}}}
    \hspace{1mm}
    \rotatebox{90}{{\hspace{6.25mm} \centering \small \textsc{Juvenile}}}
    \includegraphics[width=0.85\textwidth]{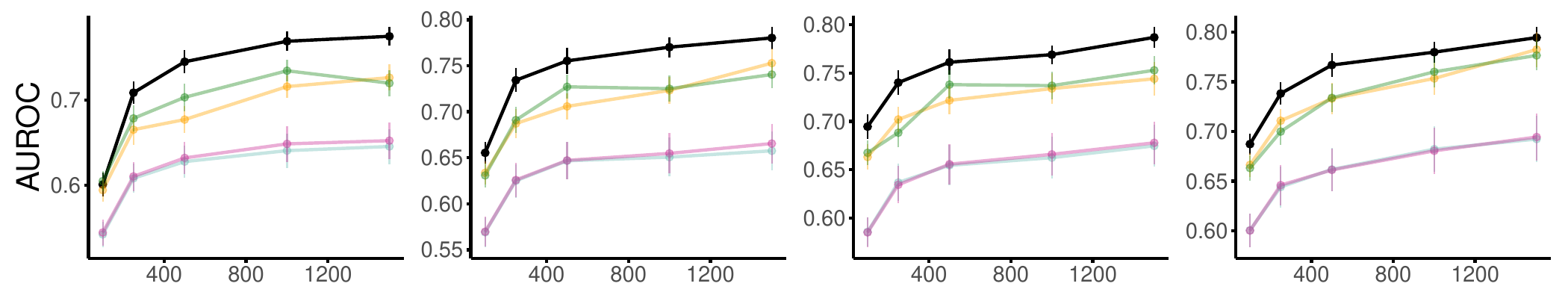}
\end{subfigure}
\begin{subfigure}{\textwidth}
    \centering
    \vspace{-2mm}
    \hspace{2mm}
    \rotatebox{90}{{\centering \small \phantom{\textbf{(A) LRg}}}}
    \hspace{1mm}
    \rotatebox{90}{{\hspace{10mm} \centering \small \textsc{Splicing}}}
    \includegraphics[width=0.85\textwidth]{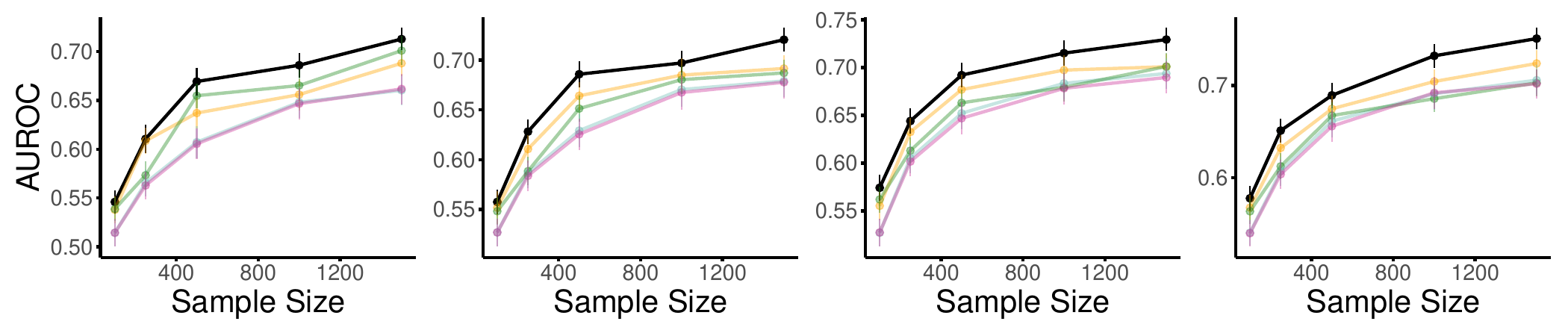}
\end{subfigure}
\caption{\method\;(ridge) outperforms other feature importance methods under the (A) polynomial interaction regression and (B) linear + LSS regression settings described in Section~\ref{subsec:sim_set_up}. This pattern is evident across various datasets (specified by row), proportions of variance explained (specified by column), and sample sizes (on the $x$-axis). In all subplots, the AUROC has been averaged across 50 experimental replicates, and error bars represent $\pm$ 1SE.}
\label{fig:reg_linear_lss_appendix}
\end{figure}

\paragraph{Regression Simulations.} Following the simulation setup from Sections \ref{subsec:sim_set_up} and \ref{subsec:regression_results}, we provide the results for all datasets and regression functions in the regression setting. We show the results for all datasets under the linear, LSS, polynomial interaction, and linear $+$ LSS response functions in Figures~\ref{fig:reg_linear_appendix} - \ref{fig:reg_linear_lss_appendix}.

\begin{figure}[h!]
\begin{subfigure}{\textwidth}
    \centering
    \vspace{-5mm}
    \begin{tabularx}{.95\textwidth}{@{}l *5{>{\centering\arraybackslash}X}@{}}
    \begin{minipage}{.12\textwidth}
    \phantom{}
    \end{minipage}%
    & {\small $25\%$ Corrupted} & {\small $15\%$ Corrupted} & {\small $5\%$ Corrupted} & {\small $0\%$ Corrupted}
    \end{tabularx}\\
    \vspace{-1mm}
    \hspace{2mm}
    \rotatebox{90}{{\hspace{-48mm} \centering \small \textbf{(A) Classification: Logistic}}}
    \hspace{1mm}
    \rotatebox{90}{{\hspace{6mm} \centering \small \textsc{Enhancer}}}
    \includegraphics[width=0.85\textwidth]{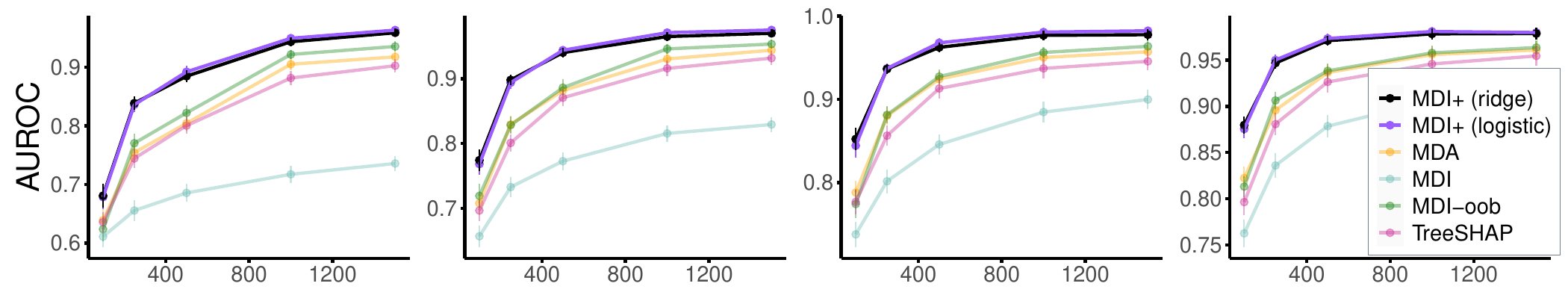}
\end{subfigure}
\begin{subfigure}{\textwidth}
    \centering
    \vspace{-2mm}
    \hspace{2mm}
    \rotatebox{90}{{\hspace{-35mm} \centering \small \phantom{(A) Linear}}}
    \hspace{1mm}
    \rotatebox{90}{{\hspace{8.5mm} \centering \small \textsc{CCLE}}}
    \includegraphics[width=0.85\textwidth]{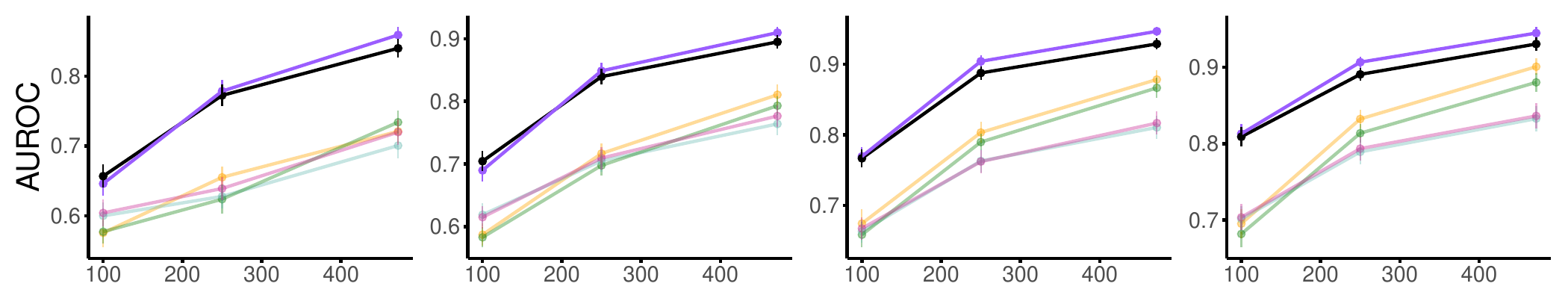}
\end{subfigure}
\begin{subfigure}{\textwidth}
    \centering
    \vspace{-2mm}
    \hspace{2mm}
    \rotatebox{90}{{\hspace{-35mm} \centering \small \phantom{(A) Linear}}}
    \hspace{1mm}
    \rotatebox{90}{{\hspace{6.25mm} \centering \small \textsc{Juvenile}}}
    \includegraphics[width=0.85\textwidth]{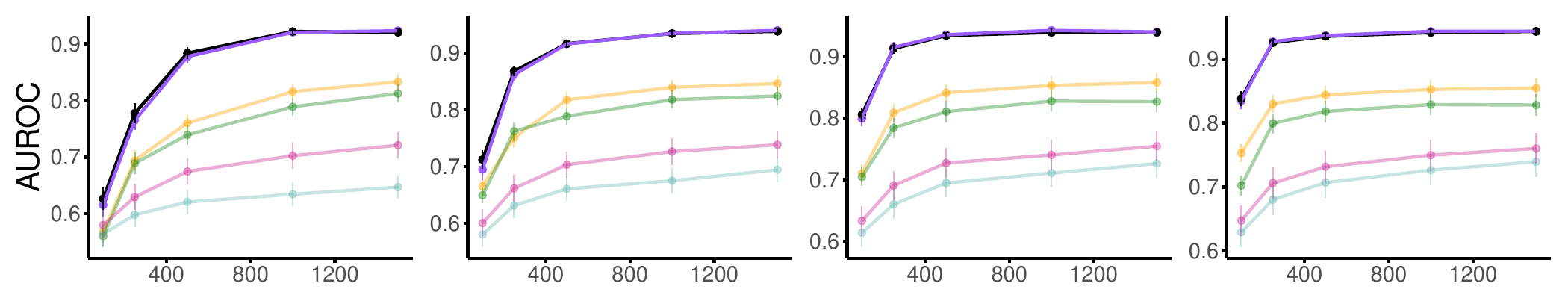}
\end{subfigure}
\begin{subfigure}{\textwidth}
    \centering
    \vspace{-2mm}
    \hspace{2mm}
    \rotatebox{90}{{\hspace{-35mm} \centering \small \phantom{(A) Linear}}}
    \hspace{1mm}
    \rotatebox{90}{{\hspace{10mm} \centering \small \textsc{Splicing}}}
    \includegraphics[width=0.85\textwidth]{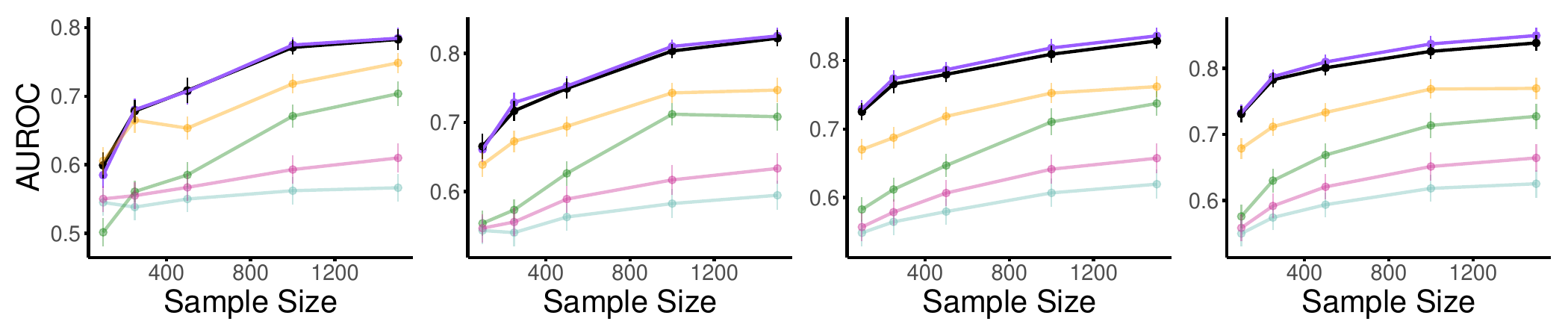}
    \vspace{-5mm}
\end{subfigure}
\noindent\makebox[\textwidth]{\hfill\rule{0.95\textwidth}{0.4pt}\hfill}
\begin{subfigure}{\textwidth}
    \centering
    \begin{tabularx}{1\textwidth}{@{}l *5{>{\centering\arraybackslash}X}@{}}
    \begin{minipage}{.075\textwidth}
    \phantom{}
    \end{minipage}%
    \end{tabularx}\\
    \vspace{-5mm}
    \hspace{2mm}
    \rotatebox{90}{{\hspace{-52mm} \centering \small \textbf{(B) Classification: Logistic LSS}}}
    \hspace{1mm}
    \rotatebox{90}{{\hspace{6mm} \centering \small \textsc{Enhancer}}}
    \includegraphics[width=0.85\textwidth]{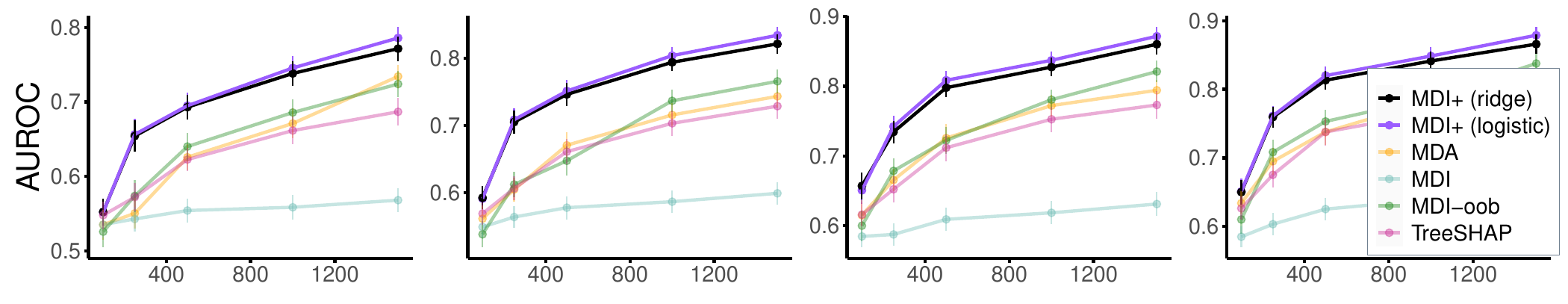}
\end{subfigure}
\begin{subfigure}{\textwidth}
    \centering
    \vspace{-2mm}
    \hspace{2mm}
    \rotatebox{90}{{\hspace{-50mm} \centering \small \phantom{(A) Linear}}}
    \hspace{1mm}
    \rotatebox{90}{{\hspace{8.5mm} \centering \small \textsc{CCLE}}}
    \includegraphics[width=0.85\textwidth]{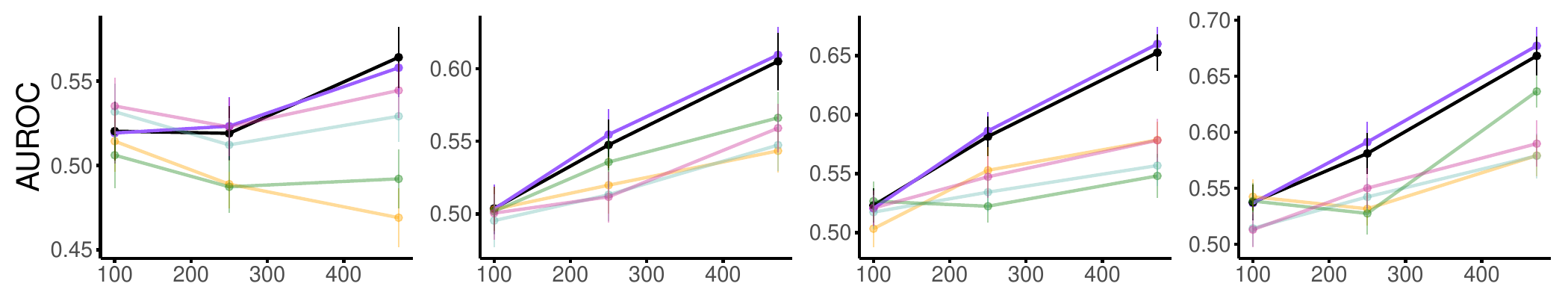}
\end{subfigure}
\begin{subfigure}{\textwidth}
    \centering
    \vspace{-2mm}
    \hspace{2mm}
    \rotatebox{90}{{\hspace{-35mm} \centering \small \phantom{(A) Linear}}}
    \hspace{1mm}
    \rotatebox{90}{{\hspace{6.25mm} \centering \small \textsc{Juvenile}}}
    \includegraphics[width=0.85\textwidth]{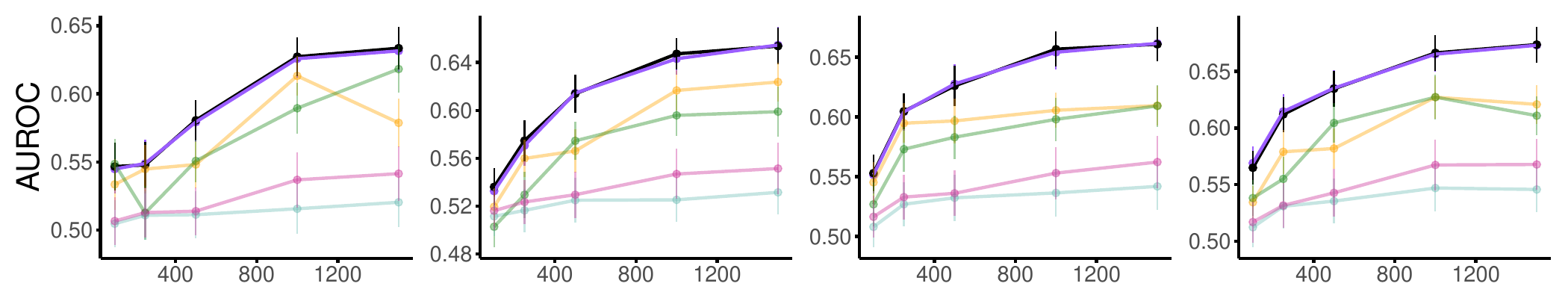}
\end{subfigure}
\begin{subfigure}{\textwidth}
    \centering
    \vspace{-2mm}
    \hspace{2mm}
    \rotatebox{90}{{\hspace{-35mm} \centering \small \phantom{(A) Linear}}}
    \hspace{1mm}
    \rotatebox{90}{{\hspace{10mm} \centering \small \textsc{Splicing}}}
    \includegraphics[width=0.85\textwidth]{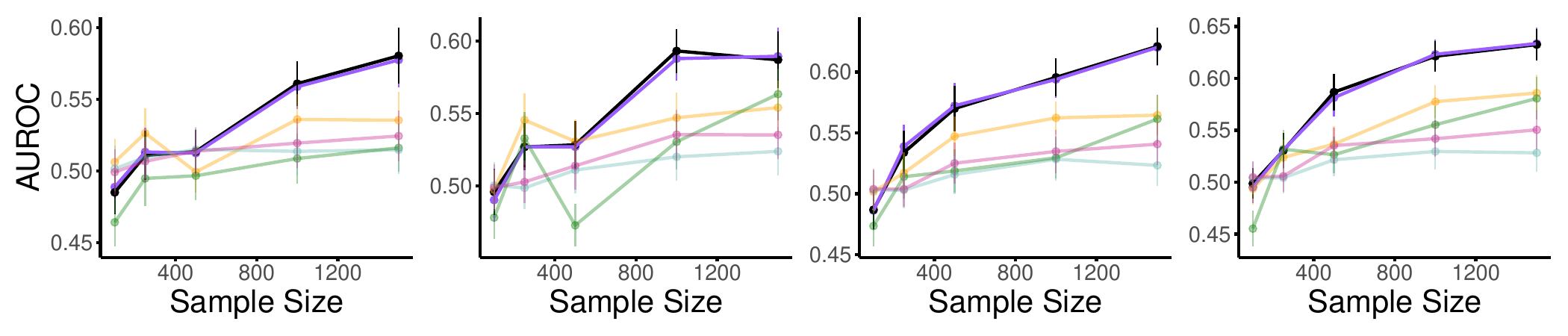}
\end{subfigure}
\caption{\method\;(ridge) and \method\;(logistic) outperform all other feature importance methods under the (A) logistic and (B) logistic LSS classification settings described in Section \ref{subsec:sim_set_up}.  This pattern is evident across datasets with different covariate structures (specified by row), proportions of corrupted labels (specified by column), and sample sizes (on the $x$-axis). Furthermore, \method\;(logistic) often slightly outperforms \method\;(ridge). In all subplots, the AUROC has been averaged across 50 experimental replicates, and error bars represent $\pm$ 1SE.}
\label{fig:class_linear_appendix}
\end{figure}

\begin{figure}[h!]
\begin{subfigure}{\textwidth}
    \centering
    \vspace{-5mm}
    \begin{tabularx}{.95\textwidth}{@{}l *5{>{\centering\arraybackslash}X}@{}}
    \begin{minipage}{.12\textwidth}
    \phantom{}
    \end{minipage}%
    & {\small $25\%$ Corrupted} & {\small $15\%$ Corrupted} & {\small $5\%$ Corrupted} & {\small $0\%$ Corrupted}
    \end{tabularx}\\
    \vspace{-1mm}
    \hspace{2mm}
    \rotatebox{90}{{\hspace{-68mm} \centering \small \textbf{(A) Classification: Logistic Polynomial Interaction}}}
    \hspace{1mm}
    \rotatebox{90}{{\hspace{6mm} \centering \small \textsc{Enhancer}}}
    \includegraphics[width=0.85\textwidth]{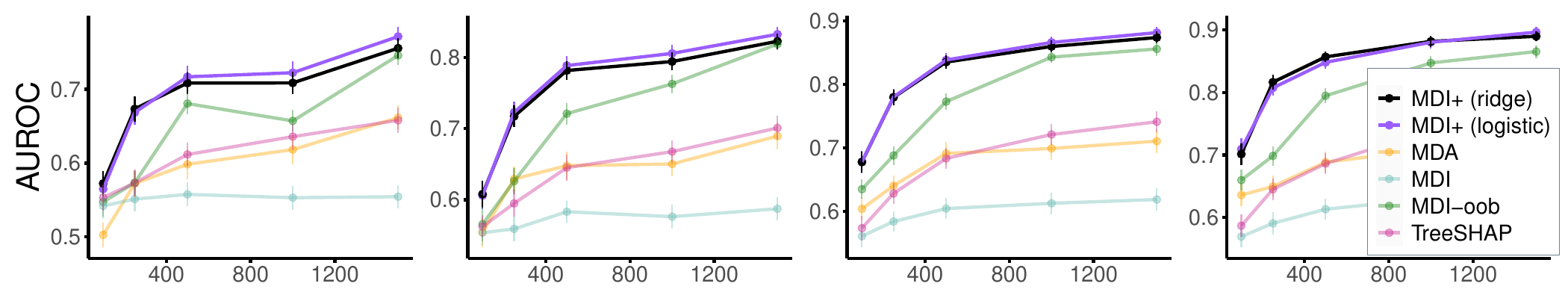}
\end{subfigure}
\begin{subfigure}{\textwidth}
    \centering
    \vspace{-2mm}
    \hspace{2mm}
    \rotatebox{90}{{\hspace{-50mm} \centering \small \phantom{CLPg}}}
    \hspace{1mm}
    \rotatebox{90}{{\hspace{8.5mm} \centering \small \textsc{CCLE}}}
    \includegraphics[width=0.85\textwidth]{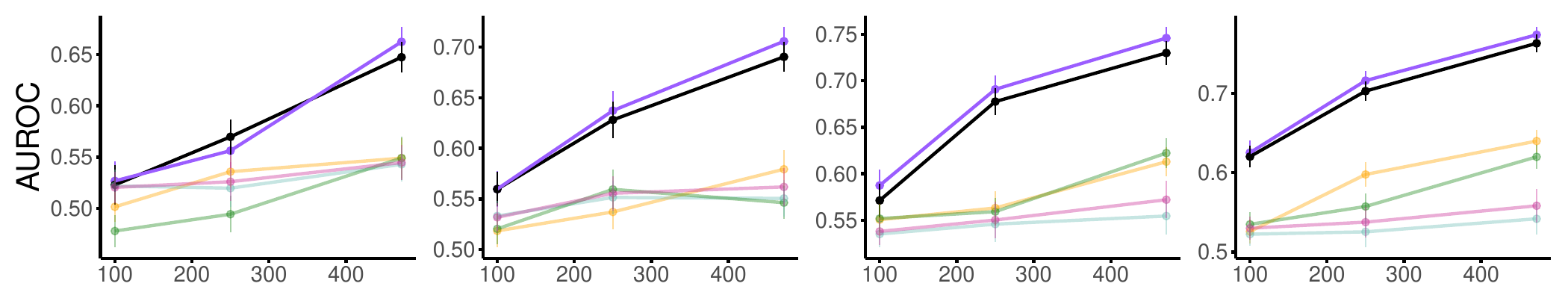}
\end{subfigure}
\begin{subfigure}{\textwidth}
    \centering
    \vspace{-2mm}
    \hspace{2mm}
    \rotatebox{90}{{\hspace{-50mm} \centering \small \phantom{CLPg}}}
    \hspace{1mm}
    \rotatebox{90}{{\hspace{6.25mm} \centering \small \textsc{Juvenile}}}
    \includegraphics[width=0.85\textwidth]{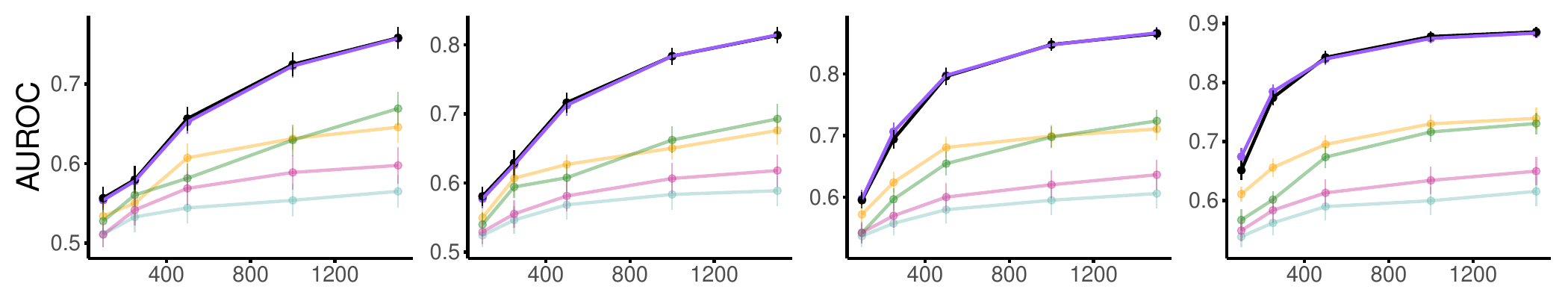}
\end{subfigure}
\begin{subfigure}{\textwidth}
    \centering
    \vspace{-2mm}
    \hspace{2mm}
    \rotatebox{90}{{\hspace{-50mm} \centering \small \phantom{CLPg}}}
    \hspace{1mm}
    \rotatebox{90}{{\hspace{10mm} \centering \small \textsc{Splicing}}}
    \includegraphics[width=0.85\textwidth]{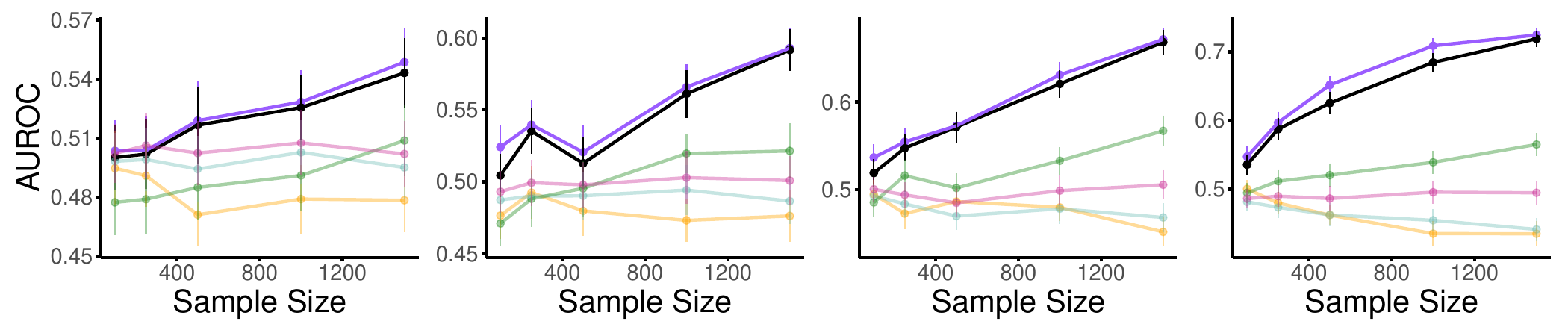}
    \vspace{-5mm}
\end{subfigure}
\noindent\makebox[\textwidth]{\hfill\rule{0.95\textwidth}{0.4pt}\hfill}
\begin{subfigure}{\textwidth}
    \centering    
    \begin{tabularx}{1\textwidth}{@{}l *5{>{\centering\arraybackslash}X}@{}}
    \begin{minipage}{.075\textwidth}
    \phantom{}
    \end{minipage}%
    \end{tabularx}\\
    \vspace{-5mm}
    \hspace{2mm}
    \rotatebox{90}{{\hspace{-54mm} \centering \small \textbf{(B) Classification: Logistic + LSS}}}
    \hspace{1mm}
    \rotatebox{90}{{\hspace{6mm} \centering \small \textsc{Enhancer}}}
    \includegraphics[width=0.85\textwidth]{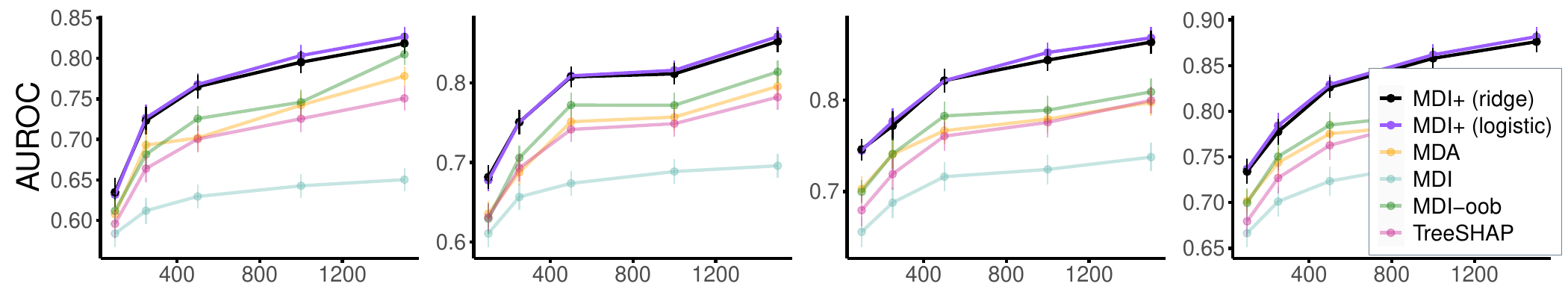}
\end{subfigure}
\begin{subfigure}{\textwidth}
    \centering
    \vspace{-2mm}
    \hspace{2mm}
    \rotatebox{90}{{\hspace{-50mm} \centering \small \phantom{CLPg}}}
    \hspace{1mm}
    \rotatebox{90}{{\hspace{8.5mm} \centering \small \textsc{CCLE}}}
    \includegraphics[width=0.85\textwidth]{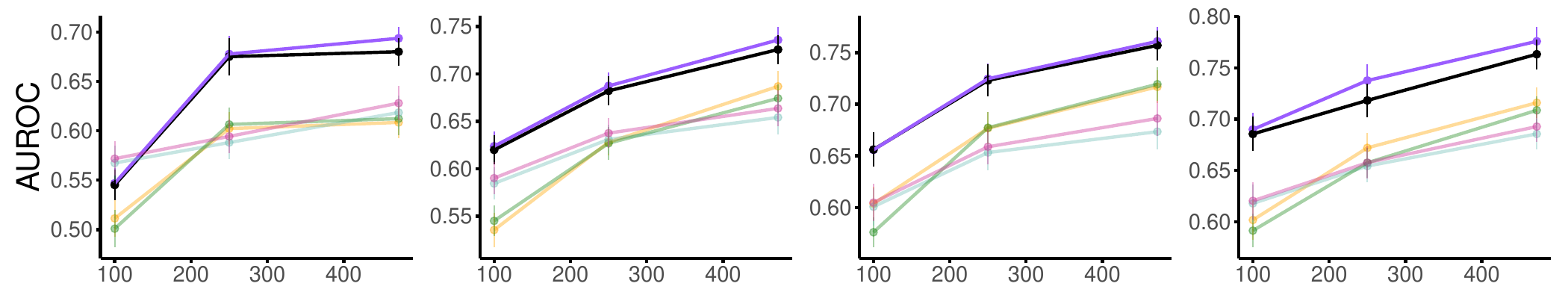}
\end{subfigure}
\begin{subfigure}{\textwidth}
    \centering
    \vspace{-2mm}
    \hspace{2mm}
    \rotatebox{90}{{\hspace{-50mm} \centering \small \phantom{CLPg}}}
    \hspace{1mm}
    \rotatebox{90}{{\hspace{6.25mm} \centering \small \textsc{Juvenile}}}
    \includegraphics[width=0.85\textwidth]{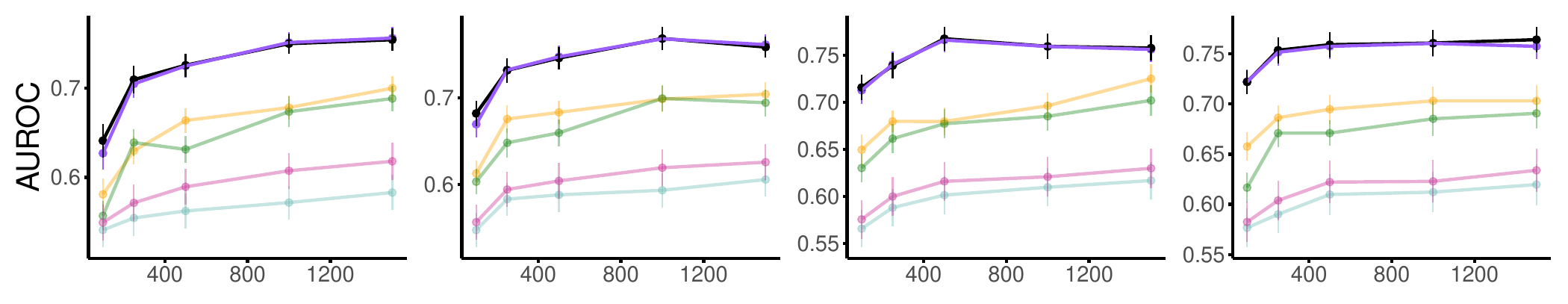}
\end{subfigure}
\begin{subfigure}{\textwidth}
    \centering
    \vspace{-2mm}
    \hspace{2mm}
    \rotatebox{90}{{\hspace{-50mm} \centering \small \phantom{CLPg}}}
    \hspace{1mm}
    \rotatebox{90}{{\hspace{10mm} \centering \small \textsc{Splicing}}}
    \includegraphics[width=0.85\textwidth]{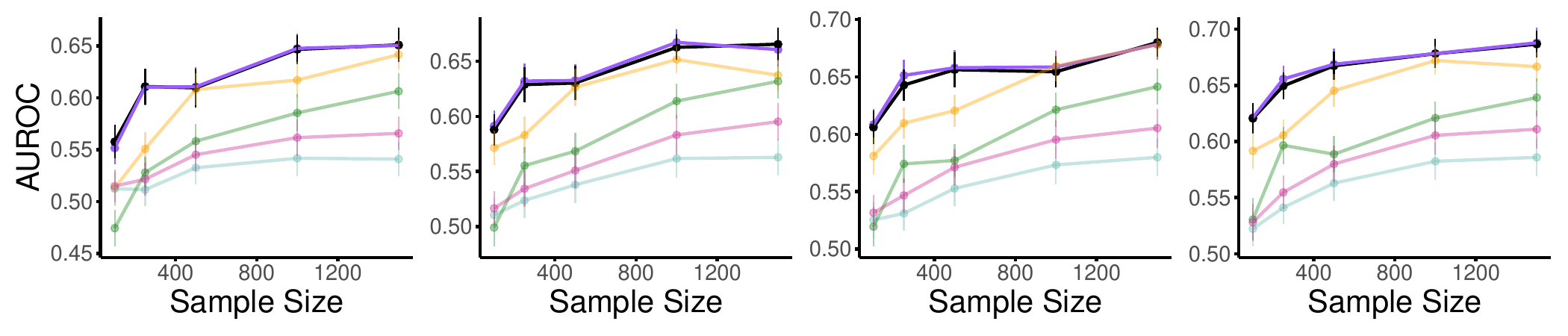}
\end{subfigure}
\caption{\method\;(ridge) and \method\;(logistic) outperform all other feature importance methods under the (A) logistic polynomial interaction and (B) logistic + LSS classification settings described in Section \ref{subsec:sim_set_up}.  This pattern is evident across datasets with different covariate structures (specified by row), proportions of corrupted labels (specified by column), and sample sizes (on the $x$-axis). Furthermore, \method\;(logistic) often slightly outperforms \method\;(ridge). In all subplots, the AUROC has been averaged across 50 experimental replicates, and error bars represent $\pm$ 1SE.}
\label{fig:class_linear_lss_appendix}
\end{figure}

\paragraph{Classification Simulations.}
Following the simulation setup from Sections \ref{subsec:sim_set_up} and \ref{subsec:classification_results}, we provide the results for all datasets and regression functions in the classification setting. We show the results for all datasets under the linear, LSS, polynomial interaction, and linear $+$ LSS response functions in Figures~\ref{fig:class_linear_appendix} - \ref{fig:class_linear_lss_appendix}. All regression functions in the classification setting were passed through the logistic link function to obtain binary responses.

\begin{figure}[h!]
\begin{subfigure}{\textwidth}
    \centering
    \vspace{-5mm}
    \begin{tabularx}{.95\textwidth}{@{}l *5{>{\centering\arraybackslash}X}@{}}
    \begin{minipage}{.12\textwidth}
    \phantom{}
    \end{minipage}%
    & {\small  $5\%$ Outliers} & {\small $2.5\%$ Outliers} & {\small $1\%$ Outliers} & {\small $0\%$ Outliers}
    \end{tabularx}\\
    \vspace{-1mm}
    \rotatebox{90}{{\hspace{-63mm} \centering \small \textbf{(A) Robust Regression: Linear with Outliers}}}
    \hspace{1mm}
    \rotatebox{90}{{\hspace{5.5mm} \centering \small \textsc{Enhancer}}}
    \hspace{1mm}
    \rotatebox{90}{{\hspace{5mm} \centering \small $\mu_{corrupt}= 10$}}  \includegraphics[width=0.85\textwidth]{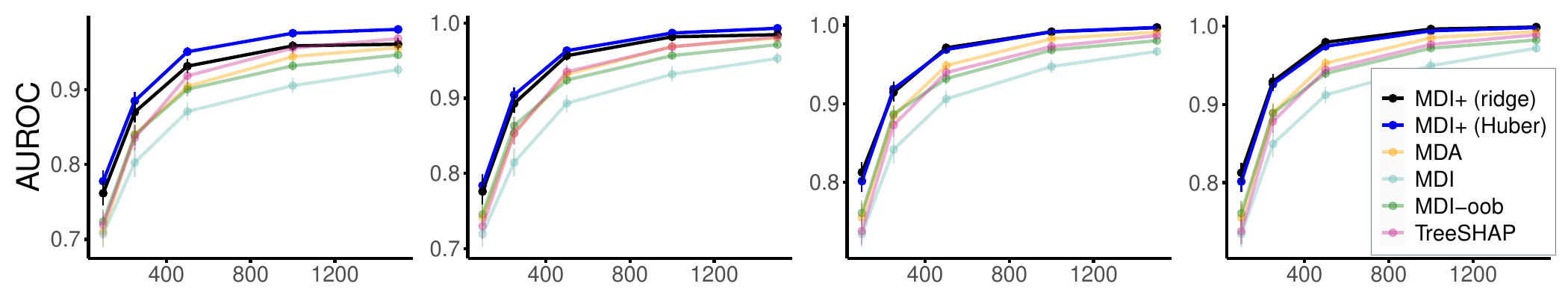}
\end{subfigure}
\begin{subfigure}{\textwidth}
    \centering
    \vspace{-2mm}
    \rotatebox{90}{{\centering \small \phantom{(A) Rgl}}}
    \hspace{1mm}
    \rotatebox{90}{{\hspace{8.5mm} \centering \small \textsc{Enhancer}}}
    \hspace{1mm}
    \rotatebox{90}{{\hspace{8mm} \centering \small $\mu_{corrupt}= 25$}}
    \includegraphics[width=0.85\textwidth]{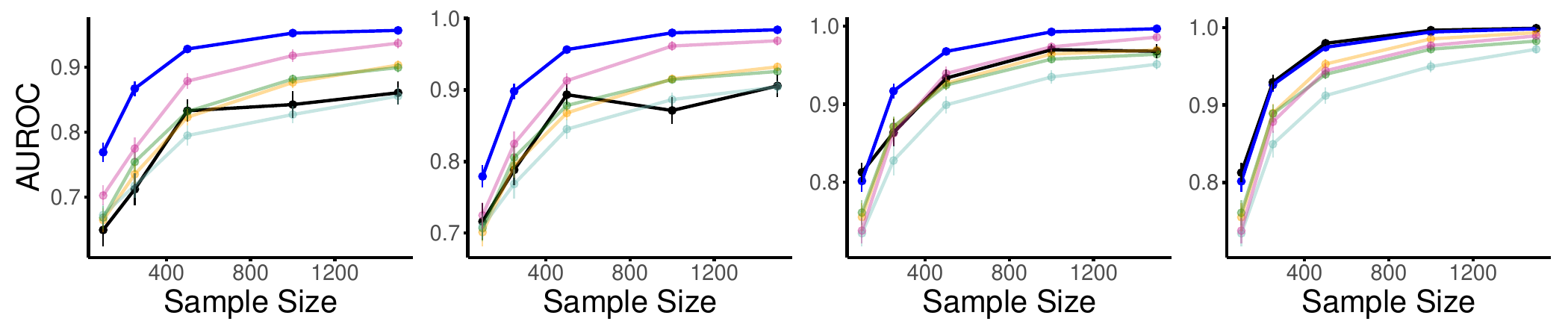}
\end{subfigure}
\begin{subfigure}{\textwidth}
    \centering
    \vspace{-5mm}
    \rotatebox{90}{{\centering \small \phantom{(A) Rgl}}}
    \hspace{1mm}
    \rotatebox{90}{{\hspace{8.5mm} \centering \small \textsc{CCLE}}}
    \hspace{1mm}
    \rotatebox{90}{{\hspace{5.5mm} \centering \small $\mu_{corrupt}= 10$}}  \includegraphics[width=0.85\textwidth]{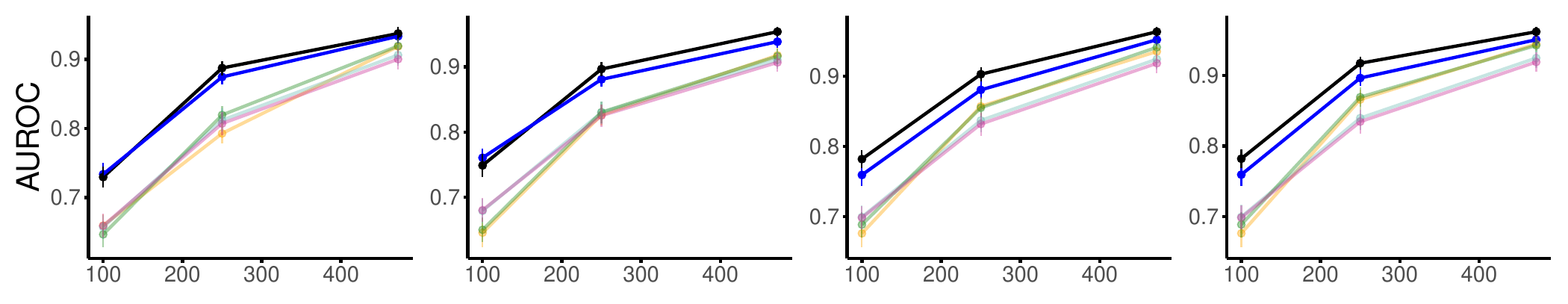}
\end{subfigure}
\begin{subfigure}{\textwidth}
    \centering
    \vspace{-2mm}
    \rotatebox{90}{{\centering \small \phantom{(A) Rgl}}}
    \hspace{1mm}
    \rotatebox{90}{{\hspace{11mm} \centering \small \textsc{CCLE}}}
    \hspace{1mm}
    \rotatebox{90}{{\hspace{8mm} \centering \small $\mu_{corrupt}= 25$}}
    \includegraphics[width=0.85\textwidth]{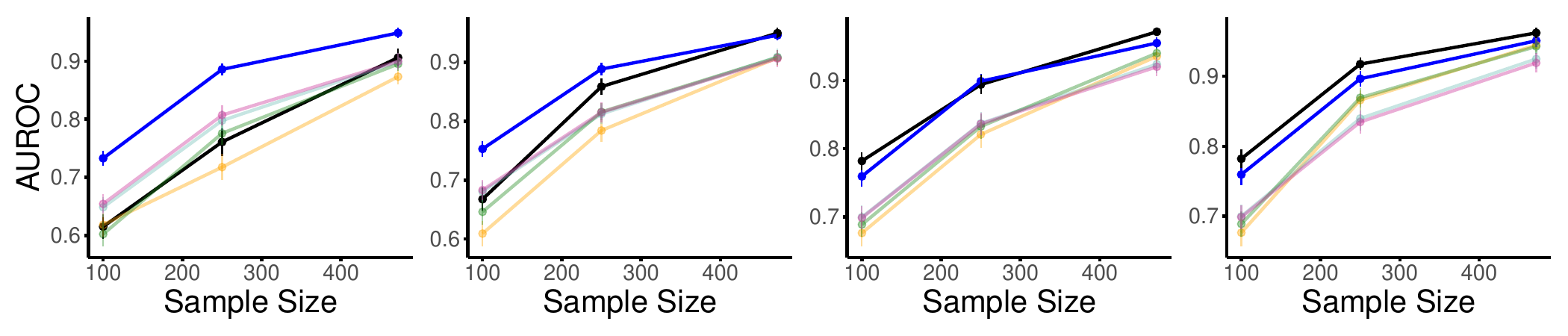}
    \vspace{-5mm}
\end{subfigure}
\noindent\makebox[\textwidth]{\hfill\rule{0.95\textwidth}{0.4pt}\hfill}
\begin{subfigure}{\textwidth}
    \centering
    \begin{tabularx}{1\textwidth}{@{}l *5{>{\centering\arraybackslash}X}@{}}
    \begin{minipage}{.092\textwidth}
    \phantom{}
    \end{minipage}%
    \end{tabularx}\\
    \vspace{-5mm}
    \rotatebox{90}{{\hspace{-60mm} \centering \small \textbf{(B) Robust Regression: LSS with Outliers}}}
    \hspace{1mm}
    \rotatebox{90}{{\hspace{5.5mm} \centering \small \textsc{Enhancer}}}
    \hspace{1mm}
    \rotatebox{90}{{\hspace{5mm} \centering \small $\mu_{corrupt}= 10$}}  \includegraphics[width=0.85\textwidth]{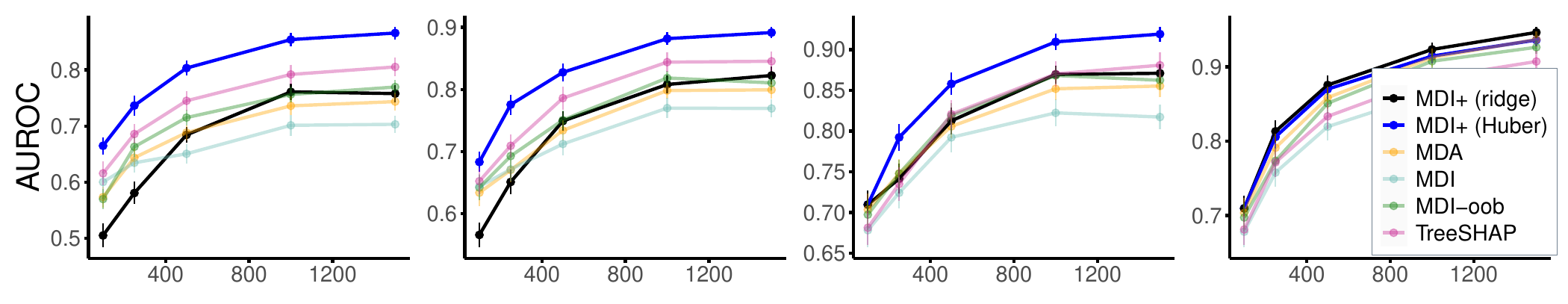}
\end{subfigure}
\begin{subfigure}{\textwidth}
    \centering
    \vspace{-2mm}
    \rotatebox{90}{{\centering \small \phantom{(A) Rgl}}}
    \hspace{1mm}
    \rotatebox{90}{{\hspace{8.5mm} \centering \small \textsc{Enhancer}}}
    \hspace{1mm}
    \rotatebox{90}{{\hspace{8mm} \centering \small $\mu_{corrupt}= 25$}}
    \includegraphics[width=0.85\textwidth]{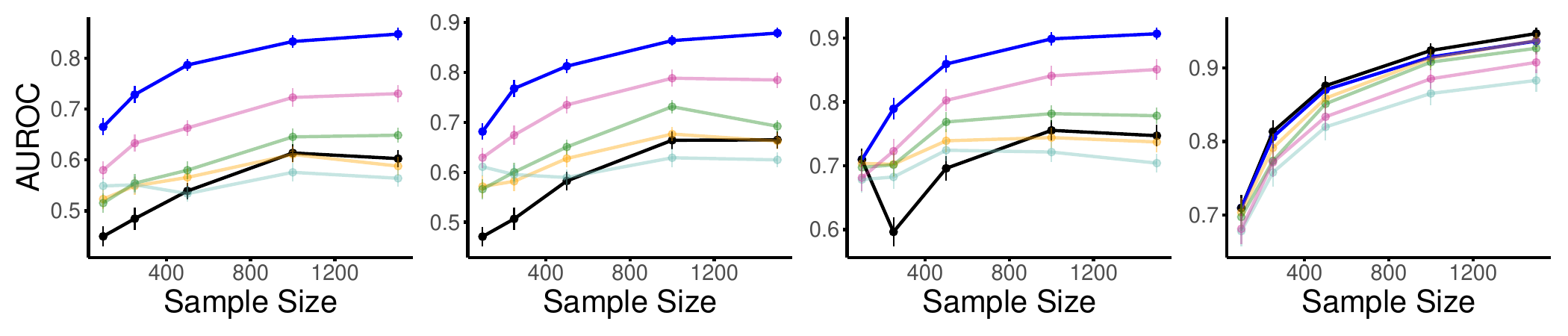}
\end{subfigure}
\begin{subfigure}{\textwidth}
    \centering
    \vspace{-5mm}
    \rotatebox{90}{{\centering \small \phantom{(A) Rgl}}}
    \hspace{1mm}
    \rotatebox{90}{{\hspace{8.5mm} \centering \small \textsc{CCLE}}}
    \hspace{1mm}
    \rotatebox{90}{{\hspace{5.5mm} \centering \small $\mu_{corrupt}= 10$}}  \includegraphics[width=0.85\textwidth]{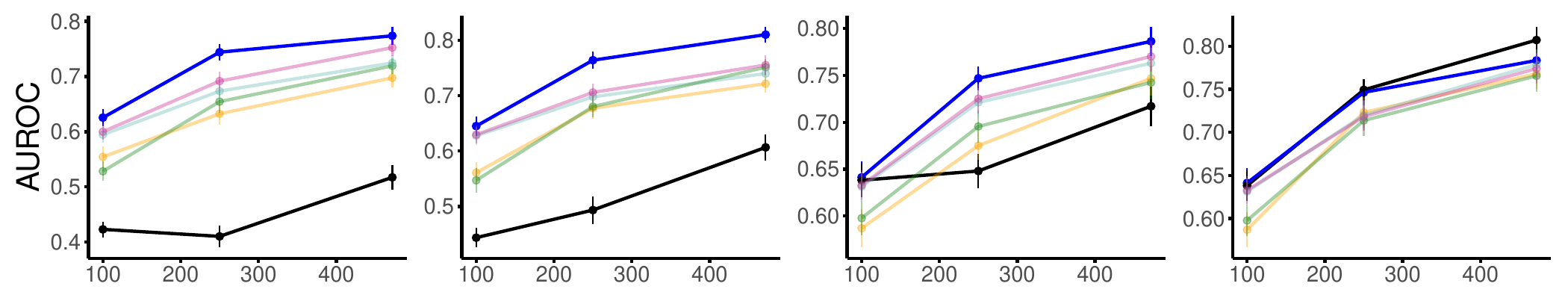}
\end{subfigure}
\begin{subfigure}{\textwidth}
    \centering
    \vspace{-2mm}
    \rotatebox{90}{{\centering \small \phantom{(A) Rgl}}}
    \hspace{1mm}
    \rotatebox{90}{{\hspace{11mm} \centering \small \textsc{CCLE}}}
    \hspace{1mm}
    \rotatebox{90}{{\hspace{8mm} \centering \small $\mu_{corrupt}= 25$}}
    \includegraphics[width=0.85\textwidth]{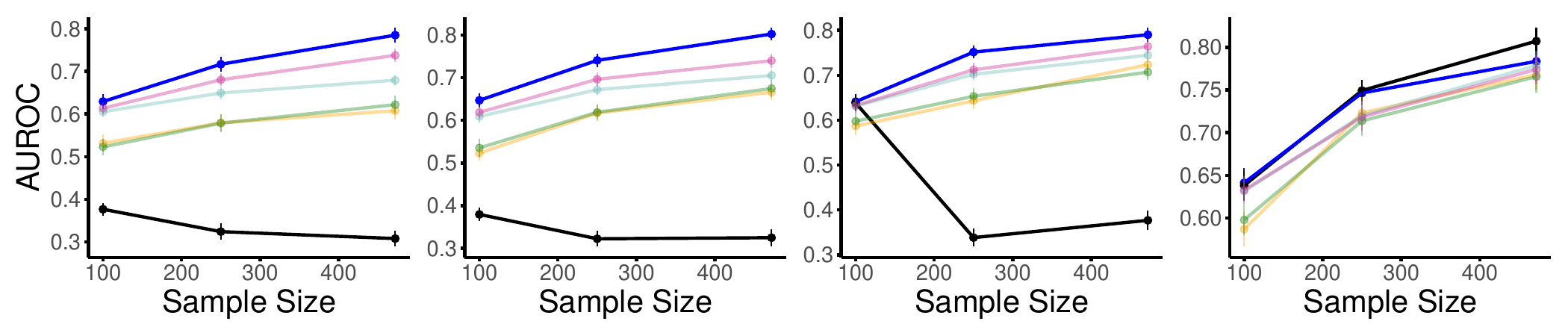}
\end{subfigure}
\caption{Under the (A) linear with outliers and (B) LSS with outliers regression setting (see Section~\ref{subsec:robust_results} for details), \method\;(Huber)'s performance remains suffers far less than other methods including \method\;(Ridge) as the level of corruption $\mu_{corrupt}$ (specified by row) and the proportion of outliers (specified by column) grow. This pattern holds across datasets (specified by row) and sample sizes (on the $x$-axis). In all subplots, the AUROC has been averaged across 50 experimental replicates, and error bars represent $\pm$ 1SE.}
\label{fig:robust_linear_appendix}
\end{figure}

\begin{figure}[h!]
\begin{subfigure}{\textwidth}
    \centering
    \vspace{-5mm}
    \begin{tabularx}{.95\textwidth}{@{}l *5{>{\centering\arraybackslash}X}@{}}
    \begin{minipage}{.12\textwidth}
    \phantom{}
    \end{minipage}%
    & {\small  $5\%$ Outliers} & {\small $2.5\%$ Outliers} & {\small $1\%$ Outliers} & {\small $0\%$ Outliers}
    \end{tabularx}\\
    \vspace{-1mm}
    \rotatebox{90}{{\hspace{-78mm} \centering \small \textbf{(A) Robust Regression: Polynomial Interaction with Outliers}}}
    \hspace{1mm}
    \rotatebox{90}{{\hspace{5.5mm} \centering \small \textsc{Enhancer}}}
    \hspace{1mm}
    \rotatebox{90}{{\hspace{5mm} \centering \small $\mu_{corrupt}= 10$}}  
    \includegraphics[width=0.85\textwidth]{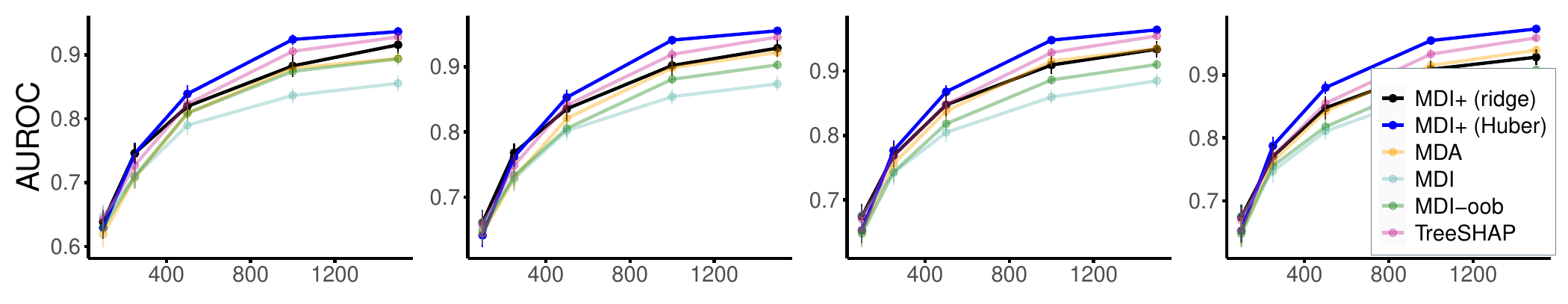}
\end{subfigure}
\begin{subfigure}{\textwidth}
    \centering
    \vspace{-2mm}
    \rotatebox{90}{{\centering \small \phantom{(A) Rgl}}}
    \hspace{1mm}
    \rotatebox{90}{{\hspace{8.5mm} \centering \small \textsc{Enhancer}}}
    \hspace{1mm}
    \rotatebox{90}{{\hspace{8mm} \centering \small $\mu_{corrupt}= 25$}}
    \includegraphics[width=0.85\textwidth]{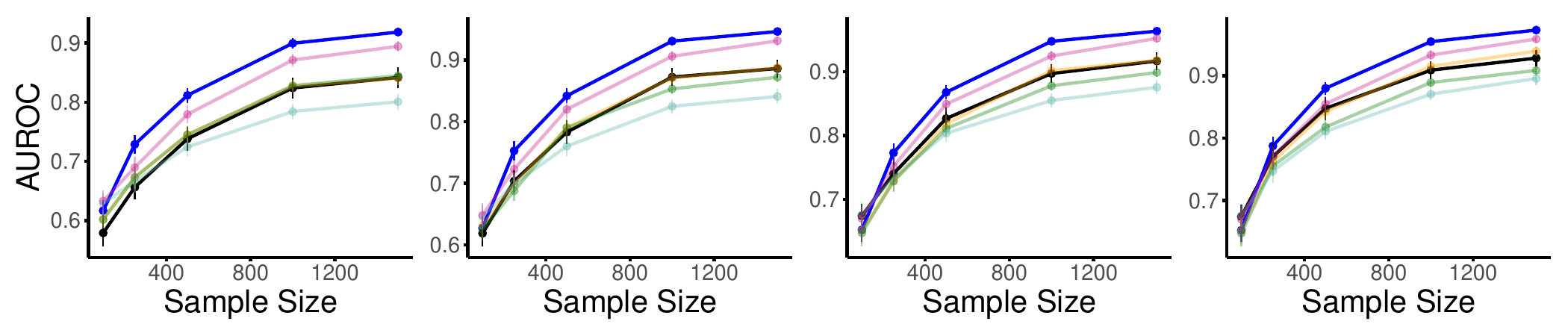}
\end{subfigure}
\begin{subfigure}{\textwidth}
    \centering
    \vspace{-5mm}
    \rotatebox{90}{{\centering \small \phantom{(A) Rgl}}}
    \hspace{1mm}
    \rotatebox{90}{{\hspace{8.5mm} \centering \small \textsc{CCLE}}}
    \hspace{1mm}
    \rotatebox{90}{{\hspace{5.5mm} \centering \small $\mu_{corrupt}= 10$}}  \includegraphics[width=0.85\textwidth]{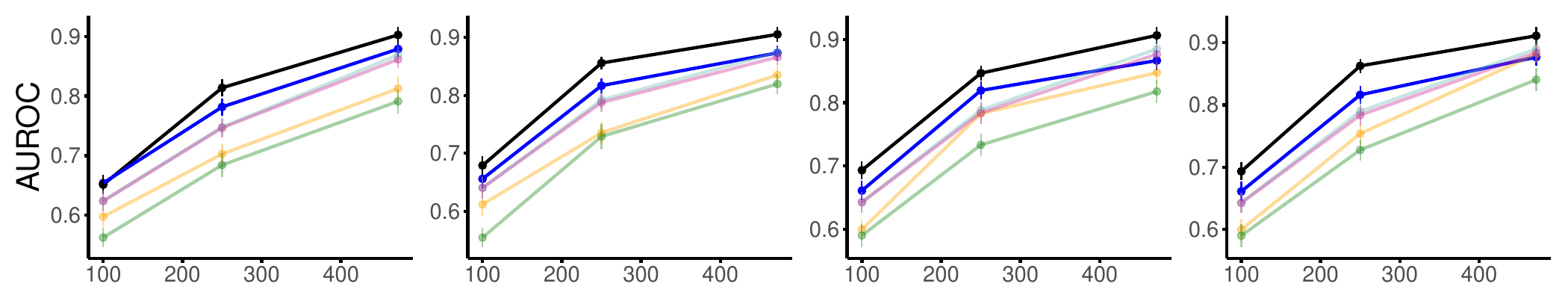}
\end{subfigure}
\begin{subfigure}{\textwidth}
    \centering
    \vspace{-2mm}
    \rotatebox{90}{{\centering \small \phantom{(A) Rgl}}}
    \hspace{1mm}
    \rotatebox{90}{{\hspace{11mm} \centering \small \textsc{CCLE}}}
    \hspace{1mm}
    \rotatebox{90}{{\hspace{8mm} \centering \small $\mu_{corrupt}= 25$}}
    \includegraphics[width=0.85\textwidth]{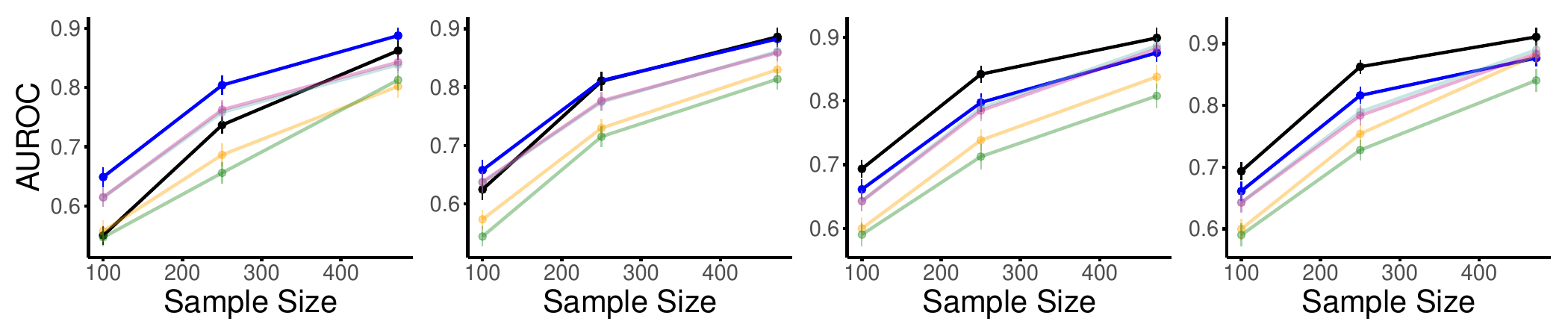}
    \vspace{-5mm}
\end{subfigure}
\noindent\makebox[\textwidth]{\hfill\rule{0.95\textwidth}{0.4pt}\hfill}
\begin{subfigure}{\textwidth}
    \centering
    \begin{tabularx}{1\textwidth}{@{}l *5{>{\centering\arraybackslash}X}@{}}
    \begin{minipage}{.092\textwidth}
    \phantom{}
    \end{minipage}%
    \end{tabularx}\\
    \vspace{-5mm}
    \rotatebox{90}{{\hspace{-68mm} \centering \small \textbf{(B) Robust Regression: Linear + LSS with Outliers}}}
    \hspace{1mm}
    \rotatebox{90}{{\hspace{5.5mm} \centering \small \textsc{Enhancer}}}
    \hspace{1mm}
    \rotatebox{90}{{\hspace{5mm} \centering \small $\mu_{corrupt}= 10$}}  \includegraphics[width=0.85\textwidth]{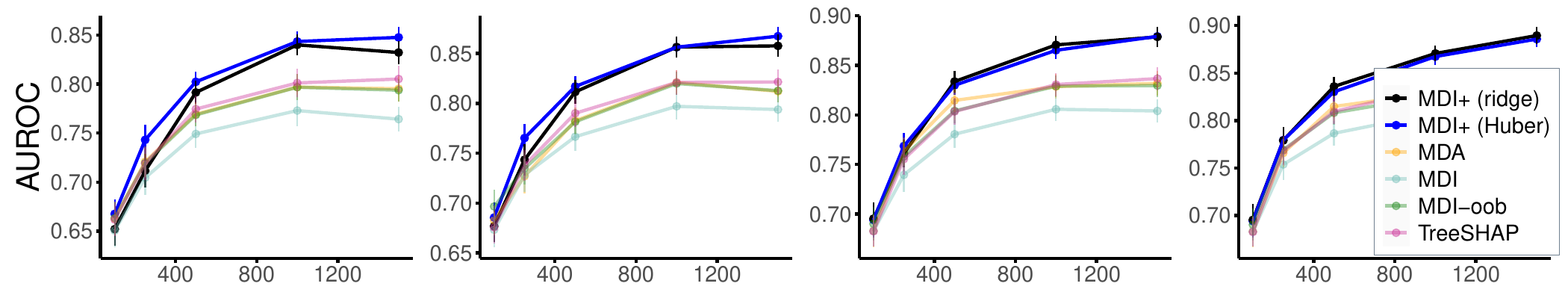}
\end{subfigure}
\begin{subfigure}{\textwidth}
    \centering
    \vspace{-2mm}
    \rotatebox{90}{{\centering \small \phantom{(A) Rgl}}}
    \hspace{1mm}
    \rotatebox{90}{{\hspace{8.5mm} \centering \small \textsc{Enhancer}}}
    \hspace{1mm}
    \rotatebox{90}{{\hspace{8mm} \centering \small $\mu_{corrupt}= 25$}}
    \includegraphics[width=0.85\textwidth]{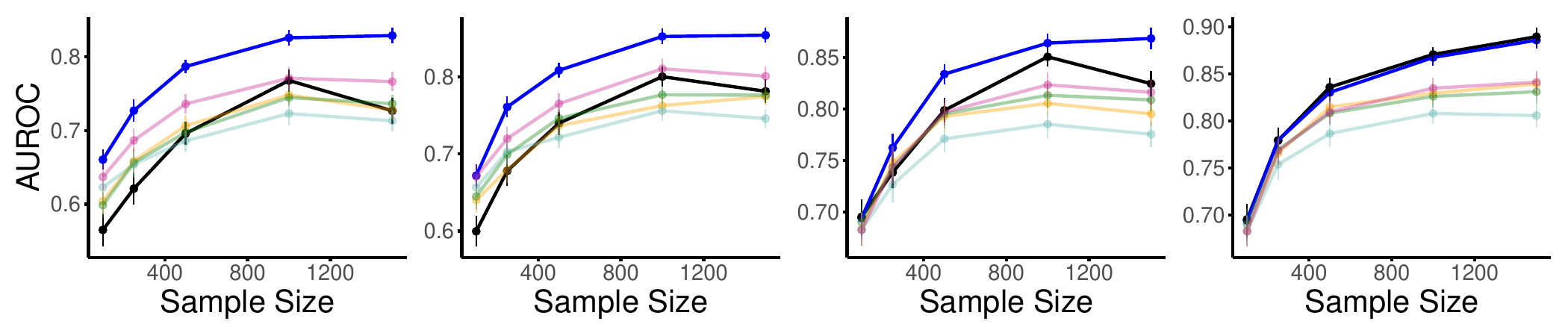}
\end{subfigure}
\begin{subfigure}{\textwidth}
    \centering
    \vspace{-5mm}
    \rotatebox{90}{{\centering \small \phantom{(A) Rgl}}}
    \hspace{1mm}
    \rotatebox{90}{{\hspace{8.5mm} \centering \small \textsc{CCLE}}}
    \hspace{1mm}
    \rotatebox{90}{{\hspace{5.5mm} \centering \small $\mu_{corrupt}= 10$}}  \includegraphics[width=0.85\textwidth]{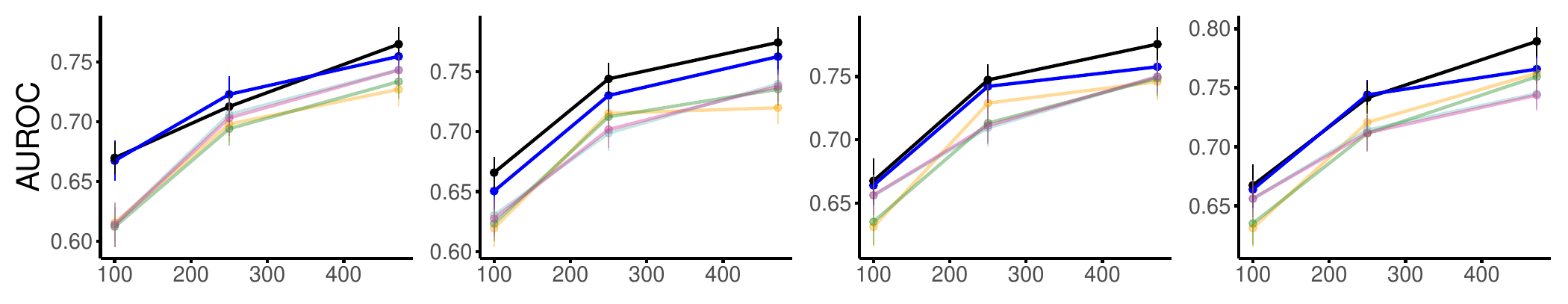}
\end{subfigure}
\begin{subfigure}{\textwidth}
    \centering
    \vspace{-2mm}
    \rotatebox{90}{{\centering \small \phantom{(A) Rgl}}}
    \hspace{1mm}
    \rotatebox{90}{{\hspace{11mm} \centering \small \textsc{CCLE}}}
    \hspace{1mm}
    \rotatebox{90}{{\hspace{8mm} \centering \small $\mu_{corrupt}= 25$}}
    \includegraphics[width=0.85\textwidth]{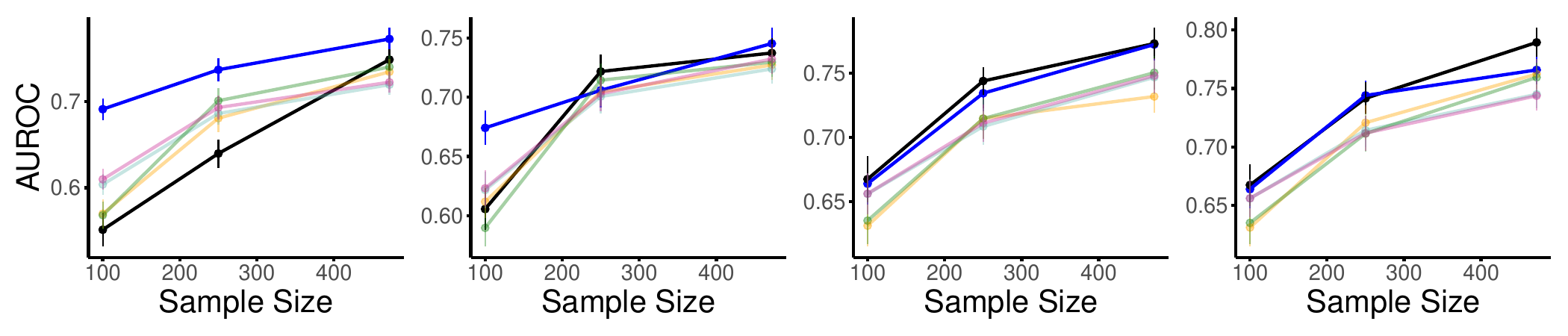}
\end{subfigure}
\caption{Under the (A) polynomial interaction with outliers and (B) linear + LSS with outliers regression setting (see Section~\ref{subsec:robust_results} for details), \method\;(Huber)'s performance remains suffers far less than other methods including \method\;(Ridge) as the level of corruption $\mu_{corrupt}$ (specified by row) and the proportion of outliers (specified by column) grow. This pattern also holds across datasets (specified by row) and sample sizes (on the $x$-axis). In all subplots, the AUROC has been averaged across 50 experimental replicates, and error bars represent $\pm$ 1SE.}
\label{fig:robust_linear_lss_appendix}
\end{figure}

\paragraph{Robust Regression Simulations.}
Following the simulation setup from Sections \ref{subsec:sim_set_up} and \ref{subsec:robust_results}, we provide the results for the Enhancer and CCLE datasets under the linear, LSS, polynomial interaction, and linear $+$ LSS response functions with outliers in Figures \ref{fig:robust_linear_appendix} - \ref{fig:robust_linear_lss_appendix}. 

\clearpage
\section{Additional Data-Inspired Feature Ranking Simulations}
\label{supp:additional_feature_ranking}
In this section, we describe other data-inspired simulations --- specifically, under a misspecificied setting, varying levels of sparsity of the generating function, and varying number of features in the covariate matrix.

\subsection{Misspecified Regression Simulations}
\label{supp:misspecified}

In practice, we are often unable to observe all covariates that are relevant for the response. To investigate \method\;under this type of misspecified model scenario, we consider the following simulation setup.

\paragraph{Experimental details.} We simulate data according to the four regression functions described in Section \ref{subsec:sim_set_up} (i.e., linear, LSS, polynomial interaction, and linear+LSS) but omit the first two signal features (i.e., $X_1,\; X_2$) from the covariate matrix $\bX$ before fitting the RF and feature importance method under study. For example, in the linear regression simulation, we simulate the response $y$ via $Y = \sum_{j=1}^{5}X_j + \varepsilon$, where $\varepsilon \stackrel{iid}{\sim} N(0, \sigma^2)$; however, we fit the RF and compute the feature importance measure using only $y$ and $X_3, \dots, X_p$. The rest of the experimental details are identical to those described previously. 

\begin{figure}[h!]
\begin{subfigure}{\textwidth}
    \centering
    \vspace{-5mm}
    \begin{tabularx}{0.95\textwidth}{@{}l *5{>{\centering\arraybackslash}X}@{}}
    \begin{minipage}{.12\textwidth}
    \phantom{}
    \end{minipage}%
    & {\small $PVE=0.1$} & {\small $PVE=0.2$} & {\small $PVE=0.4$} & {\small $PVE=0.8$}
    \end{tabularx}\\
    \vspace{-1mm}
    \hspace{2mm}
    \rotatebox{90}{{\hspace{-45mm} \centering \small \textbf{(A) Misspecified Linear}}}
    \hspace{1mm}
    \rotatebox{90}{{\hspace{6mm} \centering \small \textsc{Enhancer}}}
    \includegraphics[width=0.85\textwidth]{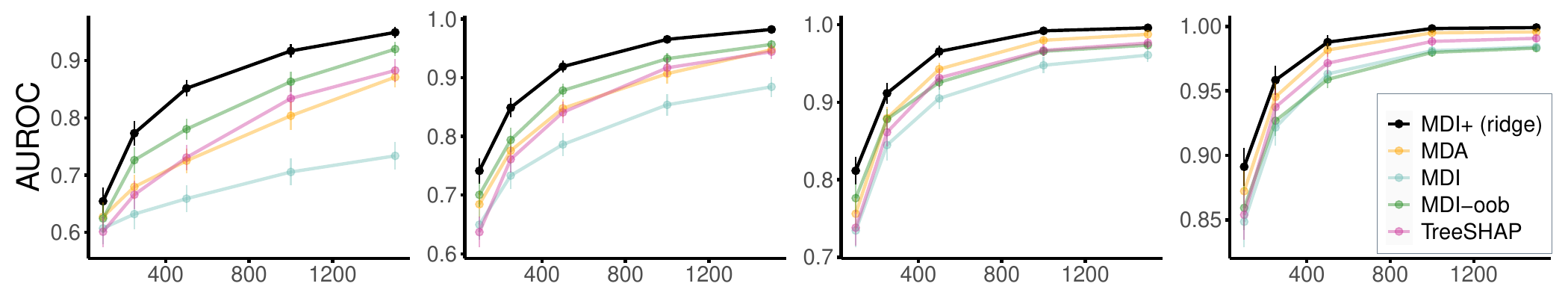}
\end{subfigure}
\begin{subfigure}{\textwidth}
    \centering
    \vspace{-2mm}
    \hspace{2mm}
    \rotatebox{90}{{\centering \small \phantom{\textbf{(A) LRg}}}}
    \hspace{1mm}
    \rotatebox{90}{{\hspace{8.5mm} \centering \small \textsc{CCLE}}}
    \includegraphics[width=0.85\textwidth]{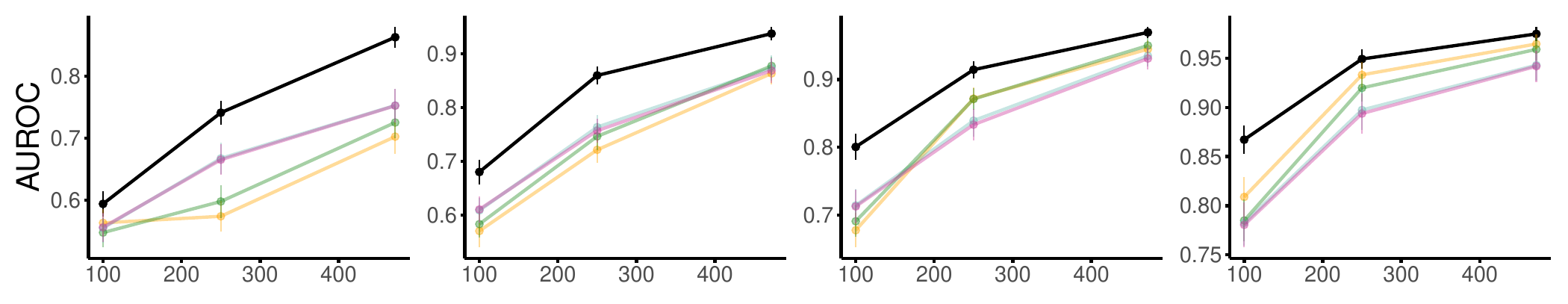}
\end{subfigure}
\begin{subfigure}{\textwidth}
    \centering
    \vspace{-2mm}
    \hspace{2mm}
    \rotatebox{90}{{\centering \small \phantom{\textbf{(A) LRg}}}}
    \hspace{1mm}
    \rotatebox{90}{{\hspace{6.25mm} \centering \small \textsc{Juvenile}}}
    \includegraphics[width=0.85\textwidth]{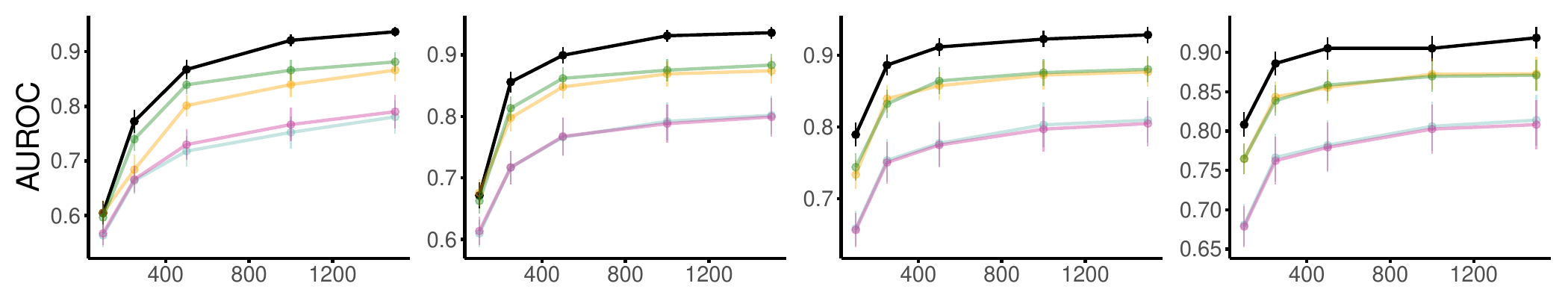}
\end{subfigure}
\begin{subfigure}{\textwidth}
    \centering
    \vspace{-2mm}
    \hspace{2mm}
    \rotatebox{90}{{\centering \small \phantom{\textbf{(A) LRg}}}}
    \hspace{1mm}
    \rotatebox{90}{{\hspace{10mm} \centering \small \textsc{Splicing}}}
    \includegraphics[width=0.85\textwidth]{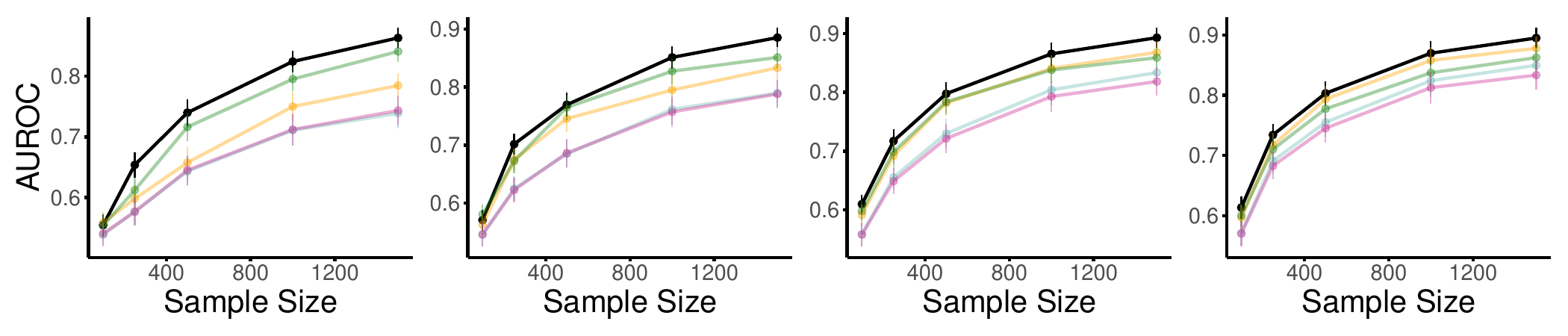}
    \vspace{-5mm}
\end{subfigure}
\noindent\makebox[\textwidth]{\hfill\rule{0.95\textwidth}{0.4pt}\hfill}
\begin{subfigure}{\textwidth}
    \centering
    \begin{tabularx}{0.91\textwidth}{@{}l *5{>{\centering\arraybackslash}X}@{}}
    \begin{minipage}{.07\textwidth}
    \phantom{}
    \end{minipage}%
    \end{tabularx}\\
    \vspace{-5mm}
    \hspace{2mm}
    \rotatebox{90}{{\hspace{-43mm} \centering \small \textbf{(B) Misspecified LSS}}}
    \hspace{1mm}
    \rotatebox{90}{{\hspace{6mm} \centering \small \textsc{Enhancer}}}
    \includegraphics[width=0.85\textwidth]{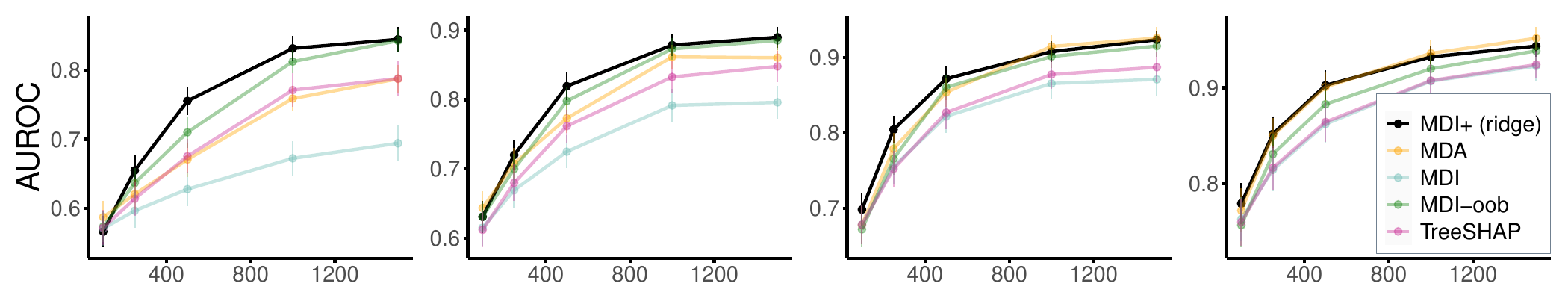}
\end{subfigure}
\begin{subfigure}{\textwidth}
    \centering
    \vspace{-2mm}
    \hspace{2mm}
    \rotatebox{90}{{\centering \small \phantom{\textbf{(A) LRg}}}}
    \hspace{1mm}
    \rotatebox{90}{{\hspace{8.5mm} \centering \small \textsc{CCLE}}}
    \includegraphics[width=0.85\textwidth]{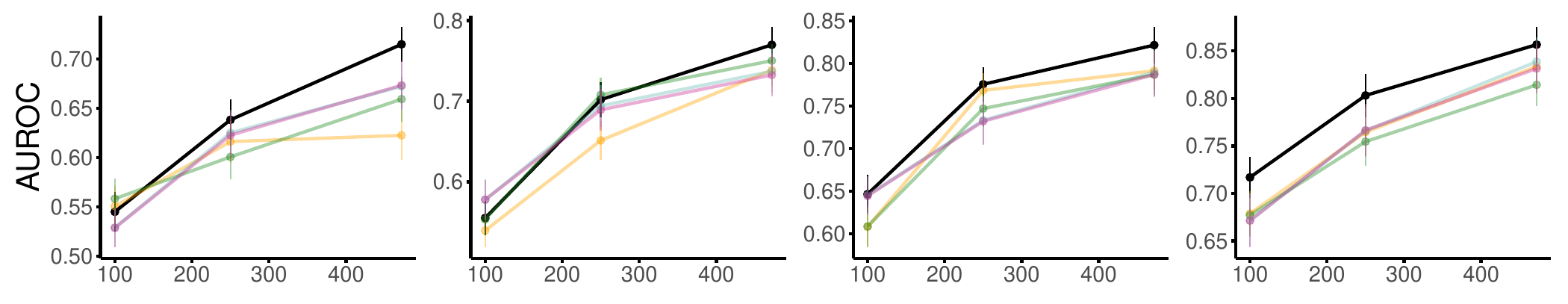}
\end{subfigure}
\begin{subfigure}{\textwidth}
    \centering
    \vspace{-2mm}
    \hspace{2mm}
    \rotatebox{90}{{\centering \small \phantom{\textbf{(A) LRg}}}}
    \hspace{1mm}
    \rotatebox{90}{{\hspace{6.25mm} \centering \small \textsc{Juvenile}}}
    \includegraphics[width=0.85\textwidth]{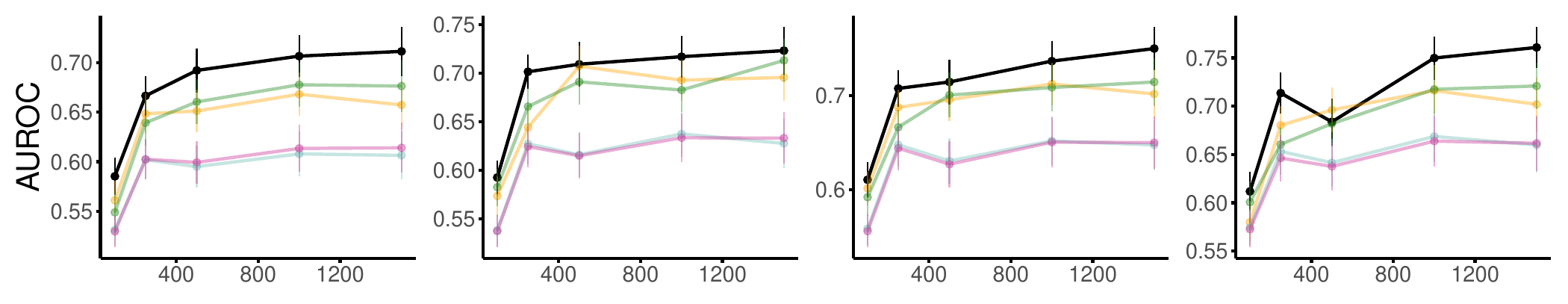}
\end{subfigure}
\begin{subfigure}{\textwidth}
    \centering
    \vspace{-2mm}
    \hspace{2mm}
    \rotatebox{90}{{\centering \small \phantom{\textbf{(A) LRg}}}}
    \hspace{1mm}
    \rotatebox{90}{{\hspace{10mm} \centering \small \textsc{Splicing}}}
    \includegraphics[width=0.85\textwidth]{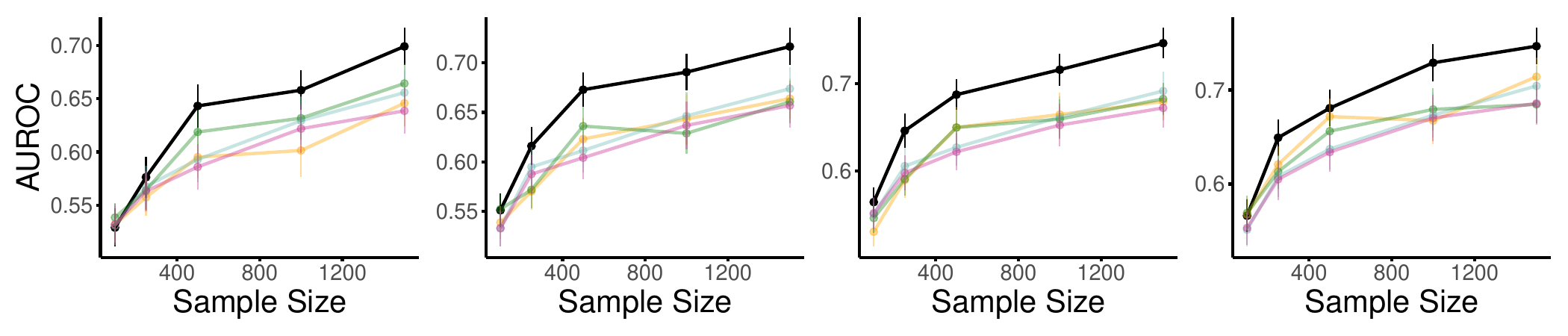}
\end{subfigure}
\caption{\method\;(ridge) outperforms other feature importance methods under the (A) misspecified linear and (B) misspecified LSS regression settings described in Appendix~\ref{supp:misspecified}. This pattern is evident across various datasets with different covariate structures (specified by row), proportions of variance explained (specified by column), and sample sizes (on the $x$-axis). In all subplots, the AUROC has been averaged across 50 experimental replicates, and error bars represent $\pm$ 1SE.}
\label{fig:misspecified_reg_linear_appendix}
\end{figure}

\begin{figure}[h!]
\begin{subfigure}{\textwidth}
    \centering
    \vspace{-5mm}
    \begin{tabularx}{0.95\textwidth}{@{}l *5{>{\centering\arraybackslash}X}@{}}
    \begin{minipage}{.12\textwidth}
    \phantom{}
    \end{minipage}%
    & {\small $PVE=0.1$} & {\small $PVE=0.2$} & {\small $PVE=0.4$} & {\small $PVE=0.8$}
    \end{tabularx}\\
    \vspace{-1mm}
    \hspace{2mm}
    \rotatebox{90}{{\hspace{-55mm} \centering \small \textbf{(A) Misspecified Polynomial Interaction}}}
    \hspace{1mm}
    \rotatebox{90}{{\hspace{6mm} \centering \small \textsc{Enhancer}}}
    \includegraphics[width=0.85\textwidth]{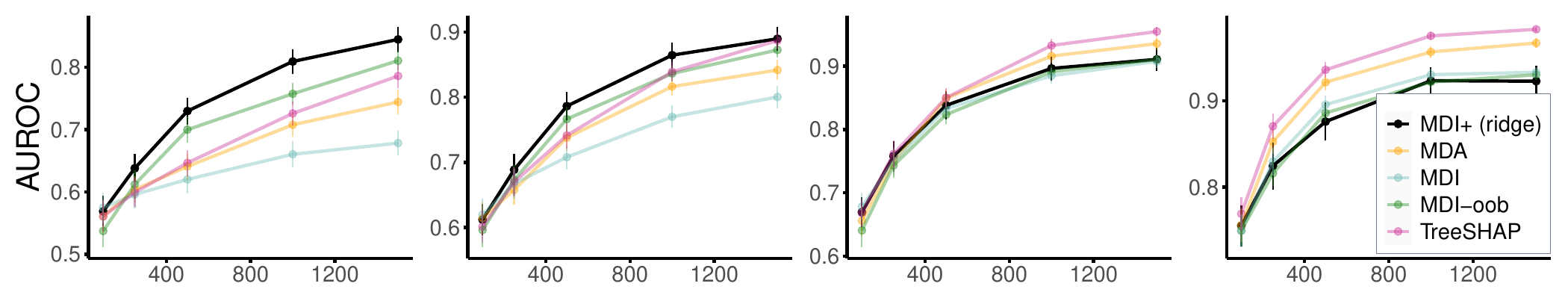}
\end{subfigure}
\begin{subfigure}{\textwidth}
    \centering
    \vspace{-2mm}
    \hspace{2mm}
    \rotatebox{90}{{\centering \small \phantom{\textbf{(A) LRg}}}}
    \hspace{1mm}
    \rotatebox{90}{{\hspace{8.5mm} \centering \small \textsc{CCLE}}}
    \includegraphics[width=0.85\textwidth]{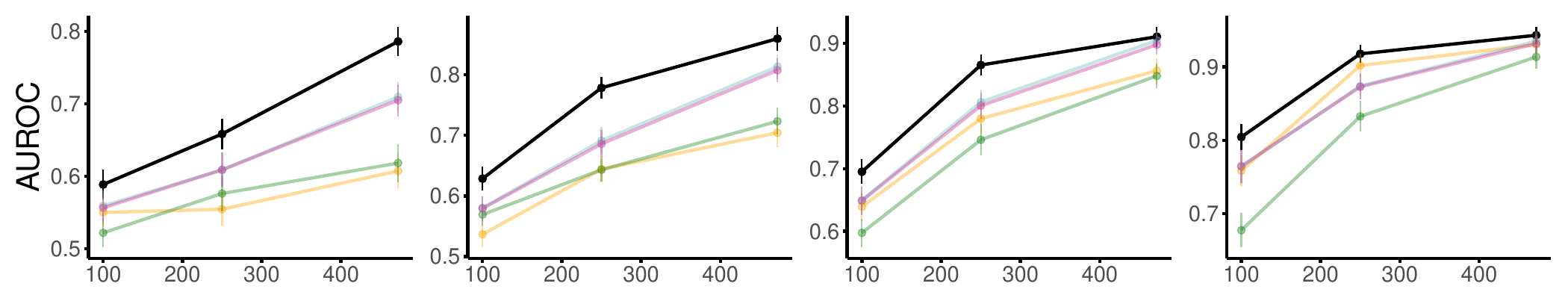}
\end{subfigure}
\begin{subfigure}{\textwidth}
    \centering
    \vspace{-2mm}
    \hspace{2mm}
    \rotatebox{90}{{\centering \small \phantom{\textbf{(A) LRg}}}}
    \hspace{1mm}
    \rotatebox{90}{{\hspace{6.25mm} \centering \small \textsc{Juvenile}}}
    \includegraphics[width=0.85\textwidth]{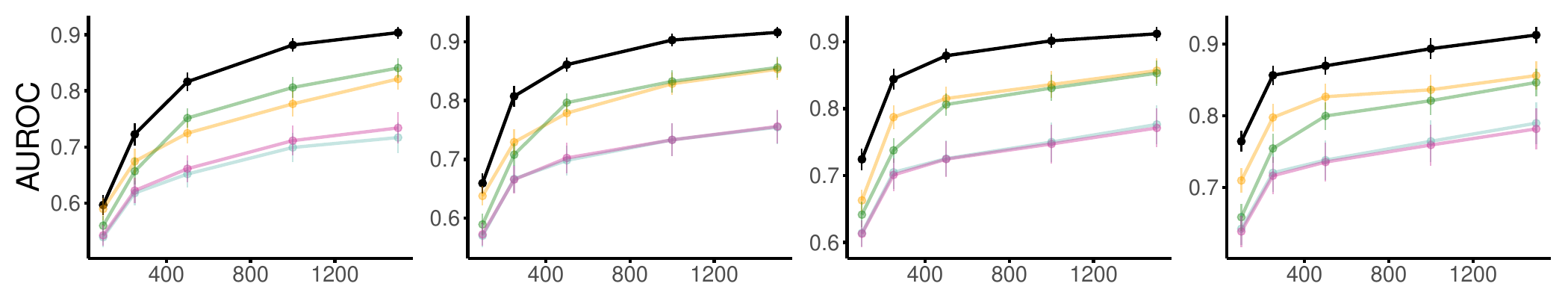}
\end{subfigure}
\begin{subfigure}{\textwidth}
    \centering
    \vspace{-2mm}
    \hspace{2mm}
    \rotatebox{90}{{\centering \small \phantom{\textbf{(A) LRg}}}}
    \hspace{1mm}
    \rotatebox{90}{{\hspace{10mm} \centering \small \textsc{Splicing}}}
    \includegraphics[width=0.85\textwidth]{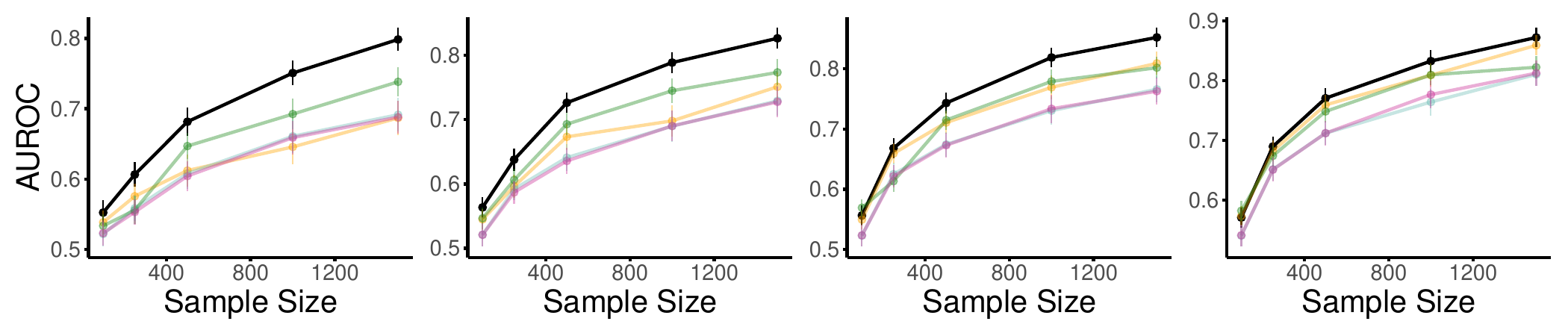}
    \vspace{-5mm}
\end{subfigure}
\noindent\makebox[\textwidth]{\hfill\rule{0.95\textwidth}{0.4pt}\hfill}
\begin{subfigure}{\textwidth}
    \centering
    \begin{tabularx}{1\textwidth}{@{}l *5{>{\centering\arraybackslash}X}@{}}
    \begin{minipage}{.075\textwidth}
    \phantom{}
    \end{minipage}%
    \end{tabularx}\\
    \vspace{-5mm}
    \hspace{2mm}
    \rotatebox{90}{{\hspace{-51mm} \centering \small \textbf{(B) Misspecified Linear + LSS}}}
    \hspace{1mm}
    \rotatebox{90}{{\hspace{6mm} \centering \small \textsc{Enhancer}}}
    \includegraphics[width=0.85\textwidth]{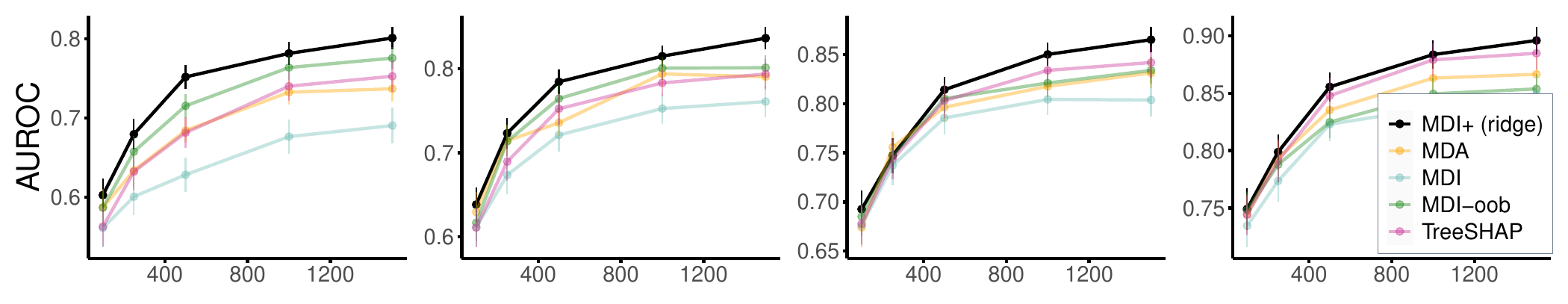}
\end{subfigure}
\begin{subfigure}{\textwidth}
    \centering
    \vspace{-2mm}
    \hspace{2mm}
    \rotatebox{90}{{\centering \small \phantom{\textbf{(A) LRg}}}}
    \hspace{1mm}
    \rotatebox{90}{{\hspace{8.5mm} \centering \small \textsc{CCLE}}}
    \includegraphics[width=0.85\textwidth]{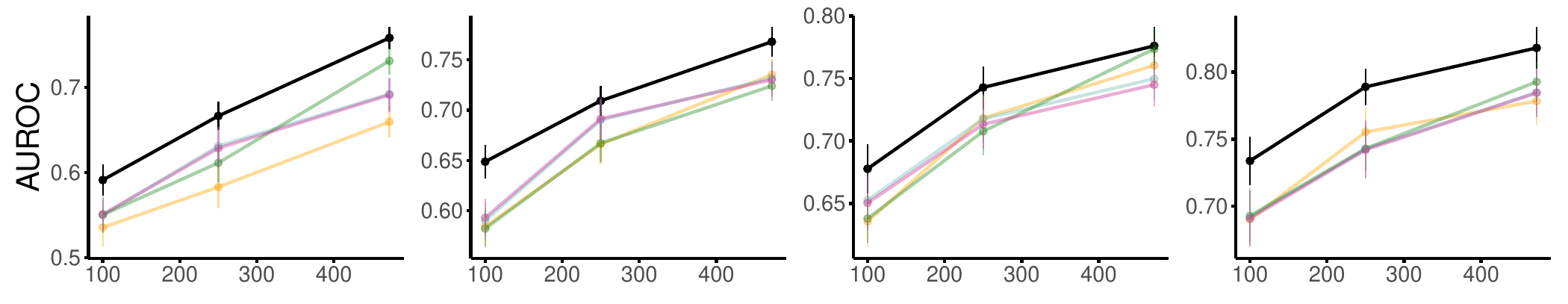}
\end{subfigure}
\begin{subfigure}{\textwidth}
    \centering
    \vspace{-2mm}
    \hspace{2mm}
    \rotatebox{90}{{\centering \small \phantom{\textbf{(A) LRg}}}}
    \hspace{1mm}
    \rotatebox{90}{{\hspace{6.25mm} \centering \small \textsc{Juvenile}}}
    \includegraphics[width=0.85\textwidth]{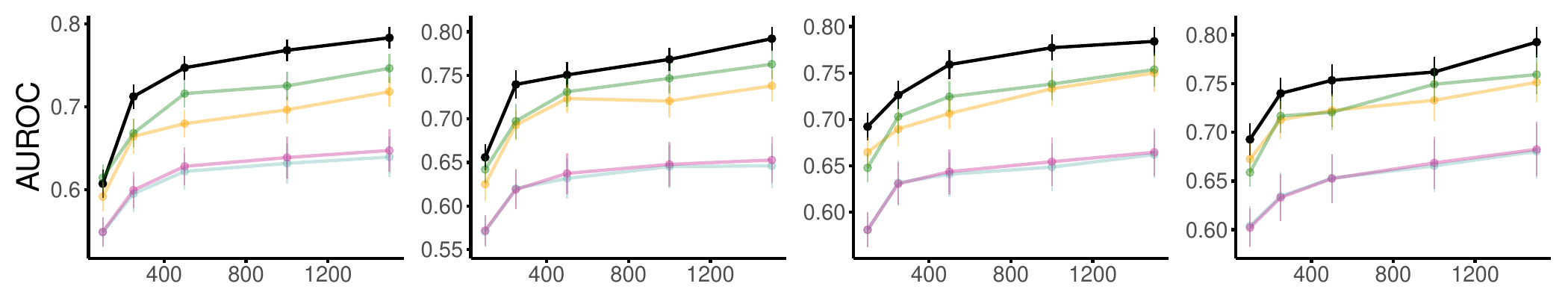}
\end{subfigure}
\begin{subfigure}{\textwidth}
    \centering
    \vspace{-2mm}
    \hspace{2mm}
    \rotatebox{90}{{\centering \small \phantom{\textbf{(A) LRg}}}}
    \hspace{1mm}
    \rotatebox{90}{{\hspace{10mm} \centering \small \textsc{Splicing}}}
    \includegraphics[width=0.85\textwidth]{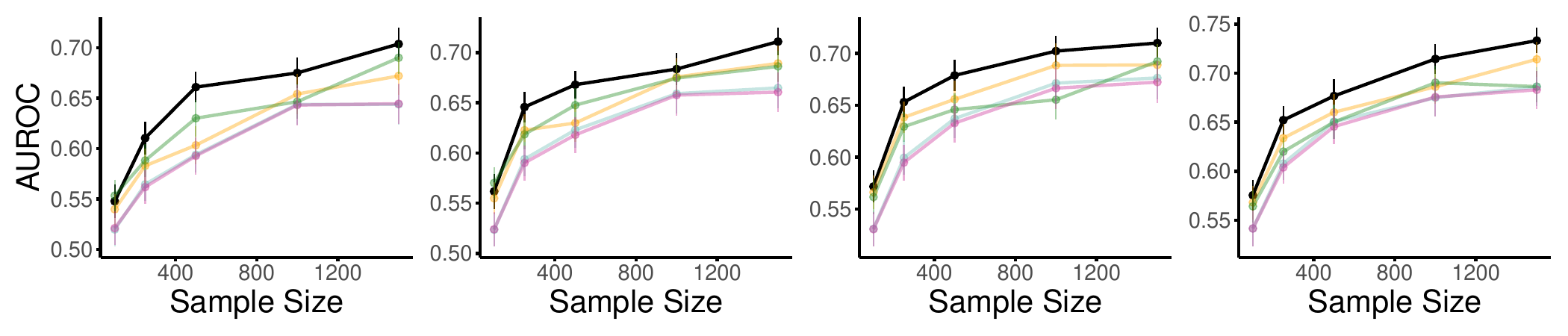}
\end{subfigure}
\caption{\method\;(ridge) outperforms other feature importance methods under the (A) misspecified polynomial interaction and (B) misspecified linear + LSS regression settings described in Appendix~\ref{supp:misspecified}. This pattern is evident across various datasets with different covariate structures (specified by row), proportions of variance explained (specified by column), and sample sizes (on the $x$-axis). In all subplots, the AUROC has been averaged across 50 experimental replicates, and error bars represent $\pm$ 1SE.}
\label{fig:misspecified_reg_linear_LSS_appendix}
\end{figure}

\paragraph{Results.} For each of the four regression functions, the AUROC results under the misspecified model regime are summarized in Figures~\ref{fig:misspecified_reg_linear_appendix} - \ref{fig:misspecified_reg_linear_LSS_appendix}. In terms of the AUROC, \method\;improves the ranking performance compared to the existing methods under this misspecified model scenario across a variety of regression functions, datasets with different covariate structures, proportions of variance explained, and sample sizes. Note here that the AUROC is computed with respect to only the observed covariates and ignore the omitted variables $X_1$ and $X_2$.

\subsection{Varying Sparsity Simulations} 
\label{supp:sparsity}
In this section, we examine the performance of \method\;as we vary the sparsity level of the regression function.

\paragraph{Experimental details.} Using the Juvenile and Splicing datasets as the covariate matrices $\bX$, we simulate the responses $\by$ according to the four regression functions described in Section \ref{subsec:sim_set_up} (i.e., linear, LSS, polynomial interaction, and linear+LSS), and we vary the sparsity level of the regression function. Specifically, for the linear regression function, we vary $S$, which denotes the number of signal features used in the regression function. For the LSS, polynomial interaction, and linear+LSS models, we vary $M$, which represents the number of interaction terms. For the simulations described in this section, we take $n=1000$ samples and evaluate the performance across varying proportions of variance explained ($PVE = 0.1,\; 0.2,\; 0.4,\; 0.8$).

\begin{figure}[h!]
\begin{subfigure}{\textwidth}
    \centering
    \vspace{-12mm}
    \begin{tabularx}{0.95\textwidth}{@{}l *5{>{\centering\arraybackslash}X}@{}}
    \begin{minipage}{.12\textwidth}
    \phantom{}
    \end{minipage}%
    & {\small $PVE=0.1$} & {\small $PVE=0.2$} & {\small $PVE=0.4$} & {\small $PVE=0.8$}
    \end{tabularx}\\
    \vspace{-1mm}
    \hspace{2mm}
    \rotatebox{90}{{\hspace{-8mm} \centering \small \textbf{(A) Linear}}}
    \hspace{1mm}
    \rotatebox{90}{{\hspace{8.5mm} \centering \small \textsc{Splicing}}}
    \includegraphics[width=0.85\textwidth]{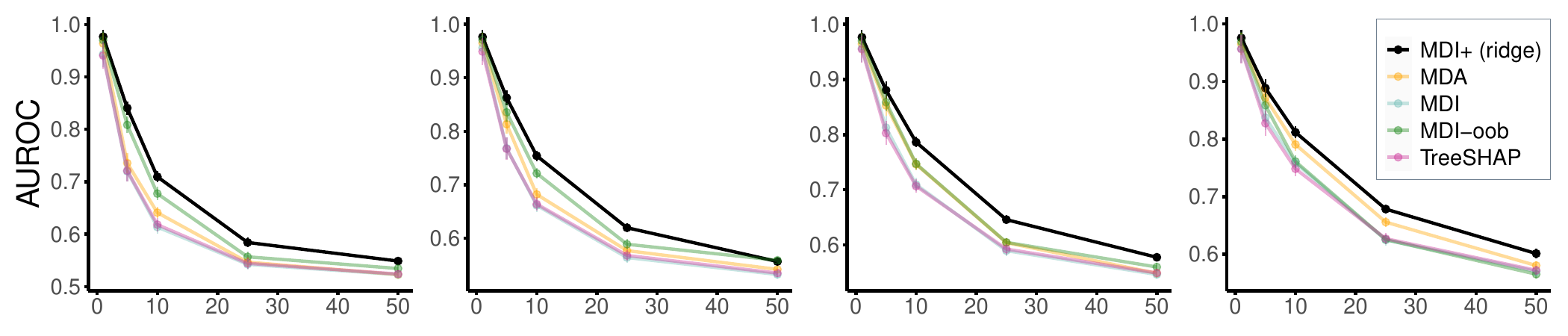}
\end{subfigure}
\begin{subfigure}{\textwidth}
    \centering
    \vspace{-2mm}
    \hspace{2mm}
    \rotatebox{90}{{\centering \small \phantom{\textbf{(A) Linear}}}}
    \hspace{1mm}
    \rotatebox{90}{{\hspace{9mm} \centering \small \textsc{Juvenile}}}
    \includegraphics[width=0.85\textwidth]{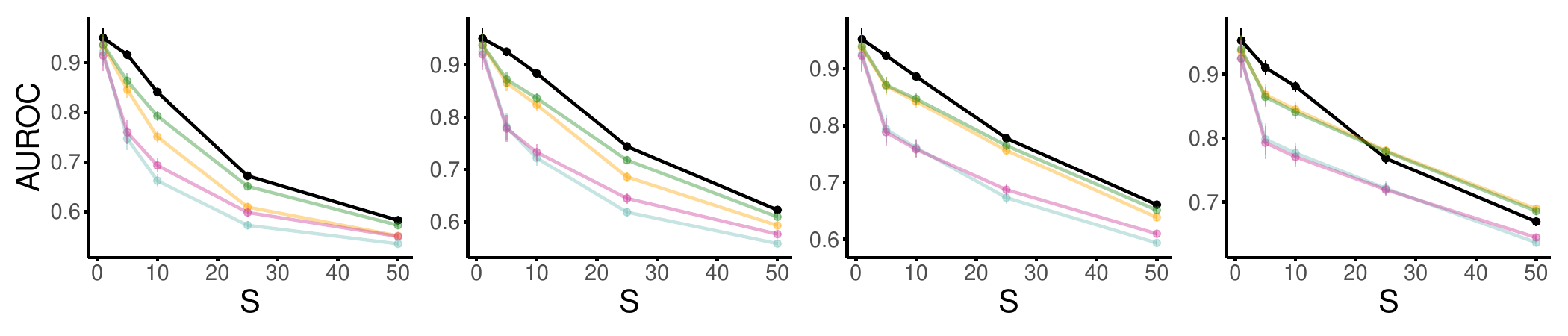}
    \vspace{-5mm}
\end{subfigure}
\noindent\makebox[\textwidth]{\hfill\rule{0.95\textwidth}{0.4pt}\hfill}
\begin{subfigure}{\textwidth}
    \centering
    \hspace{2mm}
    \rotatebox{90}{{\hspace{-6mm} \centering \small \textbf{(B) LSS}}}
    \hspace{1mm}
    \rotatebox{90}{{\hspace{8.5mm} \centering \small \textsc{Splicing}}} \includegraphics[width=0.85\textwidth]{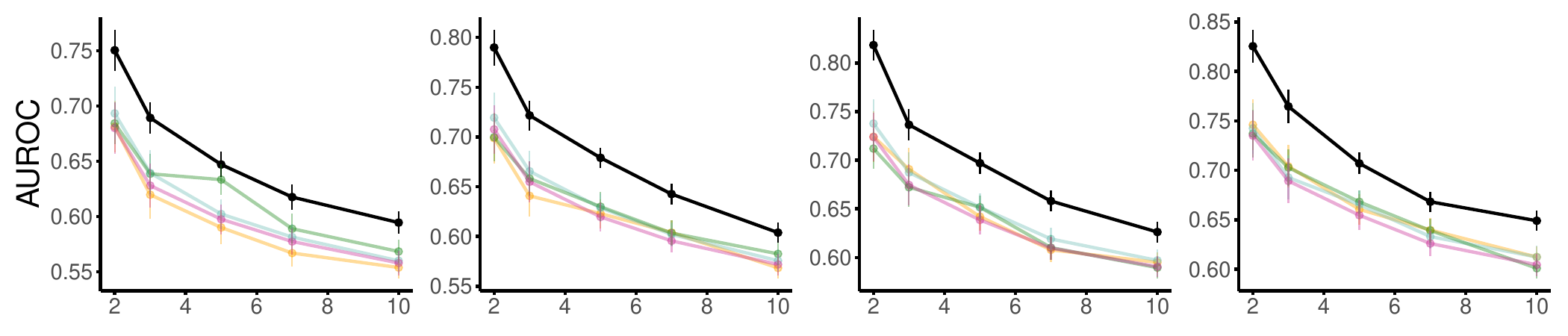}
\end{subfigure}
\begin{subfigure}{\textwidth}
    \centering
    \vspace{-2mm}
    \hspace{2mm}
    \rotatebox{90}{{\centering \small \phantom{\textbf{(A) LSS}}}}
    \hspace{1mm}
    \rotatebox{90}{{\hspace{9mm} \centering \small \textsc{Juvenile}}}
    \includegraphics[width=0.85\textwidth]{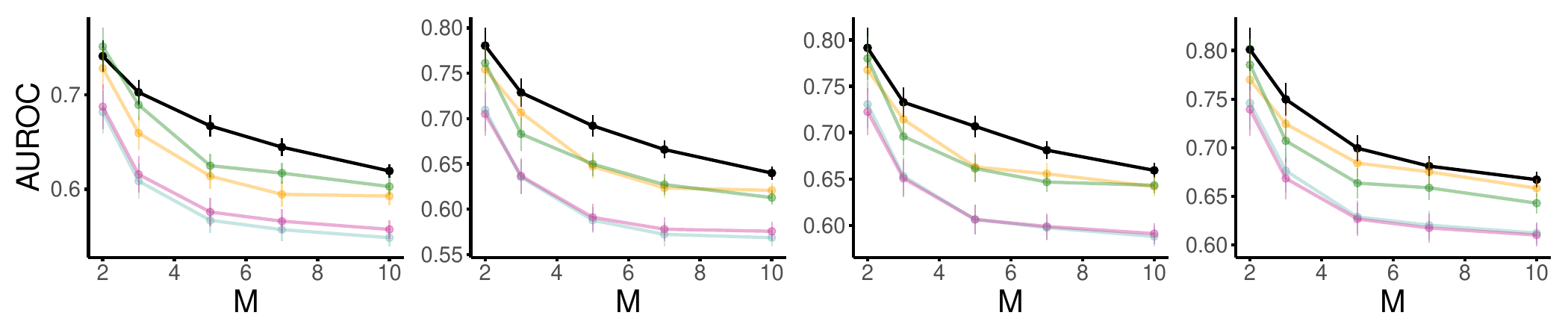}
    \vspace{-5mm}
\end{subfigure}
\noindent\makebox[\textwidth]{\hfill\rule{0.95\textwidth}{0.4pt}\hfill}
\begin{subfigure}{\textwidth}
    \centering
    \hspace{2mm}
    \rotatebox{90}{{\hspace{-22mm} \centering \small \textbf{(C) Polynomial Intearction}}}
    \hspace{1mm}
    \rotatebox{90}{{\hspace{8.5mm} \centering \small \textsc{Splicing}}}
    \includegraphics[width=0.85\textwidth]{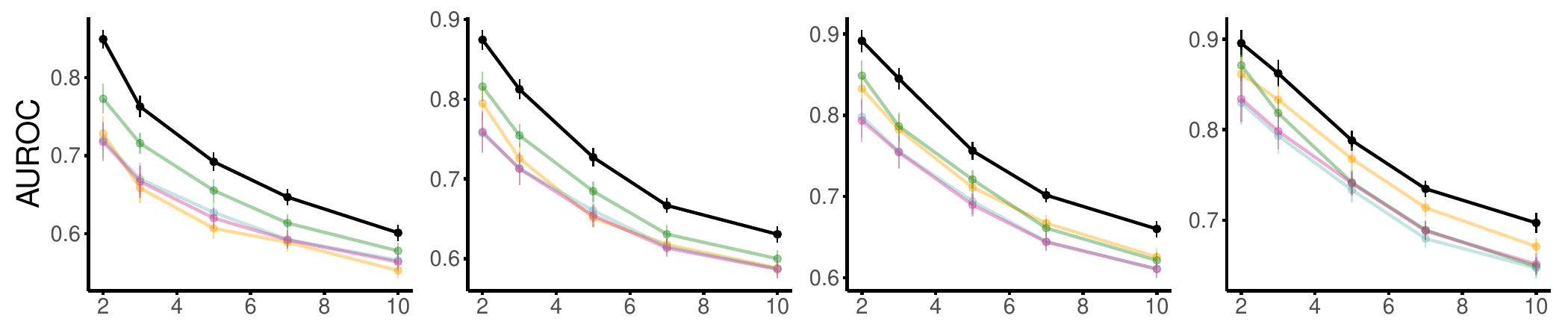}
\end{subfigure}
\begin{subfigure}{\textwidth}
    \centering
    \vspace{-2mm}
    \hspace{2mm}
    \rotatebox{90}{{\centering \small \phantom{\textbf{(A) Polynomial }}}}
    \hspace{1mm}
    \rotatebox{90}{{\hspace{9mm} \centering \small \textsc{Juvenile}}}
    \includegraphics[width=0.85\textwidth]{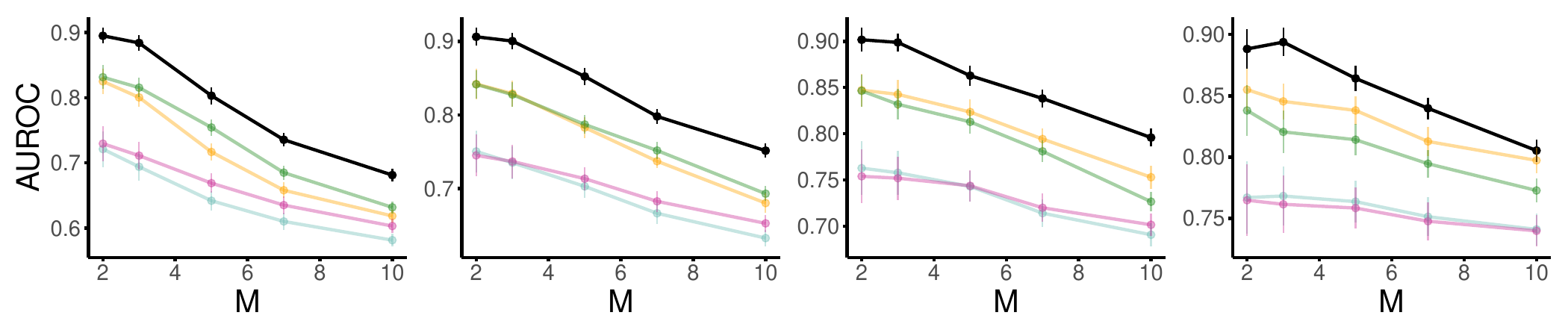}
    \vspace{-5mm}
\end{subfigure}
\noindent\makebox[\textwidth]{\hfill\rule{0.95\textwidth}{0.4pt}\hfill}
\begin{subfigure}{\textwidth}
    \centering
    \hspace{2mm}
    \rotatebox{90}{{\hspace{-14mm} \centering \small \textbf{(D) Linear + LSS}}}
    \hspace{1mm}
    \rotatebox{90}{{\hspace{8.5mm} \centering \small \textsc{Splicing}}}
    \includegraphics[width=0.85\textwidth]{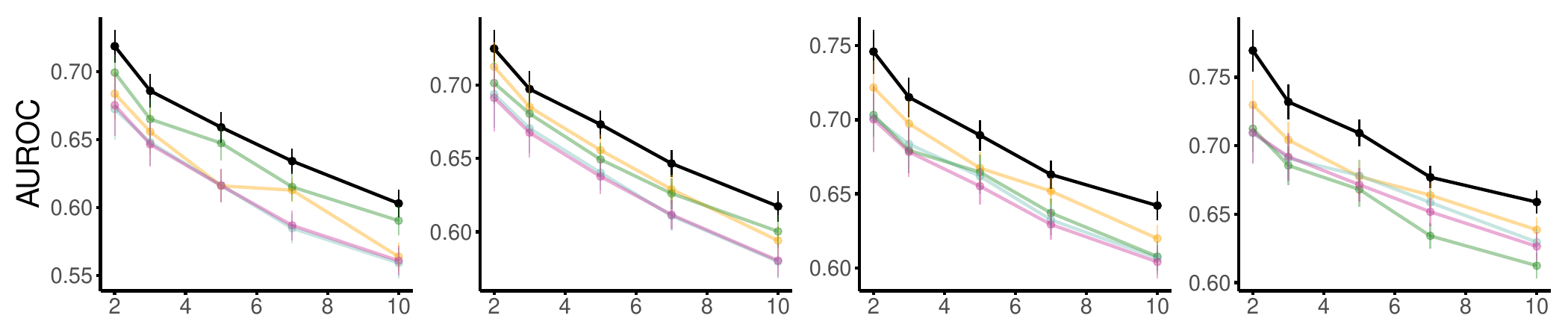}
\end{subfigure}
   \begin{subfigure}{\textwidth}
    \centering
    \vspace{-2mm}
    \hspace{2mm}
    \rotatebox{90}{{\centering \small \phantom{\textbf{(A) Linear + LSS}}}}
    \hspace{1mm}
    \rotatebox{90}{{\hspace{9mm} \centering \small \textsc{Juvenile}}}
    \includegraphics[width=0.85\textwidth]{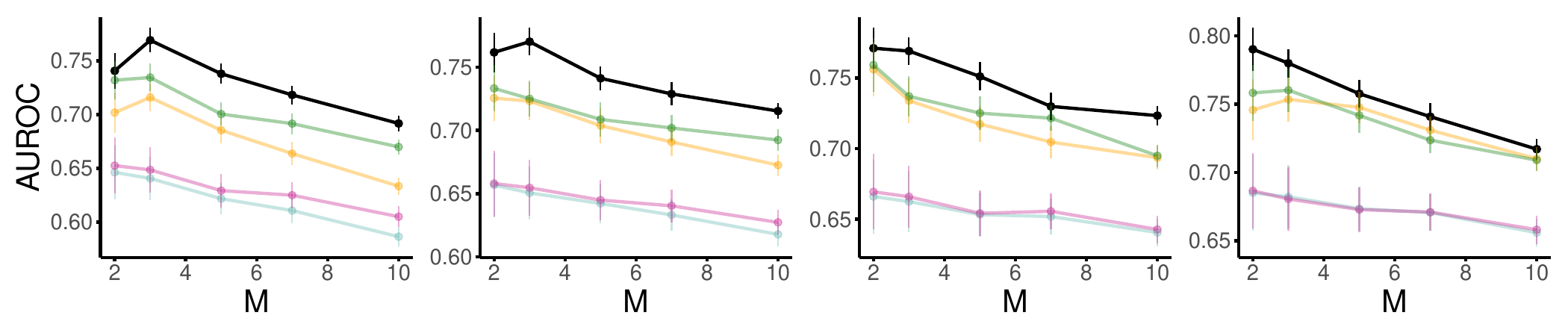}
    \vspace{-5mm}
\end{subfigure}
\caption{In the regression simulations described in Appendix~\ref{supp:sparsity}, \method\;outperforms other feature importance methods across a variety of sparsity levels (specified on the $x$-axis). This pattern is evident across various regression functions (specified by panel), datasets with different covariate structures (specified by row), and proportions of variance explained (specified by column).}
\label{fig:sparsity_sims_appendix}
\end{figure}

\paragraph{Results.} Our results are summarized in Figure~\ref{fig:sparsity_sims_appendix}, which shows that \method\;significantly outperforms competitors across various sparsity levels, as measured by AUROC. 

\subsection{Varying Number of Features Simulations}
 \label{supp:vary_p}
 
In this section, we examine \method\;as we vary the number of features in the covariate matrix $\bX$.

\paragraph{Experimental details.} Using the CCLE gene expression dataset as the covariate matrix, we simulate the responses $\by$ according to the four regression functions described in Section \ref{subsec:sim_set_up} (i.e., linear, LSS, polynomial interaction, and linear+LSS), and we vary the number of features $p$ in the covariate matrix $\bX$. Specifically, for each choice of $p = 10,\; 25,\; 50,\; 100,\; 250,\; 500,\; 1000,\; 2000$, we subsample the desired number of columns from the full CCLE gene expression dataset, which originally consists of $50,114$ genes (or features). For the simulations described in this section, we take the max number of samples in the CCLE dataset ($n=472$) and evaluate the performance across varying proportions of variance explained ($PVE = 0.1,\; 0.2,\; 0.4,\; 0.8$).

\begin{figure}[h!]
\begin{subfigure}{\textwidth}
    \centering
    \textbf{CCLE} \\
    \begin{tabularx}{1\textwidth}{@{}l *5{>{\centering\arraybackslash}X}@{}}
    \begin{minipage}{.075\textwidth}
    \phantom{}
    \end{minipage}%
    & {\small $PVE=0.1$} & {\small $PVE=0.2$} & {\small $PVE=0.4$} & {\small $PVE=0.8$}
    \end{tabularx}\\
    \rotatebox{90}{{\hspace{10mm} \centering \small \textsc{Linear}}}
    \hspace{.25mm}
    \includegraphics[width=0.95\textwidth]{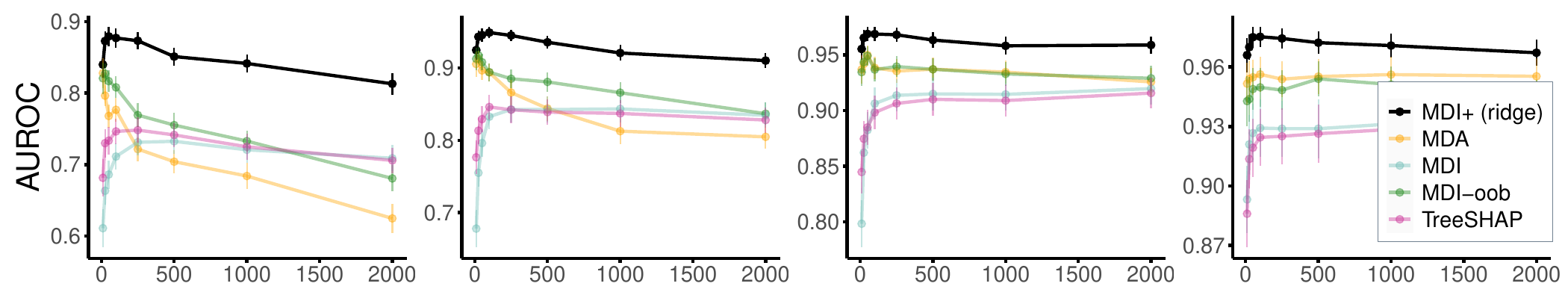}
\end{subfigure}
\begin{subfigure}{\textwidth}
    \centering
    \rotatebox{90}{{\hspace{12.5mm} \centering \small \textsc{LSS}}}
    \hspace{.25mm}
    \includegraphics[width=0.95\textwidth]{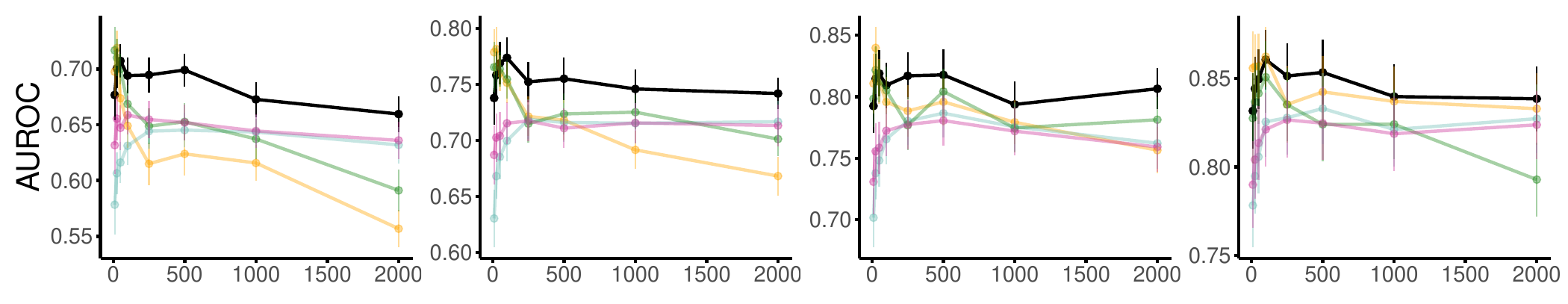}
\end{subfigure}
\begin{subfigure}{\textwidth}
    \centering
    \rotatebox{90}{{\hspace{8mm} \centering \small \textsc{Poly. Int.}}}
    \hspace{.25mm}
    \includegraphics[width=0.95\textwidth]{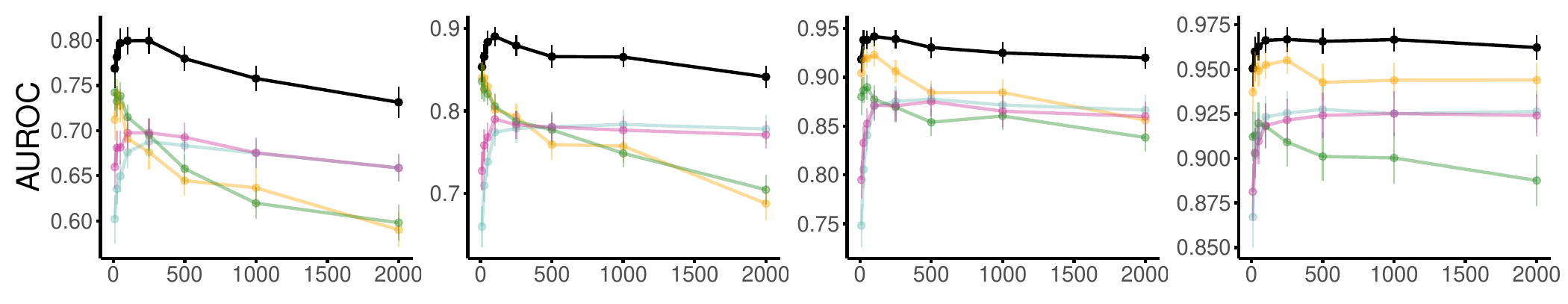}
\end{subfigure}
\begin{subfigure}{\textwidth}
    \centering
    \rotatebox{90}{{\hspace{6mm} \centering \small \textsc{Linear + LSS}}}
    \hspace{.25mm}
    \includegraphics[width=0.95\textwidth]{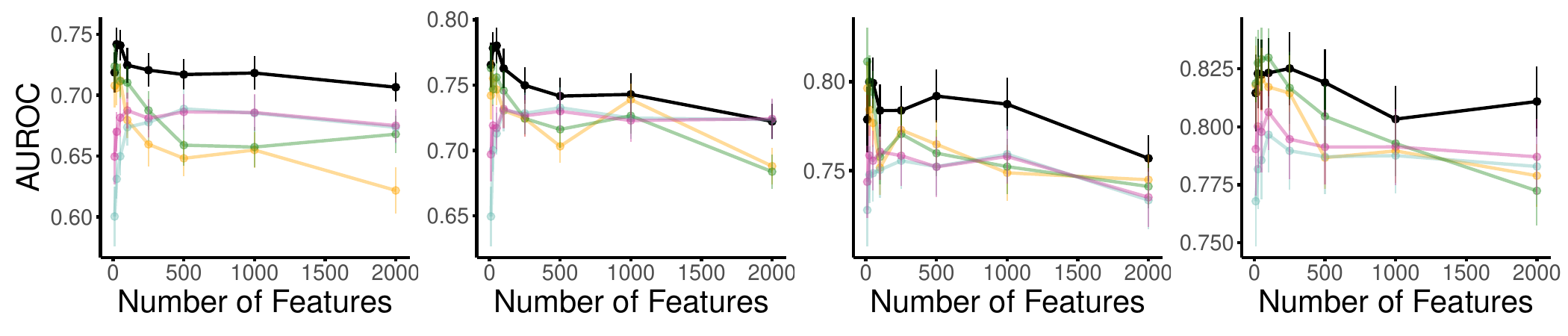}
\end{subfigure}
\caption{In the regression simulations described in Appendix~\ref{supp:vary_p}, \method\;(ridge) outperforms other feature importance methods regardless of the number of features in the covariate matrix $\bX$ (specified on the $x$-axis). This pattern is evident across various regression functions (specified by row) and proportions of variance explained (specified by column).}
\label{fig:varying_p}
\end{figure}

\paragraph{Results.} Our results are summarized in Figure~\ref{fig:varying_p}, which shows that \method\;significantly outperforms competitors in terms of AUROC, regardless of the number of features in the covariate matrix $\bX$. 

\clearpage

\section{Justifying \method\;Framework Construction}
\label{supp:modeling_choices}

We next conduct simulations, showing how each compnent in the \method\;framework (i.e., using regularization, including the raw feature, and evaluating predictions via LOO) improves the feature ranking AUROC.  

\begin{figure}[htbp]
\begin{subfigure}{\textwidth}
    \centering
    \vspace{-10mm}
    \begin{tabularx}{0.96\textwidth}{@{}l *5{>{\centering\arraybackslash}X}@{}}
    \begin{minipage}{.135\textwidth}
    \phantom{}
    \end{minipage}%
    & {\small $PVE=0.1$} & {\small $PVE=0.2$} & {\small $PVE=0.4$} & {\small $PVE=0.8$}
    \end{tabularx}\\
    \hspace{2mm}
    \rotatebox{90}{{\hspace{-12mm} \centering \small \textbf{(A) Linear with}}}
    \rotatebox{90}{{\hspace{-20mm} \centering \small \textbf{\textit{min\_samples\_per\_leaf}=1}}}
    \hspace{1mm}
    \rotatebox{90}{{\hspace{6mm} \centering \small \textsc{Enhancer}}}
    \includegraphics[width=0.85\textwidth]{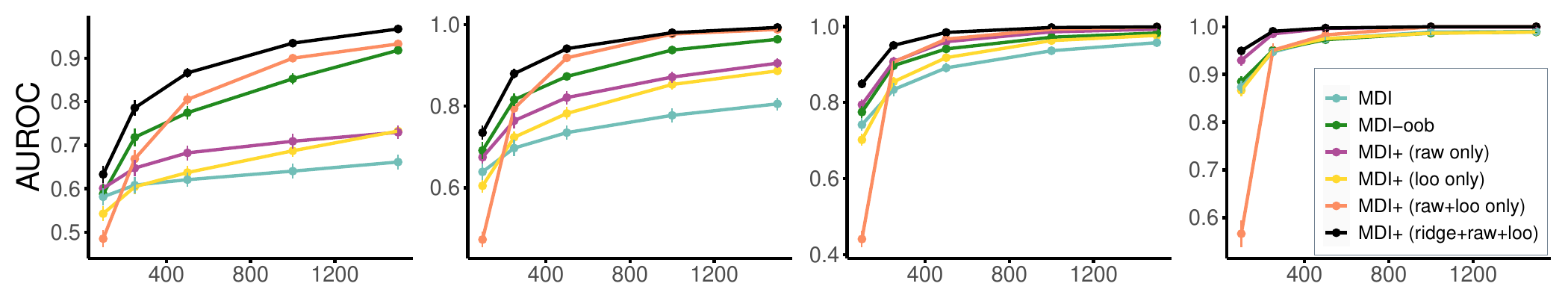}
\end{subfigure}
\begin{subfigure}{\textwidth}
    \centering
    \vspace{-2mm}
    \hspace{2mm}
    \rotatebox{90}{{\hspace{-12mm} \centering \small \phantom{(A) Linear with}}}
    \rotatebox{90}{{\hspace{-20mm} \centering \small \phantom{\textit{min\_samples\_per\_leaf}=1}}}
    \hspace{1mm}
    \rotatebox{90}{{\hspace{12mm} \centering \small \textsc{CCLE}}}
    \includegraphics[width=0.85\textwidth]{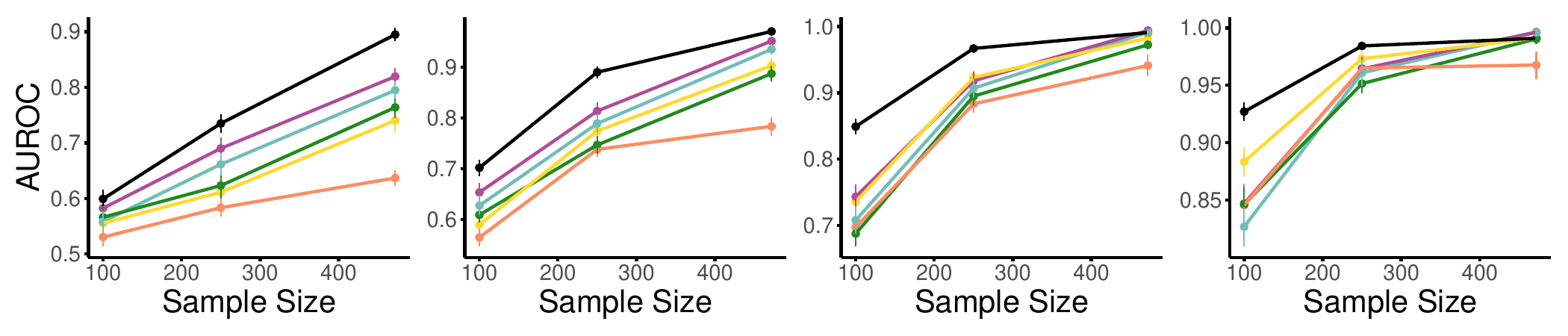}
    \vspace{-5mm}
\end{subfigure}
\noindent\makebox[\textwidth]{\hfill\rule{0.95\textwidth}{0.4pt}\hfill}
\begin{subfigure}{\textwidth}
    \centering
    \hspace{2mm}
    \rotatebox{90}{{\hspace{-29mm} \centering \small \textbf{(B) Polynomial Interaction with}}}
    \rotatebox{90}{{\hspace{-21mm} \centering \small \textbf{\textit{min\_samples\_per\_leaf}=1}}}
    \hspace{1mm}
    \rotatebox{90}{{\hspace{6mm} \centering \small \textsc{Enhancer}}} \includegraphics[width=0.85\textwidth]{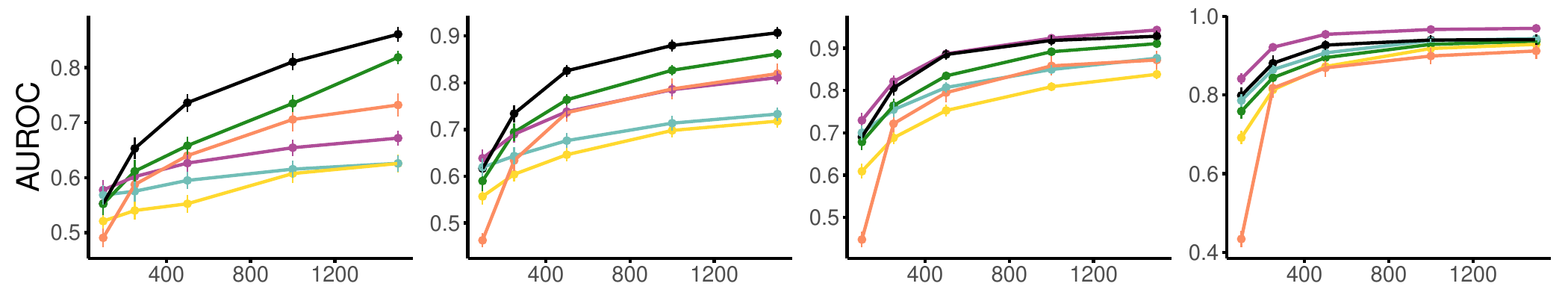}
\end{subfigure}
\begin{subfigure}{\textwidth}
    \centering
    \vspace{-2mm}
    \hspace{2mm}
    \rotatebox{90}{{\hspace{-29mm} \centering \small \phantom{(B) Polynomial Interaction with}}}
    \rotatebox{90}{{\hspace{-21mm} \centering \small \phantom{\textit{min\_samples\_per\_leaf}=1}}}
    \hspace{1mm}
    \rotatebox{90}{{\hspace{12mm} \centering \small \textsc{CCLE}}}
    \includegraphics[width=0.85\textwidth]{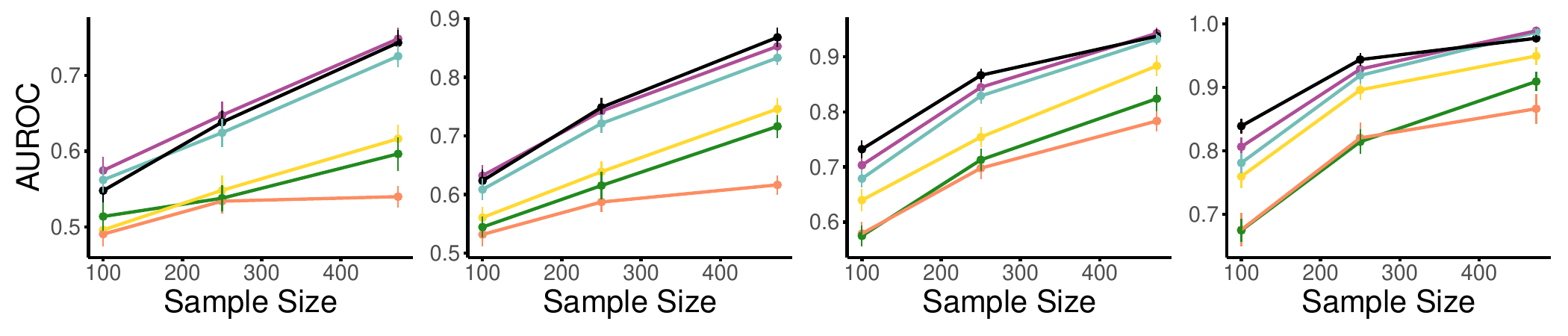}
    \vspace{-5mm}
\end{subfigure}
\noindent\makebox[\textwidth]{\hfill\rule{0.95\textwidth}{0.4pt}\hfill}
\begin{subfigure}{\textwidth}
    \centering
    \hspace{2mm}
    \rotatebox{90}{{\hspace{-12mm} \centering \small \textbf{(C) Linear with}}}
    \rotatebox{90}{{\hspace{-20mm} \centering \small \textbf{\textit{min\_samples\_per\_leaf}=5}}}
    \hspace{1mm}
    \rotatebox{90}{{\hspace{6mm} \centering \small \textsc{Enhancer}}}
    \includegraphics[width=0.85\textwidth]{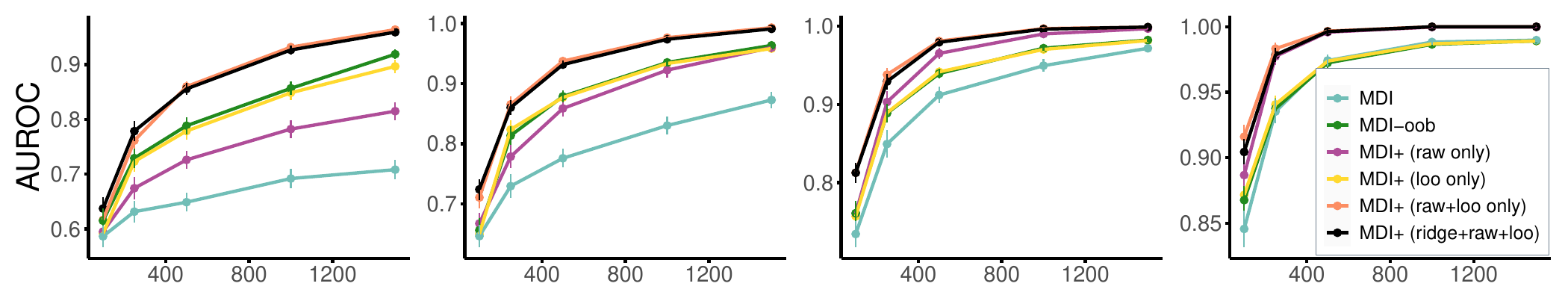}
\end{subfigure}
\begin{subfigure}{\textwidth}
    \centering
    \vspace{-2mm}
    \hspace{2mm}
    \rotatebox{90}{{\hspace{-12mm} \centering \small \phantom{(C) Linear with}}}
    \rotatebox{90}{{\hspace{-20mm} \centering \small \phantom{\textit{min\_samples\_per\_leaf}=5}}}
    \hspace{1mm}
    \rotatebox{90}{{\hspace{12mm} \centering \small \textsc{CCLE}}}
    \includegraphics[width=0.85\textwidth]{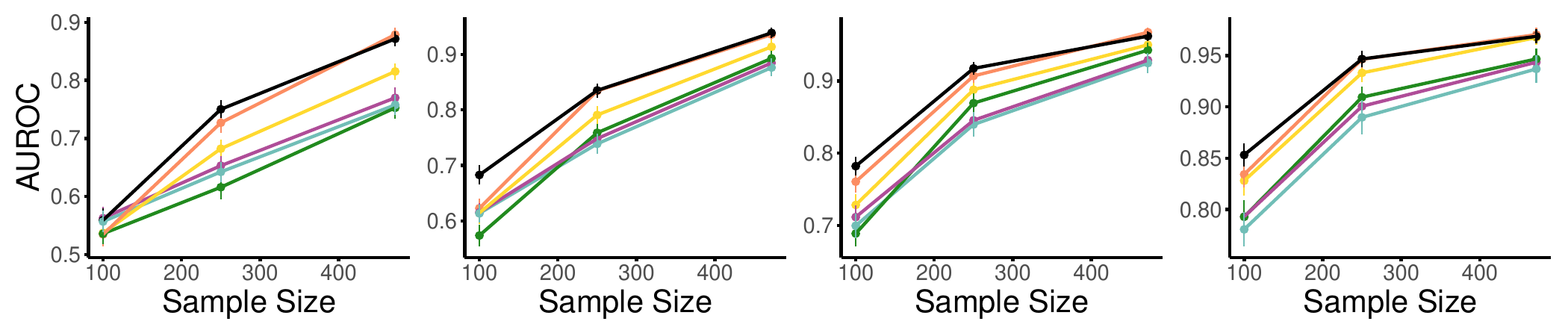}
    \vspace{-5mm}
\end{subfigure}
\noindent\makebox[\textwidth]{\hfill\rule{0.95\textwidth}{0.4pt}\hfill}
\begin{subfigure}{\textwidth}
    \centering
    \hspace{2mm}
    \rotatebox{90}{{\hspace{-29mm} \centering \small \textbf{(D) Polynomial Interaction with}}}
    \rotatebox{90}{{\hspace{-21mm} \centering \small \textbf{\textit{min\_samples\_per\_leaf}=5}}}
    \hspace{1mm}
    \rotatebox{90}{{\hspace{6mm} \centering \small \textsc{Enhancer}}} \includegraphics[width=0.85\textwidth]{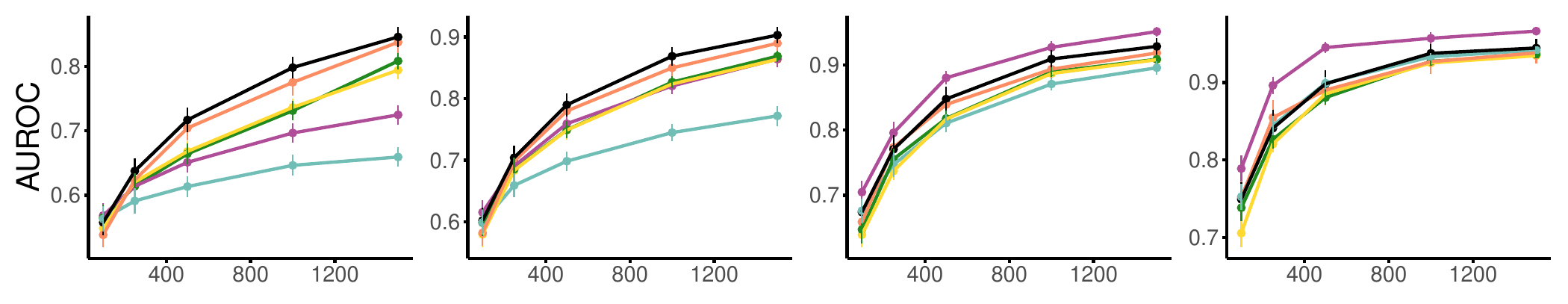}
\end{subfigure}
\begin{subfigure}{\textwidth}
    \centering
    \vspace{-2mm}
    \hspace{2mm}
    \rotatebox{90}{{\hspace{-29mm} \centering \small \phantom{(D) Polynomial Interaction with}}}
    \rotatebox{90}{{\hspace{-21mm} \centering \small \phantom{\textit{min\_samples\_per\_leaf}=5}}}
    \hspace{1mm}
    \rotatebox{90}{{\hspace{12mm} \centering \small \textsc{CCLE}}}
    \includegraphics[width=0.85\textwidth]{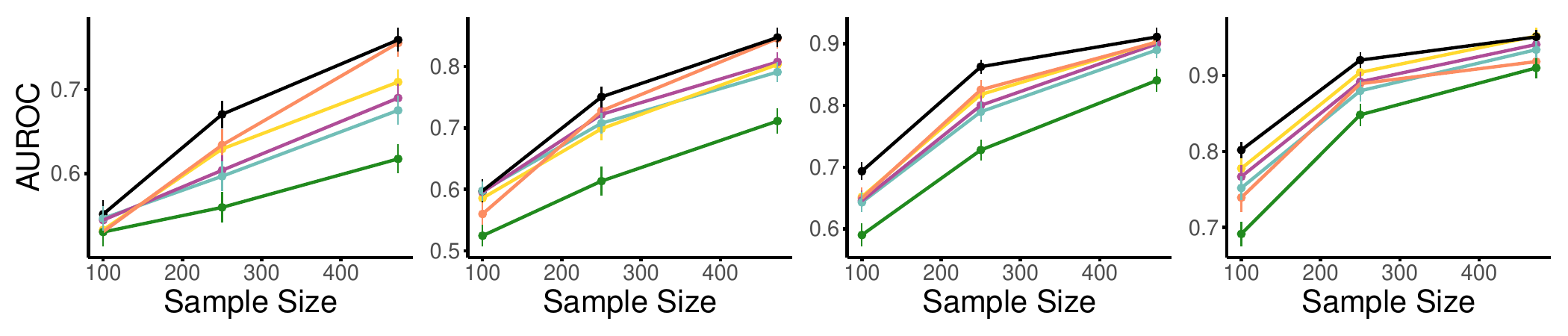}
\end{subfigure}
\caption{We illustrate the impact of various construction choices in the \method\;framework, namely, including regularization, the raw feature, and LOO evaluation, applied to an RF regressor with \textit{min\_samples\_per\_leaf}=1 or 5. \method\;with ridge, including the raw feature, and LOO evaluation (black) consistently outperforms or is on par with other \method\;construction choices across various regression functions (specified by panel), datasets (specified by row), and proportions of variance explained (specified by column).}
\label{fig:modeling_choices}
\end{figure}

\paragraph{Experimental details.} We use the Enhancer and CCLE gene expression datasets as our covariate matrix $\bX$, and simulate responses $\by$ according to the linear and polynomial interaction function described in Section \ref{subsec:sim_set_up}. As in Section~\ref{subsec:regression_results}, we vary the number of samples $n$ across $\{100,\; 250,\; 500,\; 1000,\; 1500\}$ for the Enhancer dataset and $\{100,\; 250,\; 472\}$ for the CCLE gene expression dataset. We also vary $PVE$ in $\{0.1,\; 0.2,\; 0.4,\; 0.8\}$. In order to illustrate the impact of the choices in constructing the \method\;framework, we consider the following sequence of models: 
\begin{enumerate}
    \item \textbf{MDI}: equivalent to \method\;using OLS as the GLM without the raw feature and evaluating $\rsq$ on the in-bag samples. 
    \item \textbf{MDI-oob}: equivalent to \method\;using OLS as the GLM without the raw feature and evaluating $\rsq$ on the out-of-bag samples. 
    \item \textbf{\method\;(raw only)}: \method\;using OLS as the GLM with the raw feature and in-bag evaluation.
    \item \textbf{\method\;(loo only)}: \method\;using OLS as the GLM without the raw feature but using LOO evaluation on the entire dataset.
    \item \textbf{\method\;(raw+loo only)}: \method\;using OLS as the GLM with the raw feature and LOO evaluation on the entire dataset.
    \item \textbf{\method\;(ridge+raw+loo)}: \method\;using ridge as the GLM with the raw feature and LOO evaluation (i.e., the default \method\;settings described in Section~\ref{subsec:regression_results}).
\end{enumerate}
We perform the simulation with an RF regressor using $min\_samples\_leaf = \{1,5\}$ alongside other default parameters and average the performance of each feature importance method across 50 simulation replicates. The results for $min\_samples\_leaf = \{1, 5\}$ are displayed in Figure~\ref{fig:modeling_choices}. 

\paragraph{Results.} From Figure~\ref{fig:modeling_choices}, \method\;with the default regression settings (i.e., \method\;(ridge+raw+loo) in black), most consistently outperforms MDI, MDI-oob, as well as other \method\;configurations across the various regression functions, datasets, proportions of variance explained, and choices of \textit{min\_samples\_per\_leaf}. Moreover, when using an RF regressor with \textit{min\_samples\_per\_leaf}=5 (Figure~\ref{fig:modeling_choices}C-D), we see the added benefits of including the raw feature and/or using LOO evaluation within the \method\;framework, compared to MDI which does not include the raw feature and uses in-bag evaluation. This point becomes more obscure when using an RF regressor trained to purity with \textit{min\_samples\_per\_leaf}=1 (Figure~\ref{fig:modeling_choices}A-B). Here, MDI outperforms MDI-oob, \method\;(loo only), and \method\;(raw+loo only) on the CCLE gene expression dataset, that is, in a small $n$, large $p$ setting. We hypothesize that this is due to overfitting and thus high instability, given that the trees are being grown to purity. By incorporating shrinkage using ridge regression \citep{agarwal2022hierarchical} as opposed to OLS as the GLM within the \method\;framework, we are able to mitigate this instability and regain the added benefits of including the raw feature and LOO evaluation, as illustrated by the strong performance of \method\;(ridge+raw+loo).

\section{MDI Biases Simulations}
\label{supp:mdi-bias}

\begin{figure}[h!]
    \centering
    \begin{tabularx}{.6\textwidth}{@{}l *5{>{\centering\arraybackslash}X}@{}}
    \begin{minipage}{.11\textwidth}
    \phantom{}
    \end{minipage}%
    {\small $PVE=0.1$} & {\small $PVE=0.4$}
    \end{tabularx}\\
    \includegraphics[width=.6\textwidth]{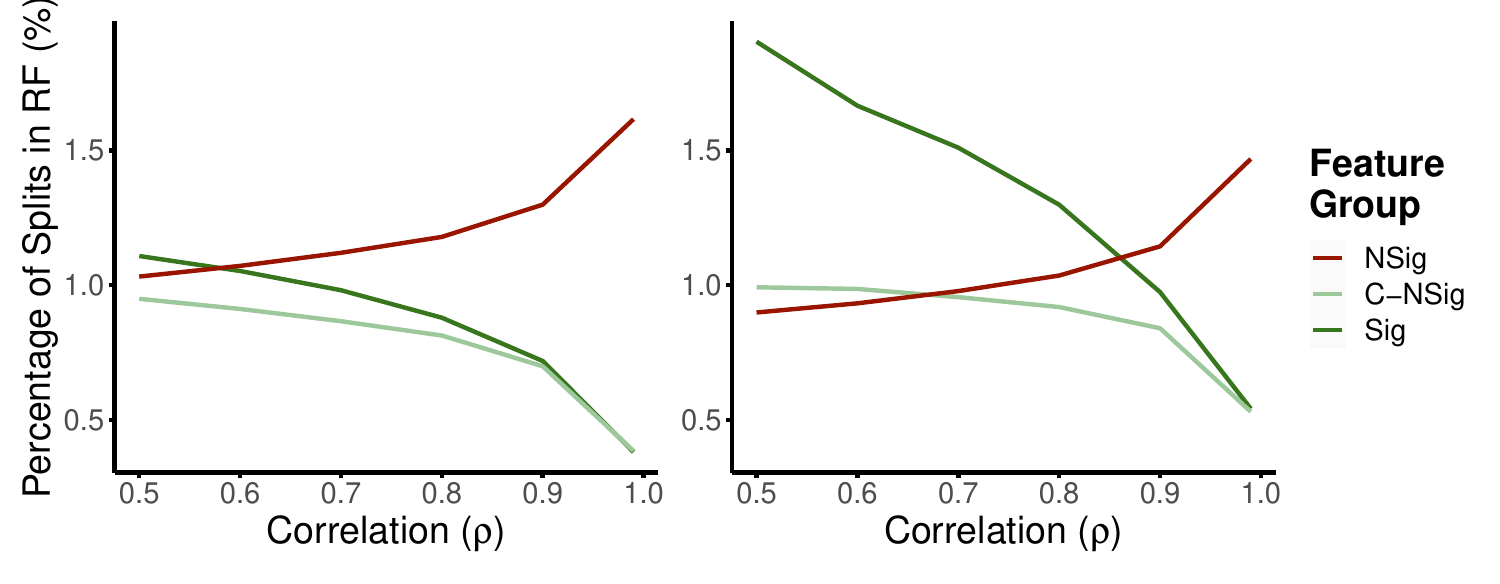}
    \caption{As the correlation increases, the percentage of splits in the RF that are made using features from the correlated group (Sig or C-NSig) decreases. This pattern is true for both $PVE = 0.1$ (left) and $PVE = 0.4$ (right) under the correlation simulation setup described in Section~\ref{sec:correlation-bias}.}
    \label{fig:correlation_splits}
\end{figure}
\begin{figure}[h!]
    \centering
    \begin{subfigure}{0.5\textwidth}
    \centering
    \small (A) $PVE=0.1$ \phantom{xxxxx}
    \vspace{1mm}
    \includegraphics[width=1\textwidth]{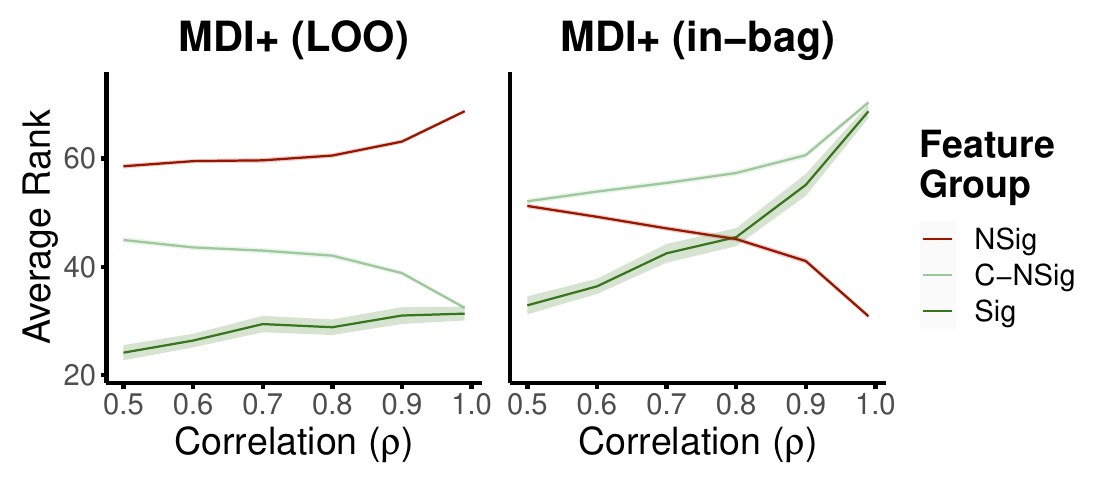}
    \end{subfigure}%
    \begin{subfigure}{0.5\textwidth}
    \centering
    \small (B) $PVE=0.4$ \phantom{xxxxx}
    \vspace{1mm}
    \includegraphics[width=1\textwidth]{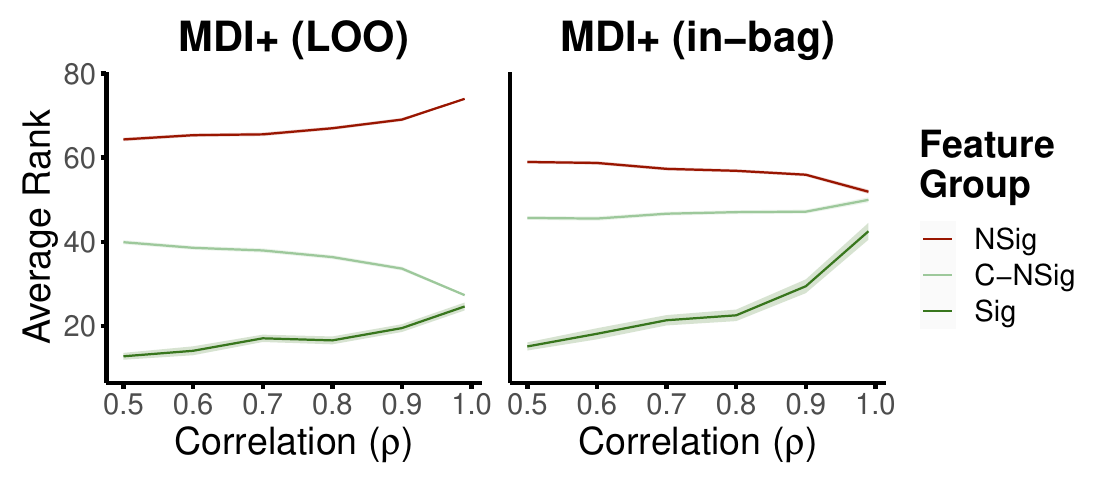}
    \end{subfigure}

    \caption{While \method\;using the LOO partial model predictions is able to mitigate the correlated feature bias, \method\;without the LOO scheme suffers from the correlated feature bias like MDI. This pattern holds for both the (A) $PVE = 0.1$ and (B) $PVE = 0.4$ simulation settings described in Section~\ref{sec:correlation-bias}.}
    \label{fig:correlation-sim-gmdi-loo}
\end{figure}

\paragraph{Correlation Bias.}
In Figure \ref{fig:correlation_splits}, we display the average percentage of RF splits per feature in each group (i.e., Sig, CSig, and NSig) under the correlation bias simulations described in Section~\ref{sec:correlation-bias}. Further, in Figure \ref{fig:correlation-sim-gmdi-loo}, we examine the performance of \method\;with and without the LOO data splitting scheme, denoted \method\;(LOO) and \method\;(in-bag), respectively. As seen in Figure \ref{fig:correlation-sim-gmdi-loo}, for both $PVE = 0.1,\; 0.4$, LOO sample-splitting overcomes the correlation bias that using in-bag samples induces. 

\begin{figure}[h!]
    \centering
    \includegraphics[width=.65\textwidth]{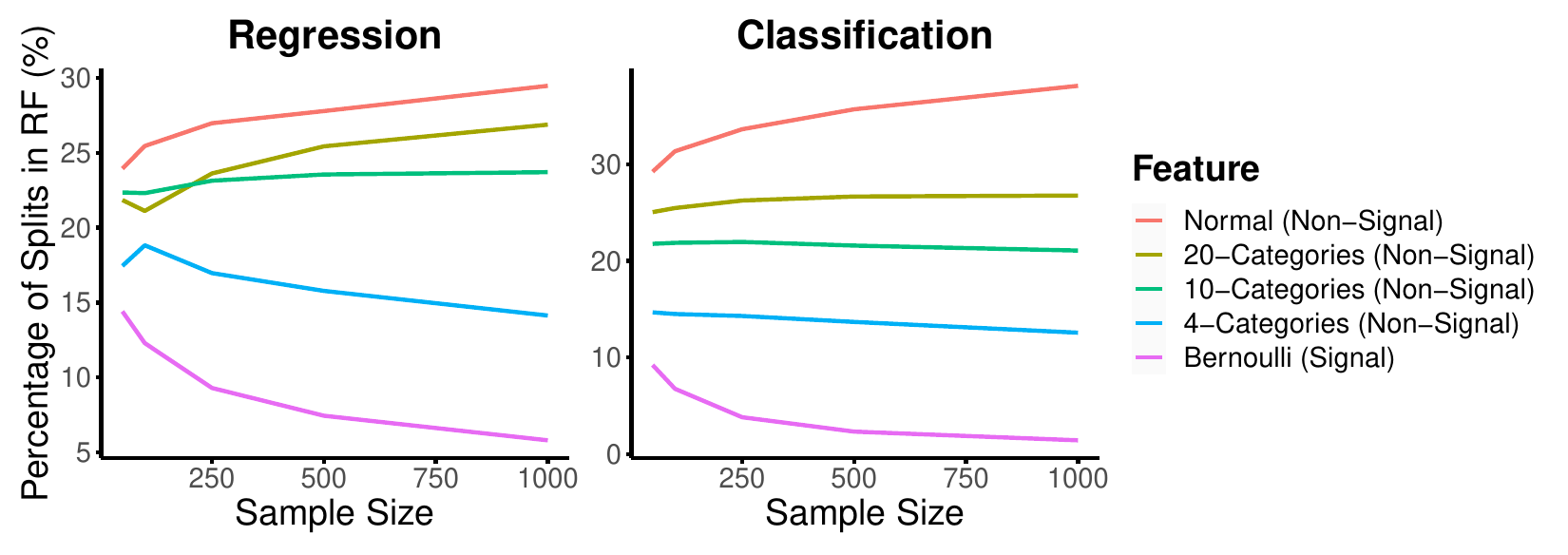}
    \caption{Noisy features with higher entropy are inherently split on more frequently in the RF. This pattern is true across various sample sizes under both the regression and classification simulation settings described in Section~\ref{sec:entropy-bias}.}
    \label{fig:entropy_splits}
\end{figure}

\begin{figure}[h!]
    \centering
    \includegraphics[width=0.8\textwidth]{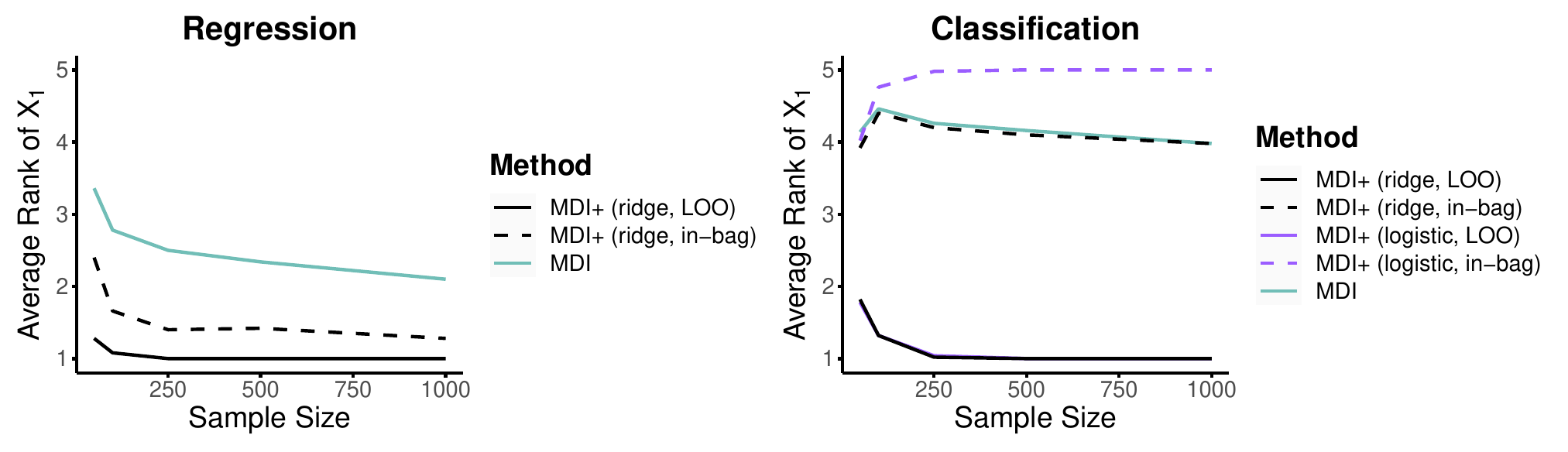}
    \caption{While \method\;using the LOO partial model predictions is able to mitigate the entropy bias, \method\;without the LOO scheme suffers from the entropy bias like MDI. This pattern holds for both the regression (left) and the classification (bottom) simulation settings described in Section~\ref{sec:entropy-bias}.}
    \label{fig:entropy_sim-gmdi-loo}
\end{figure}

\paragraph{Entropy Bias.}
In Figure \ref{fig:entropy_splits}, we display the average percentage of RF splits per feature under the entropy bias simulations described in Section~\ref{sec:entropy-bias}. Further, in Figure \ref{fig:entropy_sim-gmdi-loo}, we examine the performance of \method\;with and without ridge regularization and the LOO data splitting scheme in both the regression and classification setting. As seen in Figure \ref{fig:entropy_sim-gmdi-loo}, LOO sample-splitting overcomes the entropy bias in both settings. In the regression setting, ridge regularization also helps to mitigate the entropy bias of MDI (demonstrated by the difference between MDI and \method\;(ridge, in-bag)).

\section{PCS-Informed Model Recommendation}
\label{supp:PCS_model}
In this section, we detail the stability-driven selection and model aggregation procedure discussed in Section \ref{subsec:pcs_model_selection}. We also establish the efficacy of both approaches through a data-inspired simulation study. 

\vspace{2mm}

\noindent For both approaches, the first step is to evaluate the prediction performance of each $h \in \mathcal{H}$ on an independent test set (obtained from a train-test split). Filter out all $h$ whose predictive performance is worse than that of RF. Given this set of screened models $\{h_1, \ldots, h_M\}$, we propose the following two approaches to compute feature rankings.

\paragraph{Stability-based selection for $\tilde{\Psi}$, $\mathcal{M}$, $m$.} Let  $\mathcal{T} = \{\mathcal{S}_1, \ldots, \mathcal{S}_T\}$ denote the fitted trees of the RF trained on the data $\mathcal{D}_n$. Generate bootstrap samples of the fitted trees $\mathcal{T}_b$, $b = 1, \ldots, B$. For each bootstrap sample, $\mathcal{T}_b$, compute feature rankings $R_b(h_i)$ for the \method\;model $h_i$. Denote the set of feature rankings generated by $h_i$ as $\mathcal{R}_i = \{R_1(h_i), \ldots, R_B(h_i) \}$. To evaluate the stability of $h_i$, we measure the similarity between rankings from all pairs of the $B$ bootstraps, and take the average across these $B \choose 2$ similarity scores. To measure this similarity between two ranked lists, we use Rank-based Overlap (RBO) \citep{webber2010similarity}. RBO measures how frequently two ranked lists agree on ordering of items, weighing agreement between  
higher ranks more heavily than lower ranks. This makes it appropriate for feature importance rankings where we are often concerned with the most important features (i.e., the highest ranking features). Finally, we choose $h^{*}$ to be the $h \in \{h_1, \ldots, h_M \}$ with the highest RBO score averaged across all $B \choose 2$ pairs of bootstraps. 

\paragraph{Model aggregation.} While the approach above results in feature importance rankings that correspond to a single \method\;model, it is sometimes desirable to obtain rankings that do not depend on a single choice of \method\;model. Hence, a proposed alternative is an ensemble-based approach where we rank features based upon the median ranking of each feature $X_k$ across all prediction-screened \method\;models $h \in \{h_1, \ldots, h_M\}$. We refer to this as \method\;(ensemble). Other ways of aggregating feature importances can also be performed; we leave this to future work. 

\begin{figure}[h!]
\begin{subfigure}{\textwidth}
    \centering
    \vspace{2mm}
    \begin{tabularx}{1\textwidth}{@{}l *5{>{\centering\arraybackslash}X}@{}}
    \begin{minipage}{.07\textwidth}
    \phantom{}
    \end{minipage}%
    & {\small $PVE=0.1$} & {\small $PVE=0.2$} & {\small $PVE=0.4$} & {\small $PVE=0.8$} &
    \begin{minipage}{.02\textwidth}
    \phantom{}
    \end{minipage}%
    \end{tabularx}\\
    \rotatebox{90}{{\hspace{4mm} \centering \small \textbf{(A) CCLE RNASeq Linear}}}
    \hspace{.25mm}
    \includegraphics[width=0.95\textwidth]{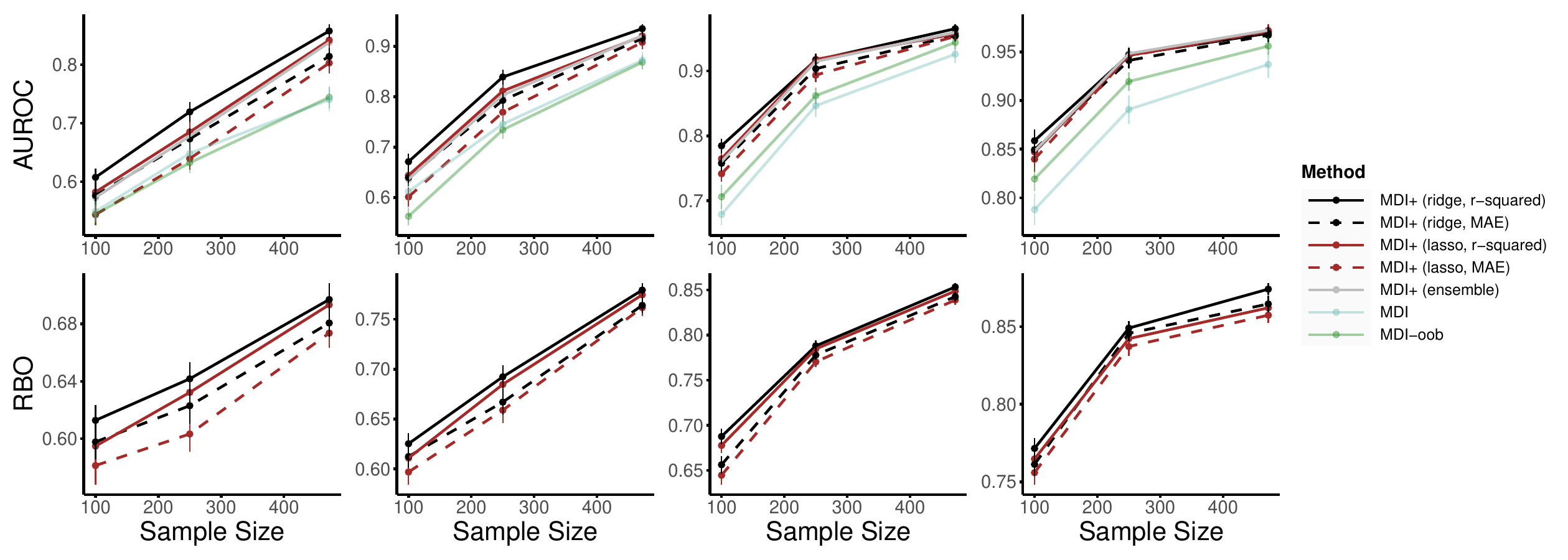}
\end{subfigure}
\hrule
\begin{subfigure}{\textwidth}
    \centering
    \vspace{-5mm}
    \begin{tabularx}{1\textwidth}{@{}l *5{>{\centering\arraybackslash}X}@{}}
    \begin{minipage}{.04\textwidth}
    \phantom{}
    \end{minipage}%
    \begin{minipage}{.02\textwidth}
    \phantom{}
    \end{minipage}%
    \end{tabularx}\\
    \rotatebox{90}{{\hspace{2mm} \centering \small \textbf{(B) CCLE RNASeq Poly. Int.}}}
    \hspace{.25mm}
    \includegraphics[width=0.95\textwidth]{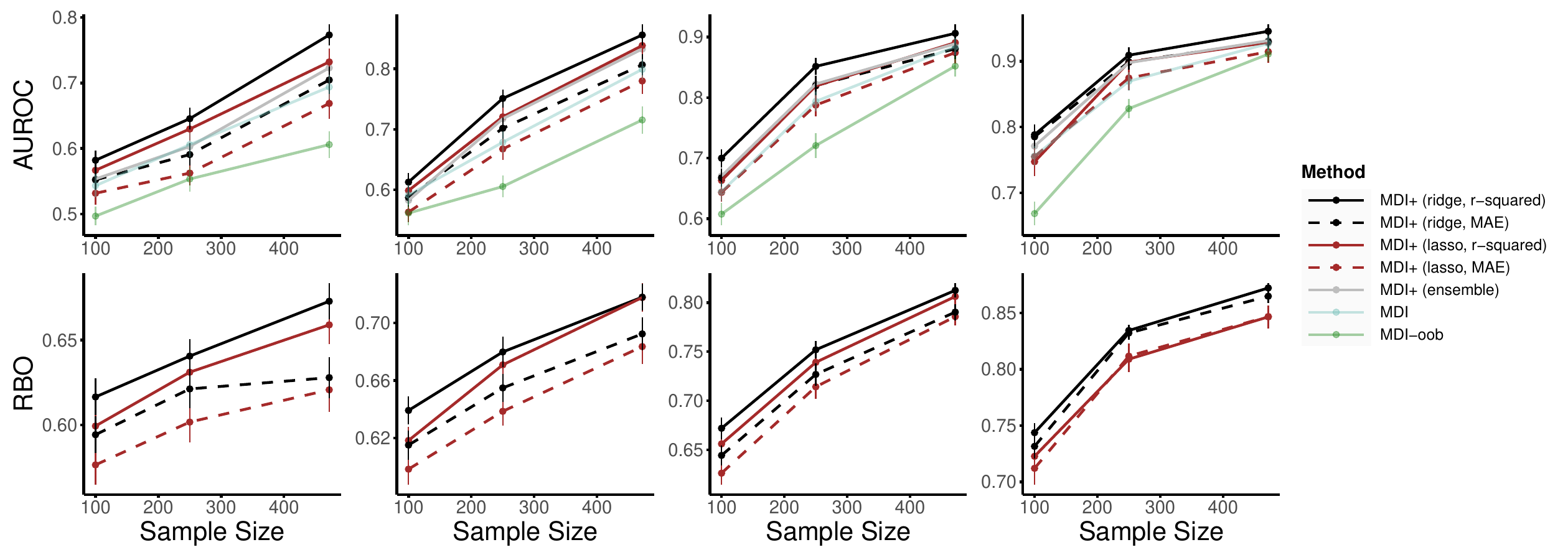}
\end{subfigure}
\caption{In each panel, the top row shows the accuracy of the feature importance rankings using various feature importance methods, and the bottom row shows the stability scores, as measured by RBO, for each choice of GLM and similarity metric used in \method. The GLM and similarity metric yielding the most stable feature rankings generally gives the most accurate feature importance rankings. \method\;(ensemble) typically falls in between the different \method\;versions and outperforms existing methods (MDI and MDI-oob) in terms of feature importance ranking accuracy. These patterns hold across various choices of sample sizes (specified on the $x$-axis) and regression functions (specified by panel).}
\label{fig:stability_sims}
\end{figure}

\paragraph{Simulation Study.} We evaluate the effectiveness of these two PCS-informed model recommendation techniques for feature importance ranking via the following simulation study. We follow the simulation setup described in Section~\ref{subsec:regression_results}. We consider ridge and LASSO as the two (regularized) GLMs, and $\rsq$ and mean absolute error (MAE) as the two metrics, producing a total of four candidate \method\;models. We also aggregate these four candidate \method\;models into an ensemble, denoted \method\;(ensemble), by taking the median feature rankings as described above. We present here the results using the CCLE dataset with responses generated by a linear and polynomial interaction model as described in Section~\ref{subsec:sim_set_up}. We note that all four \method\;models passed the predictive check. In Figure~\ref{fig:stability_sims}, we thus plot the RBO for all four \method\;models, where a higher RBO indicates greater ranking stability. Additionally, we display the feature ranking performance (as measured by AUROC) for these four \method\;models, \method\;(ensemble), as well as MDI and MDI-oob. As seen in Figure~\ref{fig:stability_sims}, \method\;(ridge, $\rsq$) has the highest stability score, which often translates to the best feature ranking performance across various sample sizes and $PVE$s. \method\;(ensemble) also performs reasonably well, typically yielding an AUROC for feature ranking accuracy in between the different \method\;versions and better than the baselines of MDI and MDI-oob. Despite the stability-based selection providing the best feature ranking accuracy in these simulations, these simulations may not capture all subtleties of applying these methods in practice, and \method\;(ensemble) may prove useful in practice. 

\section{Additional \rfmethod\;Prediction Results}
\label{supp:prediction_accuracy}

Below, we provide the full prediction results for \rfmethod\;applied to all 24 CCLE drugs in the regression setting.

\begin{figure}[h!]
    \centering
    \includegraphics[width=.5\textwidth]{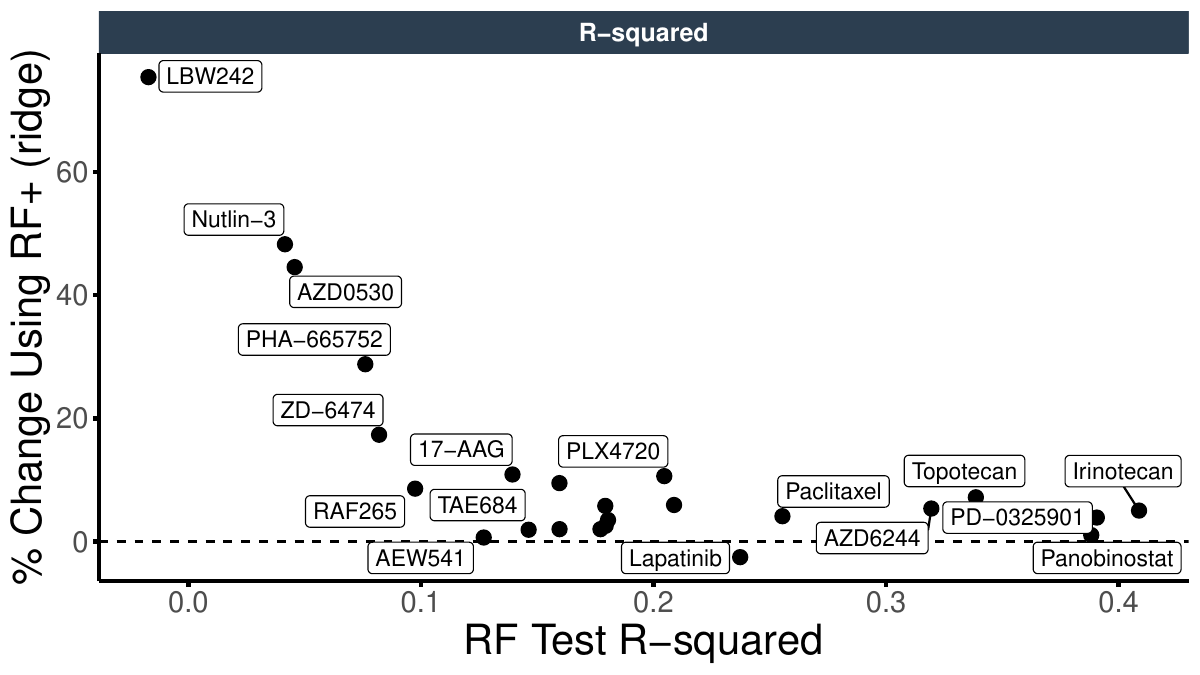}
    \caption{Relative performance of \rfmethod\;(ridge) as compared to RF for all 24 CCLE drugs in the regression setting. Here, we measure prediction performance using the test set $\rsq$. Averaged across 32 train-test splits, \rfmethod\;(ridge) yields higher test prediction performance than RF for 23 out of 24 drugs.}
    \label{fig:prediction_supp}
\end{figure}
\clearpage
\section{Case Studies} \label{supp:case_study}


\subsection{Drug Response Prediction} \label{supp:ccle_case_study}

\paragraph{Data preprocessing.}
Originally, the CCLE RNASeq gene expression data set consisted of 50114 genes. To reduce the number of features to a more manageable size for our analysis, we took the top 5000 genes with the highest variance. Screening based on these fast, marginal measurements (e.g., variance, correlation) is both common and often beneficial prior to a more in-depth analysis, especially in this ultra-high-dimensional regime. The unprocessed CCLE data can be downloaded from DepMap Public 18Q3 (\url{https://depmap.org/portal/download/}). The processed CCLE data used in our case study can be found on Zenodo at 
\if1\blind{\url{https://zenodo.org/record/8111870}}\fi
\if0\blind{\url{redacted}}\fi.

\paragraph{Details of drug response outcome variable.} In this case study, the outcome of interest is the efficacy, or response, of a drug. To obtain a measure of the drug response for each cell line, the CCLE project measured the pharmacological sensitivity of each drug in vitro across eight different dosages and quantified this using the \textit{activity area}, or area under the dose-response curve. This activity area is measured on an 8-point scale, with 0 corresponding to an inactive compound (i.e., a drug that did not inhibit the growth of the cancer cells across all 8 dosages) and 8 corresponding to a compound with 100\% inhibition of cancer cell growth at all 8 dosages. Further details on the CCLE data can be found in \citet{barretina2012cancer}.

\paragraph{Results.} To supplement the discussion in Section~\ref{sec:case_study_ccle}, we provide additional tables and figures below.

In Table~\ref{tab:ccle_top_genes}, we list the top 5 gene expression predictors for each drug according to various feature importance methods. 
Since we evaluated the feature importance methods across 32 train-test splits (details in Section~\ref{sec:case_study_ccle}), the genes are ranked according to their average feature importance ranking across the 32 splits.

\begin{figure}[h!]
    \centering
    \includegraphics[width=1\textwidth]{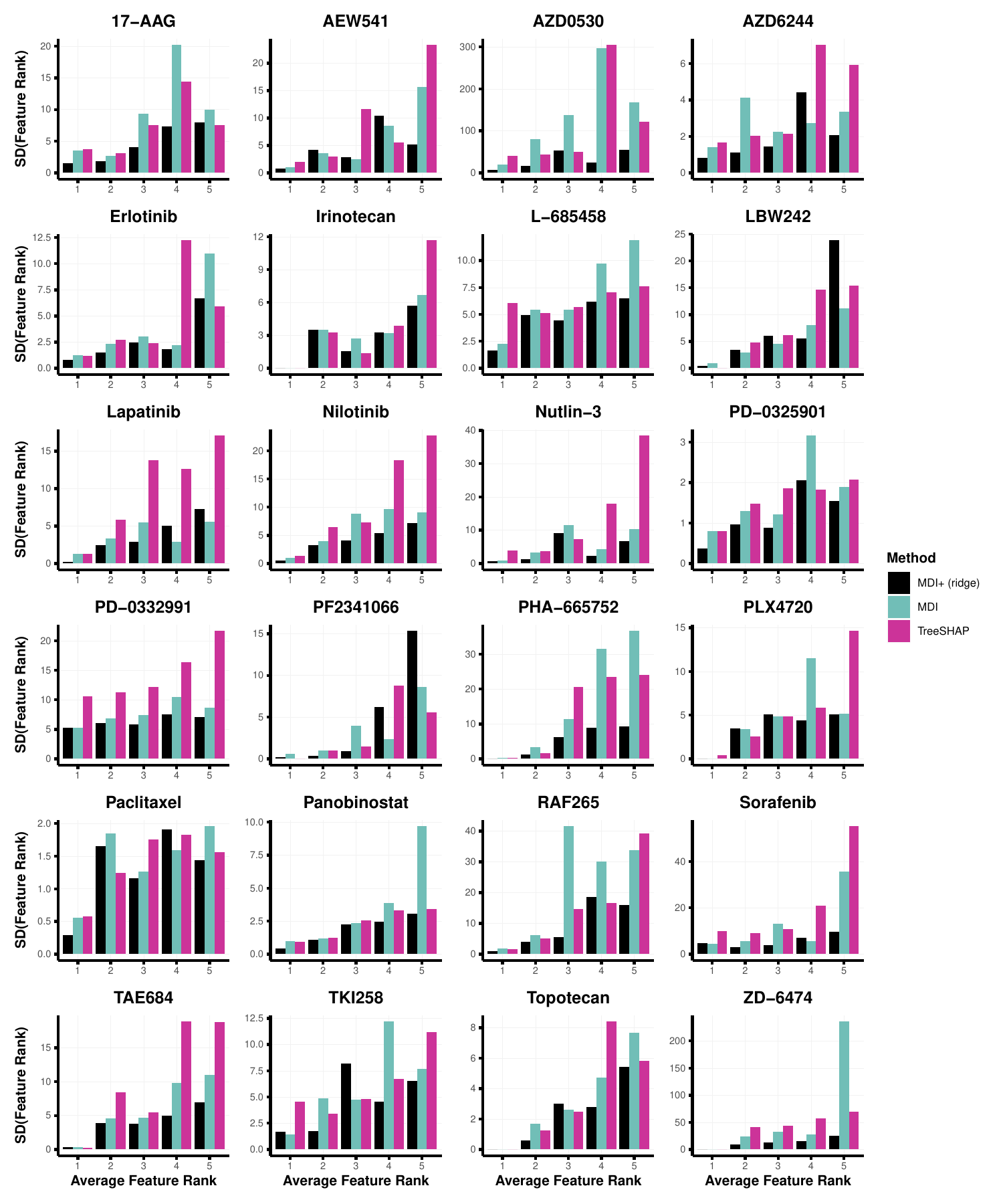}
    \caption{Stability of top 5 genes for each drug response prediction model across 32 train-test splits. The x-axis corresponds to the top 5 features for each method, ranked according to their average feature ranking across 32 train-test splits. On the y-axis, we provide one measure of stability -- namely, the standard deviation of the feature rankings across the 32 train-test splits. \method\;generally provides the most stable feature importance rankings for these top 5 genes. Results from the feature importance methods under study, excluding MDI-oob and MDA, are shown here.}
    \label{fig:ccle-stability-dist-all-zoom} 
\end{figure}

We also provide the stability distribution plots for all 24 drugs in Figure~\ref{fig:ccle-stability-dist-all-zoom}. 
Specifically, for the top 5 genes using each feature importance method under study (i.e., those listed in Table~\ref{tab:ccle_top_genes}), we plot the standard deviation of the gene's feature importance rankings across the 32 train-test splits.
A small standard deviation indicates a more stable feature importance measure.
For many drugs, MDI-oob and MDA are highly unstable (with standard deviations going far beyond the shown y-axis limits) and thus omitted from the plot for clarity.
Across all drugs, we see that \method\;generally yields the most stable rankings for the top 5 features across the 32 train-test splits.

\begin{figure}[h!]
    \centering
    \includegraphics[width=1\textwidth]{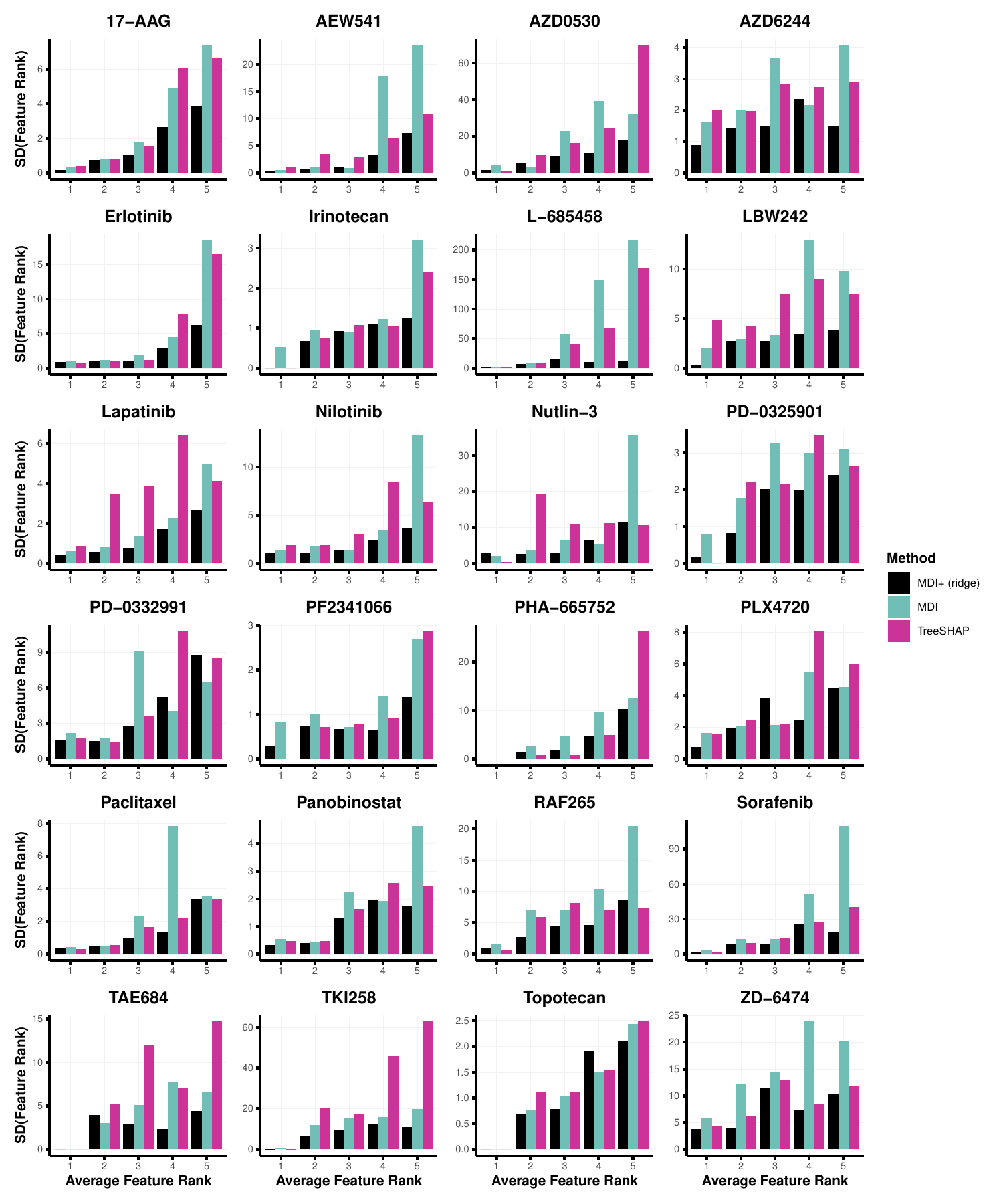}
    \caption{Stability of top 5 genes for each drug response prediction model across 32 RF fits, trained using different random seeds. The x-axis corresponds to the top 5 features for each method, ranked according to their average feature ranking across the 32 RF fits. On the y-axis, we provide one measure of stability -- namely, the standard deviation of the feature rankings across the 32 RF fits. \method\;generally provides the most stable feature importance rankings for these top 5 genes. Results from the feature importance methods under study, excluding MDI-oob and MDA, are shown here.}
    \label{fig:ccle-stability-dist-all-zoom-seed} 
\end{figure}

To further evaluate the stability of \method\;across only the randomness in the RF training, we performed an analogous stability analysis, where we varied the random seed used to train the RF but kept the training data fixed.
Specifically, we trained 32 RFs on the full CCLE data, each RF using a different random seed.
We then evaluate the stability of the feature rankings across these 32 RF fits. 
We see in Figure~\ref{fig:ccle-stability-dist-all-zoom-seed} that \method\;remains the most stable across the different RF fits. 
This highlights another practical advantage of \method, as it is highly undesirable for the feature importance rankings to change due to an arbitrary choice such as the random seed used in training the RF.

\begin{figure}[h!]
    \centering
    \includegraphics[width=1\textwidth]{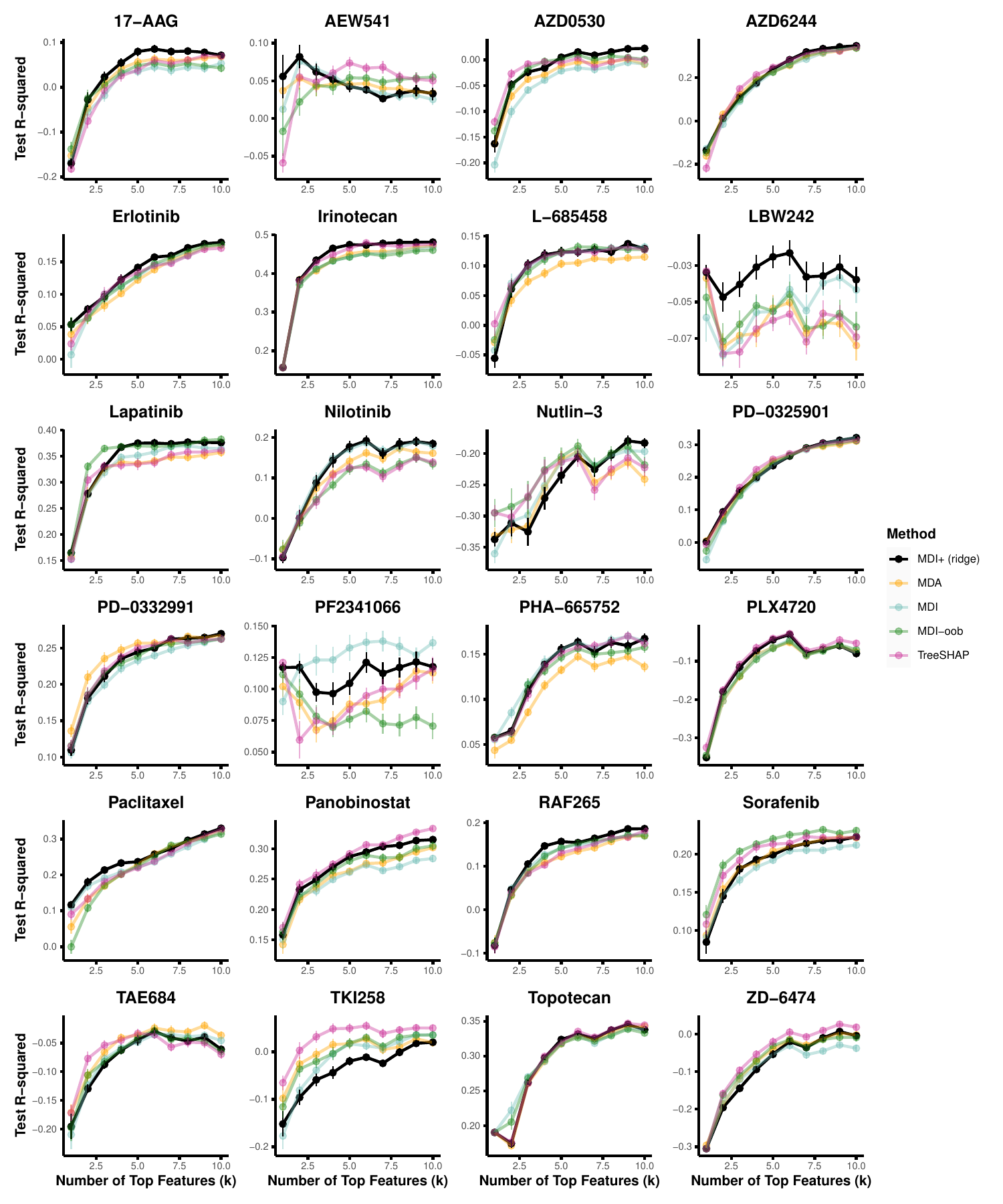}
    \caption{RF prediction performance, measured via test $\rsq$, using the top $k$ features from each feature importance methods across the 24 drugs in the CCLE case study. Results are averaged across 32 train-test data splits.}
    \label{fig:ccle-prediction} 
\end{figure}

Lastly, in Figure~\ref{fig:ccle-prediction}, we evaluated the predictive power of the top $k$ genes from each feature importance method across the 32 train-test (80-20\%) splits.
Specifically, for each $k$ and each feature importance method, we took the top $k$ ranked genes from the given feature importance method and trained an RF using the training data, restricted to only these $k$ features. 
We then evaluated the prediction accuracy of the fitted RF on the test set and average these results across the 32 train-test splits.
While evidence from the existing scientific literature remains the main source of validation, supporting the top-ranked genes from \method, the prediction accuracy of the top-ranked genes (as shown in Figure~\ref{fig:ccle-prediction}) can provide another check. 
In Table~\ref{tab:ccle-prediction}, we summarize the prediction results when $k = 10$ by counting the number of drugs, for which each feature importance method gave the best test $\rsq$, second-best test $\rsq$, etc.
We see that for 12 out of the 24 drugs, the top 10 genes from \method\;(ridge) generally had the highest prediction power compared to other methods.
In accordance with the predictability principle of the PCS framework, strong prediction performance suggests that the model (and in this case, the top-ranked features) may better capture the underlying data-generating process.

It is important to note, however, that correlated variables can pose subtle issues when interpreting these top $k$ prediction results.
For concreteness, consider the scenario where the top $k$ ranked features from Method $A$ tend to be independent while the top-ranked features from Method $B$ tend to be highly-correlated. 
It is likely that the predictive power from the top $k$ features from Method $A$ is higher than that from Method $B$ simply because the $k$ independent features from Method $A$ inherently contain more information than the $k$ highly-correlated features from Method $B$.
That is, the prediction accuracies from these two sets of $k$ features are not directly comparable and requires additional investigation into the correlation structure of $\bX$.
From preliminary explorations, this complication is typically most apparent for small values of $k$. 
Intuitively, when $k$ is small, correlation structures between two sets of features can be very different. 
As $k$ grows, the two correlation structures tend to become most similar to each other. 
This motivated our choice of $k=10$ in Table~\ref{tab:ccle-prediction}. 
However, further investigation is warranted, and we leave a deeper investigation of this phenomenon to ongoing and future work.

{\scriptsize
\setlength{\tabcolsep}{3pt}
\begin{longtable}[t]{cccccc}
\caption{Top 5 most important genes for each drug's response according to various feature importance methods. Genes are ranked by their average feature importance ranking across 32 train-test splits (shown in parentheses).}\\
\toprule
  & \textbf{\method} & \textbf{MDI} & \textbf{TreeSHAP} & \textbf{MDI-oob} & \textbf{MDA}\\
\midrule
\endfirsthead
\caption[]{ \textit{(continued)}}\\
\toprule
  & \textbf{\method} & \textbf{MDI} & \textbf{TreeSHAP} & \textbf{MDI-oob} & \textbf{MDA}\\
\midrule
\endhead

\endfoot
\bottomrule
\endlastfoot
\addlinespace[0.3em]
\multicolumn{6}{l}{\textbf{17-AAG}}\\
\hspace{1em}\cellcolor{gray!6}{1} & \cellcolor{gray!6}{PCSK1N (1.47)} & \cellcolor{gray!6}{PCSK1N (2.19)} & \cellcolor{gray!6}{PCSK1N (2.19)} & \cellcolor{gray!6}{PCSK1N (2.5)} & \cellcolor{gray!6}{PCSK1N (9.16)}\\
\hspace{1em}2 & MMP24 (3.41) & MMP24 (4.97) & MMP24 (4.06) & NQO1 (9.03) & MMP24 (194.5)\\
\hspace{1em}\cellcolor{gray!6}{3} & \cellcolor{gray!6}{RP11-109D20.2 (4.59)} & \cellcolor{gray!6}{ZSCAN18 (6.94)} & \cellcolor{gray!6}{ZSCAN18 (8.56)} & \cellcolor{gray!6}{ZNF667-AS1 (26.59)} & \cellcolor{gray!6}{ZNF667-AS1 (241.38)}\\
\hspace{1em}4 & ZSCAN18 (8.09) & RP11-109D20.2 (7.41) & RP11-109D20.2 (10.09) & ZSCAN18 (44.03) & RP11-109D20.2 (525.22)\\
\hspace{1em}\cellcolor{gray!6}{5} & \cellcolor{gray!6}{NQO1 (8.84)} & \cellcolor{gray!6}{NQO1 (9.53)} & \cellcolor{gray!6}{NQO1 (11.81)} & \cellcolor{gray!6}{TST (49.38)} & \cellcolor{gray!6}{SH3BP1 (587.94)}\\
\addlinespace[0.3em]
\multicolumn{6}{l}{\textbf{AEW541}}\\
\hspace{1em}1 & TXNDC5 (1.41) & TCEAL4 (1.62) & TXNDC5 (1.75) & TXNDC5 (1.5) & TCEAL4 (5.59)\\
\hspace{1em}\cellcolor{gray!6}{2} & \cellcolor{gray!6}{ATP8B2 (4.34)} & \cellcolor{gray!6}{TXNDC5 (3.53)} & \cellcolor{gray!6}{ATP8B2 (3.84)} & \cellcolor{gray!6}{ATP8B2 (4.69)} & \cellcolor{gray!6}{IQGAP2 (238.8)}\\
\hspace{1em}3 & VAV2 (6.03) & ATP8B2 (8.38) & VAV2 (5.41) & VAV2 (5.84) & RP11-343H19.2 (303.25)\\
\hspace{1em}\cellcolor{gray!6}{4} & \cellcolor{gray!6}{TNFRSF17 (8.53)} & \cellcolor{gray!6}{VAV2 (10.47)} & \cellcolor{gray!6}{TCEAL4 (9.44)} & \cellcolor{gray!6}{TCEAL4 (6.5)} & \cellcolor{gray!6}{TXNDC5 (312.62)}\\
\hspace{1em}5 & TCEAL4 (9.03) & PLEKHF1 (19.25) & TNFRSF17 (9.69) & TNFRSF17 (13.56) & ATP8B2 (318.59)\\
\addlinespace[0.3em]
\multicolumn{6}{l}{\textbf{AZD0530}}\\
\hspace{1em}\cellcolor{gray!6}{1} & \cellcolor{gray!6}{PRSS57 (5.16)} & \cellcolor{gray!6}{SYTL1 (17.62)} & \cellcolor{gray!6}{PRSS57 (7.69)} & \cellcolor{gray!6}{PRSS57 (12.09)} & \cellcolor{gray!6}{VTN (105.69)}\\
\hspace{1em}2 & SYTL1 (12.31) & PRSS57 (32.5) & SYTL1 (42.94) & DDAH2 (440.16) & SYTL1 (216.62)\\
\hspace{1em}\cellcolor{gray!6}{3} & \cellcolor{gray!6}{STXBP2 (15.38)} & \cellcolor{gray!6}{SFTA1P (36.5)} & \cellcolor{gray!6}{NFE2 (43.06)} & \cellcolor{gray!6}{SLC16A9 (484.34)} & \cellcolor{gray!6}{STXBP2 (245.31)}\\
\hspace{1em}4 & NFE2 (23.12) & STXBP2 (62.59) & STXBP2 (61.62) & STXBP2 (486.09) & ZBED2 (466.48)\\
\hspace{1em}\cellcolor{gray!6}{5} & \cellcolor{gray!6}{THEM4 (34.41)} & \cellcolor{gray!6}{CLDN16 (67.28)} & \cellcolor{gray!6}{SLC16A9 (61.81)} & \cellcolor{gray!6}{RAPGEF3 (514.2)} & \cellcolor{gray!6}{DDAH2 (472.06)}\\
\addlinespace[0.3em]
\multicolumn{6}{l}{\textbf{AZD6244}}\\
\hspace{1em}1 & LYZ (1.66) & TOR4A (2.31) & LYZ (1.72) & LYZ (2.53) & LYZ (2.09)\\
\hspace{1em}\cellcolor{gray!6}{2} & \cellcolor{gray!6}{SPRY2 (2.34)} & \cellcolor{gray!6}{SPRY2 (3.31)} & \cellcolor{gray!6}{RP11-1143G9.4 (3.59)} & \cellcolor{gray!6}{SPRY2 (2.59)} & \cellcolor{gray!6}{RP11-1143G9.4 (3.59)}\\
\hspace{1em}3 & RP11-1143G9.4 (2.84) & LYZ (3.69) & SPRY2 (3.91) & TOR4A (3.25) & TOR4A (3.66)\\
\hspace{1em}\cellcolor{gray!6}{4} & \cellcolor{gray!6}{ETV4 (5.22)} & \cellcolor{gray!6}{ETV4 (5.19)} & \cellcolor{gray!6}{TOR4A (6.41)} & \cellcolor{gray!6}{RP11-1143G9.4 (4.75)} & \cellcolor{gray!6}{SPRY2 (4.5)}\\
\hspace{1em}5 & TOR4A (6.41) & RP11-1143G9.4 (6.84) & RNF125 (6.66) & ETV4 (6.91) & ETV4 (6.34)\\
\addlinespace[0.3em]
\multicolumn{6}{l}{\textbf{Erlotinib}}\\
\hspace{1em}\cellcolor{gray!6}{1} & \cellcolor{gray!6}{CDH3 (1.47)} & \cellcolor{gray!6}{CDH3 (1.84)} & \cellcolor{gray!6}{CDH3 (2.03)} & \cellcolor{gray!6}{CDH3 (1.97)} & \cellcolor{gray!6}{CDH3 (1.88)}\\
\hspace{1em}2 & RP11-615I2.2 (2.28) & RP11-615I2.2 (3.28) & RP11-615I2.2 (2.88) & RP11-615I2.2 (3.53) & RP11-615I2.2 (3.16)\\
\hspace{1em}\cellcolor{gray!6}{3} & \cellcolor{gray!6}{EGFR (4.34)} & \cellcolor{gray!6}{SPRR1A (3.97)} & \cellcolor{gray!6}{EGFR (3.97)} & \cellcolor{gray!6}{SPRR1A (6.25)} & \cellcolor{gray!6}{SPRR1A (3.84)}\\
\hspace{1em}4 & SPRR1A (4.44) & SYTL1 (7.84) & SPRR1A (4.19) & GJB3 (8.78) & EGFR (8.31)\\
\hspace{1em}\cellcolor{gray!6}{5} & \cellcolor{gray!6}{GJB3 (7.44)} & \cellcolor{gray!6}{EGFR (8.69)} & \cellcolor{gray!6}{KRT16 (11.41)} & \cellcolor{gray!6}{EGFR (9.31)} & \cellcolor{gray!6}{SYTL1 (8.72)}\\
\addlinespace[0.3em]
\multicolumn{6}{l}{\textbf{Irinotecan}}\\
\hspace{1em}1 & SLFN11 (1) & SLFN11 (1) & SLFN11 (1) & SLFN11 (1) & SLFN11 \vphantom{1} (1)\\
\hspace{1em}\cellcolor{gray!6}{2} & \cellcolor{gray!6}{S100A16 (4.12)} & \cellcolor{gray!6}{S100A16 (3.75)} & \cellcolor{gray!6}{S100A16 (3.84)} & \cellcolor{gray!6}{S100A16 (3.25)} & \cellcolor{gray!6}{WWTR1 (6.03)}\\
\hspace{1em}3 & IFITM10 (4.19) & IFITM10 (4.09) & WWTR1 (4.28) & WWTR1 (4.12) & TRIM16L (150.38)\\
\hspace{1em}\cellcolor{gray!6}{4} & \cellcolor{gray!6}{WWTR1 (4.94)} & \cellcolor{gray!6}{WWTR1 (4.78)} & \cellcolor{gray!6}{IFITM10 (8.03)} & \cellcolor{gray!6}{RP11-359P5.1 (8.22)} & \cellcolor{gray!6}{IFITM10 (163.44)}\\
\hspace{1em}5 & PPIC (7.81) & PPIC (10.22) & RP11-359P5.1 (8.47) & IFITM10 (8.41) & S100A16 (182.19)\\
\addlinespace[0.3em]
\multicolumn{6}{l}{\textbf{L-685458}}\\
\hspace{1em}\cellcolor{gray!6}{1} & \cellcolor{gray!6}{PXK (2.03)} & \cellcolor{gray!6}{DEF6 (4.62)} & \cellcolor{gray!6}{PXK (2.03)} & \cellcolor{gray!6}{PXK (3)} & \cellcolor{gray!6}{PXK (2.94)}\\
\hspace{1em}2 & DEF6 (4.28) & PXK (4.94) & DEF6 (4.38) & IKZF1 (3.44) & CXorf21 (4.62)\\
\hspace{1em}\cellcolor{gray!6}{3} & \cellcolor{gray!6}{CXorf21 (4.84)} & \cellcolor{gray!6}{CXorf21 (5.44)} & \cellcolor{gray!6}{CXorf21 (5.62)} & \cellcolor{gray!6}{CXorf21 (5.75)} & \cellcolor{gray!6}{DEF6 (4.66)}\\
\hspace{1em}4 & IKZF1 (6.03) & IKZF1 (5.75) & IKZF1 (7.91) & DEF6 (10.31) & IKZF1 (5.47)\\
\hspace{1em}\cellcolor{gray!6}{5} & \cellcolor{gray!6}{RP11-359P5.1 (9.09)} & \cellcolor{gray!6}{RP11-359P5.1 (9.94)} & \cellcolor{gray!6}{RP11-359P5.1 (12.81)} & \cellcolor{gray!6}{CTNNA1 (13.88)} & \cellcolor{gray!6}{RP11-359P5.1 (10.06)}\\
\addlinespace[0.3em]
\multicolumn{6}{l}{\textbf{LBW242}}\\
\hspace{1em}1 & SERPINB6 (1.12) & SERPINB6 (1) & SERPINB6 (1.66) & SERPINB6 (1.31) & SERPINB6 (1.56)\\
\hspace{1em}\cellcolor{gray!6}{2} & \cellcolor{gray!6}{RGS14 (5.12)} & \cellcolor{gray!6}{RGS14 (6.66)} & \cellcolor{gray!6}{RGS14 (3.66)} & \cellcolor{gray!6}{GPT2 (45.62)} & \cellcolor{gray!6}{HERC5 (54.31)}\\
\hspace{1em}3 & HERC5 (5.41) & MAGEC1 (7.5) & MAGEC1 (5.53) & GBP1 (179.28) & ITGA1 (73.34)\\
\hspace{1em}\cellcolor{gray!6}{4} & \cellcolor{gray!6}{MAGEC1 (7.62)} & \cellcolor{gray!6}{ITGA1 (10.53)} & \cellcolor{gray!6}{GBP1 (5.78)} & \cellcolor{gray!6}{ZNF32 (222.03)} & \cellcolor{gray!6}{PTGS1 (296.62)}\\
\hspace{1em}5 & GBP1 (8.22) & HERC5 (12.41) & CCL2 (13.5) & IGSF3 (257.88) & GPT2 (316.12)\\
\addlinespace[0.3em]
\multicolumn{6}{l}{\textbf{Lapatinib}}\\
\hspace{1em}\cellcolor{gray!6}{1} & \cellcolor{gray!6}{ERBB2 (1.06)} & \cellcolor{gray!6}{ERBB2 (1.5)} & \cellcolor{gray!6}{ERBB2 (1.41)} & \cellcolor{gray!6}{ERBB2 (1.69)} & \cellcolor{gray!6}{ERBB2 (1.47)}\\
\hspace{1em}2 & PGAP3 (3.09) & NA (6.81) & PGAP3 (3.44) & PGAP3 (8.03) & NA (4.31)\\
\hspace{1em}\cellcolor{gray!6}{3} & \cellcolor{gray!6}{NA (5.03)} & \cellcolor{gray!6}{PGAP3 (12.41)} & \cellcolor{gray!6}{IKBIP (6.19)} & \cellcolor{gray!6}{C2orf54 (13.09)} & \cellcolor{gray!6}{PGAP3 (14.03)}\\
\hspace{1em}4 & C2orf54 (6.91) & DPYSL2 (16.16) & NA (6.22) & DPYSL2 (15.41) & PKP3 (20.31)\\
\hspace{1em}\cellcolor{gray!6}{5} & \cellcolor{gray!6}{IKBIP (8.28)} & \cellcolor{gray!6}{PKP3 (16.47)} & \cellcolor{gray!6}{C2orf54 (7.41)} & \cellcolor{gray!6}{EMP3 (20.47)} & \cellcolor{gray!6}{EMP3 (22.38)}\\
\addlinespace[0.3em]
\multicolumn{6}{l}{\textbf{Nilotinib}}\\
\hspace{1em}1 & SPN (1.25) & SPN (1.81) & SPN (1.62) & SPN (1.47) & SPN (3.59)\\
\hspace{1em}\cellcolor{gray!6}{2} & \cellcolor{gray!6}{GPC1 (3.5)} & \cellcolor{gray!6}{GPC1 (4.38)} & \cellcolor{gray!6}{GPC1 (3.5)} & \cellcolor{gray!6}{GPC1 (3.44)} & \cellcolor{gray!6}{SELPLG (9.03)}\\
\hspace{1em}3 & TRDC (6.62) & SELPLG (7.5) & TRDC (10.16) & SELPLG (9.97) & KLF13 (26.22)\\
\hspace{1em}\cellcolor{gray!6}{4} & \cellcolor{gray!6}{SELPLG (6.78)} & \cellcolor{gray!6}{KLF13 (16.97)} & \cellcolor{gray!6}{LMO2 (10.44)} & \cellcolor{gray!6}{TRDC (10.22)} & \cellcolor{gray!6}{BCL2 (51.09)}\\
\hspace{1em}5 & LMO2 (9.44) & TRDC (20.22) & CISH (11.03) & LMO2 (10.75) & GPC1 (166.19)\\
\addlinespace[0.3em]
\multicolumn{6}{l}{\textbf{Nutlin-3}}\\
\hspace{1em}\cellcolor{gray!6}{1} & \cellcolor{gray!6}{RP11-148O21.4 (1.41)} & \cellcolor{gray!6}{MET (2.53)} & \cellcolor{gray!6}{RP11-148O21.4 (1.59)} & \cellcolor{gray!6}{RP11-148O21.4 (1.72)} & \cellcolor{gray!6}{RAPGEF5 (37)}\\
\hspace{1em}2 & MET (2.53) & RP11-148O21.4 (4.75) & MET (4.06) & LRRC16A (9.56) & G6PD (63.53)\\
\hspace{1em}\cellcolor{gray!6}{3} & \cellcolor{gray!6}{BLK (5.12)} & \cellcolor{gray!6}{LAYN (6.41)} & \cellcolor{gray!6}{BLK (5.25)} & \cellcolor{gray!6}{BLK (10.97)} & \cellcolor{gray!6}{MET (147.78)}\\
\hspace{1em}4 & LRRC16A (5.16) & RPS27L (12.94) & LRRC16A (5.97) & MET (26.09) & BLK (160.06)\\
\hspace{1em}\cellcolor{gray!6}{5} & \cellcolor{gray!6}{LAT2 (7.34)} & \cellcolor{gray!6}{ADD3 (21.03)} & \cellcolor{gray!6}{LAT2 (7.78)} & \cellcolor{gray!6}{LAYN (138.84)} & \cellcolor{gray!6}{RP11-148O21.4 (164.94)}\\
\addlinespace[0.3em]
\multicolumn{6}{l}{\textbf{PD-0325901}}\\
\hspace{1em}1 & SPRY2 (1.16) & SPRY2 (1.53) & SPRY2 (1.75) & SPRY2 (1.19) & SPRY2 (1.75)\\
\hspace{1em}\cellcolor{gray!6}{2} & \cellcolor{gray!6}{LYZ (2.72)} & \cellcolor{gray!6}{ETV4 (2.88)} & \cellcolor{gray!6}{LYZ (2.09)} & \cellcolor{gray!6}{LYZ (2.88)} & \cellcolor{gray!6}{LYZ (2.62)}\\
\hspace{1em}3 & ETV4 (2.72) & LYZ (3.59) & ETV4 (3.66) & ETV4 (3.38) & ETV4 (3.47)\\
\hspace{1em}\cellcolor{gray!6}{4} & \cellcolor{gray!6}{RP11-1143G9.4 (4.34)} & \cellcolor{gray!6}{TOR4A (4.56)} & \cellcolor{gray!6}{RP11-1143G9.4 (4.53)} & \cellcolor{gray!6}{TOR4A (4.72)} & \cellcolor{gray!6}{TOR4A (4.88)}\\
\hspace{1em}5 & PLEKHG4B (5.62) & PLEKHG4B (4.66) & PLEKHG4B (5.31) & RP11-1143G9.4 (5.38) & PLEKHG4B (5.5)\\
\addlinespace[0.3em]
\multicolumn{6}{l}{\textbf{PD-0332991}}\\
\hspace{1em}\cellcolor{gray!6}{1} & \cellcolor{gray!6}{SH2D3C (4.31)} & \cellcolor{gray!6}{SH2D3C (6.81)} & \cellcolor{gray!6}{SH2D3C (4.56)} & \cellcolor{gray!6}{SH2D3C (8)} & \cellcolor{gray!6}{KRT15 (10.56)}\\
\hspace{1em}2 & FMNL1 (6.56) & FMNL1 (8.38) & HSD3B7 (6.75) & AL162151.3 (9.62) & HSD3B7 (14.5)\\
\hspace{1em}\cellcolor{gray!6}{3} & \cellcolor{gray!6}{HSD3B7 (6.59)} & \cellcolor{gray!6}{AL162151.3 (11.59)} & \cellcolor{gray!6}{FMNL1 (7.09)} & \cellcolor{gray!6}{HSD3B7 (11.03)} & \cellcolor{gray!6}{SEPT6 (16.44)}\\
\hspace{1em}4 & KRT15 (7.19) & TWF1 (12.03) & KRT15 (7.78) & KRT15 (11.97) & PPIC (17.03)\\
\hspace{1em}\cellcolor{gray!6}{5} & \cellcolor{gray!6}{AL162151.3 (8.84)} & \cellcolor{gray!6}{KRT15 (12.56)} & \cellcolor{gray!6}{TWF1 (8.97)} & \cellcolor{gray!6}{FMNL1 (16.34)} & \cellcolor{gray!6}{AL162151.3 (18.34)}\\
\addlinespace[0.3em]
\multicolumn{6}{l}{\textbf{PF2341066}}\\
\hspace{1em}1 & ENAH (1.03) & ENAH (1) & ENAH (1.31) & ENAH (1.12) & ENAH (1.06)\\
\hspace{1em}\cellcolor{gray!6}{2} & \cellcolor{gray!6}{SELPLG (2.09)} & \cellcolor{gray!6}{SELPLG (2.81)} & \cellcolor{gray!6}{SELPLG (2.06)} & \cellcolor{gray!6}{SELPLG (2.28)} & \cellcolor{gray!6}{SELPLG (2.47)}\\
\hspace{1em}3 & HGF (3.62) & HGF (3.94) & MET (5.44) & HGF (7.03) & CTD-2020K17.3 (10.31)\\
\hspace{1em}\cellcolor{gray!6}{4} & \cellcolor{gray!6}{CTD-2020K17.3 (9.72)} & \cellcolor{gray!6}{CTD-2020K17.3 (9.41)} & \cellcolor{gray!6}{HGF (6)} & \cellcolor{gray!6}{MET (10.69)} & \cellcolor{gray!6}{HGF (14.06)}\\
\hspace{1em}5 & MET (10.53) & MET (11.5) & MLKL (12) & CTD-2020K17.3 (11.88) & DOK2 (14.41)\\
\addlinespace[0.3em]
\multicolumn{6}{l}{\textbf{PHA-665752}}\\
\hspace{1em}\cellcolor{gray!6}{1} & \cellcolor{gray!6}{ARHGAP4 (1)} & \cellcolor{gray!6}{ARHGAP4 (1.06)} & \cellcolor{gray!6}{ARHGAP4 (1.06)} & \cellcolor{gray!6}{ARHGAP4 (1.09)} & \cellcolor{gray!6}{ARHGAP4 (1)}\\
\hspace{1em}2 & CTD-2020K17.3 (2.88) & CTD-2020K17.3 (2.66) & CTD-2020K17.3 (4.25) & CTD-2020K17.3 (2.56) & FMNL1 (4.22)\\
\hspace{1em}\cellcolor{gray!6}{3} & \cellcolor{gray!6}{FMNL1 (4.88)} & \cellcolor{gray!6}{FMNL1 (7.44)} & \cellcolor{gray!6}{PFN2 (10.84)} & \cellcolor{gray!6}{PFN2 (24.31)} & \cellcolor{gray!6}{CTD-2020K17.3 (5.44)}\\
\hspace{1em}4 & PFN2 (8.06) & PGPEP1 (13.28) & FMNL1 (11.78) & FMNL1 (33.12) & INHBB (216.5)\\
\hspace{1em}\cellcolor{gray!6}{5} & \cellcolor{gray!6}{PGPEP1 (10.12)} & \cellcolor{gray!6}{INHBB (18.88)} & \cellcolor{gray!6}{FDFT1 (18.66)} & \cellcolor{gray!6}{MICB (59.69)} & \cellcolor{gray!6}{PGPEP1 (335.97)}\\
\addlinespace[0.3em]
\multicolumn{6}{l}{\textbf{PLX4720}}\\
\hspace{1em}1 & RXRG (1) & RXRG (1.16) & RXRG (1) & RXRG (1) & RXRG (1.03)\\
\hspace{1em}\cellcolor{gray!6}{2} & \cellcolor{gray!6}{MMP8 (5.19)} & \cellcolor{gray!6}{MMP8 (3.97)} & \cellcolor{gray!6}{MMP8 (5.16)} & \cellcolor{gray!6}{MMP8 (5.56)} & \cellcolor{gray!6}{MMP8 (4.88)}\\
\hspace{1em}3 & RP11-164J13.1 (6.28) & RP11-599J14.2 (7.22) & MYO5A (7.09) & MYO5A (6.81) & MYO5A (8.75)\\
\hspace{1em}\cellcolor{gray!6}{4} & \cellcolor{gray!6}{RP11-599J14.2 (7)} & \cellcolor{gray!6}{AP1S2 (8.47)} & \cellcolor{gray!6}{LYST (7.97)} & \cellcolor{gray!6}{LYST (11.31)} & \cellcolor{gray!6}{AP1S2 (10.53)}\\
\hspace{1em}5 & RP4-718J7.4 (7.84) & LYST (9.34) & RP11-599J14.2 (8.41) & RP4-718J7.4 (13.22) & RP11-599J14.2 (12.69)\\
\addlinespace[0.3em]
\multicolumn{6}{l}{\textbf{Paclitaxel}}\\
\hspace{1em}\cellcolor{gray!6}{1} & \cellcolor{gray!6}{MMP24 (1.09)} & \cellcolor{gray!6}{MMP24 (1.28)} & \cellcolor{gray!6}{MMP24 (1.22)} & \cellcolor{gray!6}{MMP24 (2.38)} & \cellcolor{gray!6}{SH3BP1 (10.38)}\\
\hspace{1em}2 & AGAP2 (3.16) & SH3BP1 (2.75) & AGAP2 (3.44) & SH3BP1 (2.88) & PRODH (60.41)\\
\hspace{1em}\cellcolor{gray!6}{3} & \cellcolor{gray!6}{SH3BP1 (3.5)} & \cellcolor{gray!6}{AGAP2 (4.06)} & \cellcolor{gray!6}{SH3BP1 (3.78)} & \cellcolor{gray!6}{SLC38A5 (3.84)} & \cellcolor{gray!6}{AGAP2 (157.41)}\\
\hspace{1em}4 & SLC38A5 (4.34) & SLC38A5 (4.22) & PTTG1IP (3.91) & AGAP2 (5.88) & SLC38A5 (179.34)\\
\hspace{1em}\cellcolor{gray!6}{5} & \cellcolor{gray!6}{PTTG1IP (4.72)} & \cellcolor{gray!6}{PTTG1IP (4.34)} & \cellcolor{gray!6}{SLC38A5 (4.31)} & \cellcolor{gray!6}{PTTG1IP (6.97)} & \cellcolor{gray!6}{MMP24 (308.66)}\\
\addlinespace[0.3em]
\multicolumn{6}{l}{\textbf{Panobinostat}}\\
\hspace{1em}1 & AGAP2 (1.12) & AGAP2 (1.56) & AGAP2 (1.81) & AGAP2 (2.16) & CYR61 (2.56)\\
\hspace{1em}\cellcolor{gray!6}{2} & \cellcolor{gray!6}{CYR61 (2.44)} & \cellcolor{gray!6}{CYR61 (2.09)} & \cellcolor{gray!6}{CYR61 (2.16)} & \cellcolor{gray!6}{CYR61 (2.78)} & \cellcolor{gray!6}{AGAP2 (3.25)}\\
\hspace{1em}3 & RPL39P5 (4.19) & RPL39P5 (4.41) & RPL39P5 (3.75) & RPL39P5 (3.88) & RPL39P5 (152.44)\\
\hspace{1em}\cellcolor{gray!6}{4} & \cellcolor{gray!6}{WWTR1 (5.16)} & \cellcolor{gray!6}{WWTR1 (5.78)} & \cellcolor{gray!6}{WWTR1 (5.94)} & \cellcolor{gray!6}{WWTR1 (6.53)} & \cellcolor{gray!6}{S100A2 (316.66)}\\
\hspace{1em}5 & MYOF (6.56) & MYOF (6.16) & IKZF1 (12.41) & IKZF1 (9.72) & MYOF (366.28)\\
\addlinespace[0.3em]
\multicolumn{6}{l}{\textbf{RAF265}}\\
\hspace{1em}\cellcolor{gray!6}{1} & \cellcolor{gray!6}{CMTM3 (1.34)} & \cellcolor{gray!6}{CMTM3 (1.5)} & \cellcolor{gray!6}{CMTM3 (1.69)} & \cellcolor{gray!6}{SH2B3 (34)} & \cellcolor{gray!6}{CMTM3 (7.06)}\\
\hspace{1em}2 & SYT17 (5.69) & SYT17 (5.66) & SYT17 (7.69) & CMTM3 (155.25) & SH2B3 (470.66)\\
\hspace{1em}\cellcolor{gray!6}{3} & \cellcolor{gray!6}{SH2B3 (6.03)} & \cellcolor{gray!6}{SH2B3 (8.91)} & \cellcolor{gray!6}{SH2B3 (17.5)} & \cellcolor{gray!6}{SLC29A3 (159)} & \cellcolor{gray!6}{SYT17 (652.84)}\\
\hspace{1em}4 & EMILIN2 (11.94) & SLC29A3 (11.84) & STAT5A (19.66) & PRKCQ (235) & RGS16 (713.47)\\
\hspace{1em}\cellcolor{gray!6}{5} & \cellcolor{gray!6}{STAT5A (12.47)} & \cellcolor{gray!6}{NA (19.91)} & \cellcolor{gray!6}{SLC29A3 (22.22)} & \cellcolor{gray!6}{LCP2 (259.7)} & \cellcolor{gray!6}{AC007620.3 (1087.11)}\\
\addlinespace[0.3em]
\multicolumn{6}{l}{\textbf{Sorafenib}}\\
\hspace{1em}1 & PXK (4.47) & TP63 (6.72) & PXK (4.12) & FAM212A (6.41) & PXK (9.34)\\
\hspace{1em}\cellcolor{gray!6}{2} & \cellcolor{gray!6}{P2RX1 (4.62)} & \cellcolor{gray!6}{P2RX1 (7.78)} & \cellcolor{gray!6}{FAM212A (5.75)} & \cellcolor{gray!6}{P2RX1 (8)} & \cellcolor{gray!6}{ARHGAP9 (34.25)}\\
\hspace{1em}3 & FAM212A (4.69) & PXK (8.88) & STAC3 (7.16) & SEC31B (41.69) & P2RX1 (156.91)\\
\hspace{1em}\cellcolor{gray!6}{4} & \cellcolor{gray!6}{STAC3 (5.16)} & \cellcolor{gray!6}{FAM212A (16.97)} & \cellcolor{gray!6}{P2RX1 (7.25)} & \cellcolor{gray!6}{ARHGAP9 (43.19)} & \cellcolor{gray!6}{FAM212A (187.31)}\\
\hspace{1em}5 & ARHGAP9 (7.91) & STAC3 (20.16) & TP63 (40.97) & CXCL8 (57.72) & SEC31B (222.41)\\
\addlinespace[0.3em]
\multicolumn{6}{l}{\textbf{TAE684}}\\
\hspace{1em}\cellcolor{gray!6}{1} & \cellcolor{gray!6}{SELPLG (1.09)} & \cellcolor{gray!6}{SELPLG (1.06)} & \cellcolor{gray!6}{SELPLG (1.12)} & \cellcolor{gray!6}{SELPLG (1.03)} & \cellcolor{gray!6}{SELPLG (1.34)}\\
\hspace{1em}2 & IL6R (3.19) & ARID3A (8.12) & IL6R (3.34) & IL6R (6.41) & ARID3A (18.31)\\
\hspace{1em}\cellcolor{gray!6}{3} & \cellcolor{gray!6}{NFIL3 (6.34)} & \cellcolor{gray!6}{GALNT18 (8.62)} & \cellcolor{gray!6}{NFIL3 (6.25)} & \cellcolor{gray!6}{NFIL3 (8.06)} & \cellcolor{gray!6}{FMNL1 (25.25)}\\
\hspace{1em}4 & ARID3A (7) & IL6R (10.16) & RRAS2 (10.19) & ARID3A (16.38) & RP11-334A14.2 (144.34)\\
\hspace{1em}\cellcolor{gray!6}{5} & \cellcolor{gray!6}{RRAS2 (8.66)} & \cellcolor{gray!6}{PPP2R3A (15.69)} & \cellcolor{gray!6}{FMNL1 (10.19)} & \cellcolor{gray!6}{RRAS2 (17.97)} & \cellcolor{gray!6}{PPP2R3A (165.84)}\\
\addlinespace[0.3em]
\multicolumn{6}{l}{\textbf{TKI258}}\\
\hspace{1em}1 & TWF1 (2.44) & TWF1 (3.31) & TWF1 (2.41) & TWF1 (2.28) & LAPTM5 (20.84)\\
\hspace{1em}\cellcolor{gray!6}{2} & \cellcolor{gray!6}{SLC43A1 (2.88)} & \cellcolor{gray!6}{GPR162 (3.75)} & \cellcolor{gray!6}{PRTN3 (3.31)} & \cellcolor{gray!6}{LAPTM5 (5.34)} & \cellcolor{gray!6}{SLC43A1 (156.12)}\\
\hspace{1em}3 & PRTN3 (5.25) & SLC43A1 (6.84) & SLC43A1 (6.06) & PRTN3 (6.12) & TWF1 (163.78)\\
\hspace{1em}\cellcolor{gray!6}{4} & \cellcolor{gray!6}{LAPTM5 (5.78)} & \cellcolor{gray!6}{TTC28 (8.94)} & \cellcolor{gray!6}{LAT2 (6.62)} & \cellcolor{gray!6}{SLC43A1 (14.91)} & \cellcolor{gray!6}{GPR162 (169.66)}\\
\hspace{1em}5 & LAT2 (5.94) & LAPTM5 (11.53) & LAPTM5 (8.19) & LAT2 (17.41) & LYL1 (387.97)\\
\addlinespace[0.3em]
\multicolumn{6}{l}{\textbf{Topotecan}}\\
\hspace{1em}\cellcolor{gray!6}{1} & \cellcolor{gray!6}{SLFN11 (1)} & \cellcolor{gray!6}{SLFN11 (1)} & \cellcolor{gray!6}{SLFN11 (1)} & \cellcolor{gray!6}{SLFN11 (1)} & \cellcolor{gray!6}{SLFN11 (1)}\\
\hspace{1em}2 & HSPB8 (2.16) & HSPB8 (2.28) & HSPB8 (2.84) & HSPB8 (2.06) & HSPB8 (2.62)\\
\hspace{1em}\cellcolor{gray!6}{3} & \cellcolor{gray!6}{PPIC (5.69)} & \cellcolor{gray!6}{OSGIN1 (5.28)} & \cellcolor{gray!6}{OSGIN1 (5.31)} & \cellcolor{gray!6}{PPIC (7.81)} & \cellcolor{gray!6}{OSGIN1 (7.19)}\\
\hspace{1em}4 & OSGIN1 (5.88) & AGAP2 (8.81) & PPIC (6.81) & RP11-359P5.1 (10.75) & AGAP2 (15.25)\\
\hspace{1em}\cellcolor{gray!6}{5} & \cellcolor{gray!6}{AGAP2 (6.53)} & \cellcolor{gray!6}{PPIC (8.91)} & \cellcolor{gray!6}{RP11-359P5.1 (8.25)} & \cellcolor{gray!6}{CORO1A (12.06)} & \cellcolor{gray!6}{HMGB2 (21.53)}\\
\addlinespace[0.3em]
\multicolumn{6}{l}{\textbf{ZD-6474}}\\
\hspace{1em}1 & MAP3K12 (1) & MAP3K12 (1) & MAP3K12 (1) & MAP3K12 (1.09) & MAP3K12 (1)\\
\hspace{1em}\cellcolor{gray!6}{2} & \cellcolor{gray!6}{PIM1 (5.47)} & \cellcolor{gray!6}{CTSH (21.88)} & \cellcolor{gray!6}{PIM1 (10.28)} & \cellcolor{gray!6}{PIM1 (31.44)} & \cellcolor{gray!6}{SCD5 (339.58)}\\
\hspace{1em}3 & PRKCQ (9.03) & TIMP1 (26.31) & PRKCQ (15.28) & DYNLT3 (192.12) & ITGA10 (527.41)\\
\hspace{1em}\cellcolor{gray!6}{4} & \cellcolor{gray!6}{CTSH (12.91)} & \cellcolor{gray!6}{PRKCQ (26.47)} & \cellcolor{gray!6}{CTSH (18.16)} & \cellcolor{gray!6}{TIMP1 (211.25)} & \cellcolor{gray!6}{TIMP1 (569.81)}\\
\hspace{1em}5 & ITGA10 (20.81) & PIM1 (33.81) & ANXA5 (78.78) & EPHA1 (320.22) & CTSH (631.5)
\label{tab:ccle_top_genes}
\end{longtable}
}

\begin{table}[h!]

\caption{Summary of prediction power using the top 10 features from each feature importance method in the CCLE drug response case study. For each drug, we evaluated the average test $\rsq$ (averaged across 32 train-test splits) from an RF trained using only the top 10 features from each feature importance method. We then rank the feature importance methods by this average test $\rsq$ (1 = best test $\rsq$, 5 = worst test $\rsq$) and display the number of drugs, for which that rank was achieved. In particular, taking the top 10 genes from \method\;(ridge) gave the best prediction performance for 12 out of the 24 drugs.}
\centering
\begin{tabular}[t]{cccccc}
\toprule
\textbf{Rank} & \textbf{\method\;(ridge)} & \textbf{TreeSHAP} & \textbf{MDI} & \textbf{MDI-oob} & \textbf{MDA}\\
\midrule
\cellcolor{gray!6}{\textbf{1}} & \cellcolor{gray!6}{12} & \cellcolor{gray!6}{5} & \cellcolor{gray!6}{3} & \cellcolor{gray!6}{3} & \cellcolor{gray!6}{1}\\
\textbf{2} & 6 & 6 & 6 & 3 & 3\\
\cellcolor{gray!6}{\textbf{3}} & \cellcolor{gray!6}{2} & \cellcolor{gray!6}{6} & \cellcolor{gray!6}{6} & \cellcolor{gray!6}{4} & \cellcolor{gray!6}{6}\\
\textbf{4} & 3 & 5 & 5 & 5 & 6\\
\cellcolor{gray!6}{\textbf{5}} & \cellcolor{gray!6}{1} & \cellcolor{gray!6}{2} & \cellcolor{gray!6}{4} & \cellcolor{gray!6}{9} & \cellcolor{gray!6}{8}\\
\bottomrule
\end{tabular}
\label{tab:ccle-prediction}
\end{table}

\subsection{Breast Cancer Subtype Prediction} \label{supp:tcga_case_study}

\paragraph{Data preprocessing.} We pulled data from the TCGA breast cancer project using the \texttt{TCGAbiolinks} R package. This gene expression dataset originally consisted of 19,947 genes. As before, we reduced the number of features under consideration by taking only the top 5000 features with the highest variance. The processed TCGA data used in our case study can be found on Zenodo at
\if1\blind{\url{https://zenodo.org/record/8111870}}\fi
\if0\blind{\url{redacted}}\fi.

\paragraph{Results.} In Table~\ref{tab:tcga-prediction}, we summarize the test prediction performance for RF, \rfmethod\;(ridge), and \rfmethod\;(logistic), applied to the TCGA case study. We show the average test classification accuracy, AUROC, and area under the precision recall curve (AUPRC), averaged across 32 train-test splits. In Table~\ref{tab:tcga_top_genes}, we list the top 25 gene expression predictors according to each feature importance method. These genes are ranked according to their average feature importance ranking across the 32 train-test splits. We also show the RF predictive power of the top $k$ genes from each feature importance method in Figure~\ref{fig:ccle-prediction}. Details and discussion on this procedure were provided in the previous section. As discussed, correlated features can make it challenging to directly compare the prediction accuracy using the top $k$ genes from various feature importance methods. Still, it is worth noting that the top 10 features from \method\;(ridge) and \method\;(logistic) yield a higher test prediction performance (in terms of both AUROC and classification accuracy) than other competing methods. Moreover, the top 15 features from \method\;(ridge) and \method\;(logistic) yield a higher test prediction performance (in terms of both AUROC and classification accuracy) as the top 25 features from the other competing feature importance methods. This improvement in predictive power further supports the practical utility of \method\;for feature ranking \citep{yu2020veridical}.

\begin{table}[h!]
\footnotesize
\caption{Test prediction performance for various methods, averaged across 32 train-test splits, on the TCGA case study. Standard errors are shown in parentheses.}
\centering
\begin{tabular}[t]{ccccccc}
\toprule
\textbf{Model} & \textbf{Classification Accuracy} & \textbf{AUROC} & \textbf{AUPRC} \\
\midrule
\cellcolor{gray!6}{\rfmethod\;(logistic)} & \cellcolor{gray!6}{\textbf{0.884 (0.003)}} & \cellcolor{gray!6}{\textbf{0.981 (0.001)}} & \cellcolor{gray!6}{\textbf{0.895 (0.004)}}\\
\rfmethod\;(ridge) & 0.873 (0.003) & 0.981 (0.001) & 0.892 (0.004)\\
\cellcolor{gray!6}{RF} & \cellcolor{gray!6}{0.861 (0.003)} & \cellcolor{gray!6}{0.978 (0.001)} & \cellcolor{gray!6}{0.878 (0.005)}\\
\bottomrule
\end{tabular}
\label{tab:tcga-prediction}
\end{table}

\begin{table}[h!]
\scriptsize
\caption{Top 25 most important genes for predicting breast cancer subtype according to various feature importance methods. Genes are ranked by their average feature importance ranking across 32 train-test splits (shown in parentheses).}
\centering
\begin{tabular}[t]{cccccc}
\toprule
\textbf{Rank} & \textbf{\method\;(ridge)} & \textbf{\method\;(logistic)} & \textbf{MDA} & \textbf{TreeSHAP} & \textbf{MDI}\\
\midrule
\cellcolor{gray!6}{1} & \cellcolor{gray!6}{ESR1 (1.91)} & \cellcolor{gray!6}{ESR1 (1.91)} & \cellcolor{gray!6}{ESR1 (4.5)} & \cellcolor{gray!6}{ESR1 (7.62)} & \cellcolor{gray!6}{ESR1 (13.91)}\\
2 & FOXA1 (4.25) & GATA3 (4.5) & GATA3 (6.38) & TPX2 (10.41) & TPX2 (15.34)\\
\cellcolor{gray!6}{3} & \cellcolor{gray!6}{FOXC1 (6.12)} & \cellcolor{gray!6}{FOXA1 (5.09)} & \cellcolor{gray!6}{FOXA1 (8.11)} & \cellcolor{gray!6}{GATA3 (19.62)} & \cellcolor{gray!6}{FOXM1 (22.84)}\\
4 & GATA3 (6.97) & TPX2 (6.81) & TPX2 (10.12) & FOXM1 (20.06) & MLPH (24.97)\\
\cellcolor{gray!6}{5} & \cellcolor{gray!6}{AGR3 (7.94)} & \cellcolor{gray!6}{AGR3 (10.22)} & \cellcolor{gray!6}{MLPH (10.16)} & \cellcolor{gray!6}{FOXA1 (20.72)} & \cellcolor{gray!6}{FOXA1 (25.66)}\\
6 & MLPH (8.16) & FOXC1 (12.94) & AGR3 (12.94) & CDK1 (22.38) & GATA3 (30.44)\\
\cellcolor{gray!6}{7} & \cellcolor{gray!6}{TPX2 (11.03)} & \cellcolor{gray!6}{MLPH (15.69)} & \cellcolor{gray!6}{TBC1D9 (14.22)} & \cellcolor{gray!6}{MLPH (22.53)} & \cellcolor{gray!6}{CDK1 (31.41)}\\
8 & TBC1D9 (14.44) & FOXM1 (18.12) & FOXC1 (15.09) & AGR3 (25.88) & THSD4 (34.69)\\
\cellcolor{gray!6}{9} & \cellcolor{gray!6}{FOXM1 (18.66)} & \cellcolor{gray!6}{TBC1D9 (21.03)} & \cellcolor{gray!6}{FOXM1 (19.88)} & \cellcolor{gray!6}{PLK1 (28.47)} & \cellcolor{gray!6}{FOXC1 (35)}\\
10 & THSD4 (21.78) & THSD4 (23.66) & THSD4 (21.28) & TBC1D9 (29.84) & TBC1D9 (35.44)\\
\cellcolor{gray!6}{11} & \cellcolor{gray!6}{SPDEF (25.81)} & \cellcolor{gray!6}{CDK1 (24.44)} & \cellcolor{gray!6}{CDK1 (21.77)} & \cellcolor{gray!6}{FOXC1 (30.44)} & \cellcolor{gray!6}{AGR3 (36.44)}\\
12 & CA12 (29.44) & MYBL2 (25.34) & XBP1 (24.88) & MYBL2 (33.06) & PLK1 (36.81)\\
\cellcolor{gray!6}{13} & \cellcolor{gray!6}{CDK1 (36.09)} & \cellcolor{gray!6}{RACGAP1 (26.81)} & \cellcolor{gray!6}{KIF2C (27.95)} & \cellcolor{gray!6}{THSD4 (33.53)} & \cellcolor{gray!6}{MYBL2 (45.22)}\\
14 & GABRP (36.72) & ASPM (27.56) & PLK1 (28.58) & KIF2C (35.56) & KIF2C (46.22)\\
\cellcolor{gray!6}{15} & \cellcolor{gray!6}{PLK1 (37.97)} & \cellcolor{gray!6}{PLK1 (28.84)} & \cellcolor{gray!6}{MYBL2 (30.97)} & \cellcolor{gray!6}{ASPM (40.59)} & \cellcolor{gray!6}{ASPM (49.72)}\\
16 & FAM171A1 (38.59) & UBE2C (30.41) & GABRP (31.94) & GMPS (41.41) & GMPS (52.03)\\
\cellcolor{gray!6}{17} & \cellcolor{gray!6}{ASPM (39.5)} & \cellcolor{gray!6}{SPAG5 (31.81)} & \cellcolor{gray!6}{ASPM (35.97)} & \cellcolor{gray!6}{XBP1 (50.44)} & \cellcolor{gray!6}{SPDEF (58.88)}\\
18 & SFRP1 (40.25) & GMPS (35.34) & FAM171A1 (38.55) & CENPF (56.69) & FAM171A1 (68.56)\\
\cellcolor{gray!6}{19} & \cellcolor{gray!6}{XBP1 (40.44)} & \cellcolor{gray!6}{KIF2C (36.03)} & \cellcolor{gray!6}{CA12 (40.16)} & \cellcolor{gray!6}{MKI67 (59.69)} & \cellcolor{gray!6}{CA12 (69.59)}\\
20 & MYBL2 (43.12) & CA12 (37.31) & CDC20 (48.45) & CA12 (59.91) & UBE2C (69.78)\\
\cellcolor{gray!6}{21} & \cellcolor{gray!6}{TFF3 (43.12)} & \cellcolor{gray!6}{RRM2 (46.09)} & \cellcolor{gray!6}{SPDEF (53.05)} & \cellcolor{gray!6}{RACGAP1 (60.53)} & \cellcolor{gray!6}{TFF3 (70)}\\
22 & KIF2C (44.62) & CENPF (46.31) & UBE2C (54.27) & SPAG5 (61.94) & CENPF (70.31)\\
\cellcolor{gray!6}{23} & \cellcolor{gray!6}{PRR15 (44.88)} & \cellcolor{gray!6}{GABRP (47.59)} & \cellcolor{gray!6}{ANXA9 (55.41)} & \cellcolor{gray!6}{FAM171A1 (63)} & \cellcolor{gray!6}{RACGAP1 (70.94)}\\
24 & AGR2 (45) & SFRP1 (49.34) & C1orf64 (57.22) & KIF11 (63.56) & SPAG5 (71.81)\\
\cellcolor{gray!6}{25} & \cellcolor{gray!6}{MIA (45.31)} & \cellcolor{gray!6}{XBP1 (49.78)} & \cellcolor{gray!6}{RACGAP1 (57.97)} & \cellcolor{gray!6}{ANLN (64.28)} & \cellcolor{gray!6}{KIF11 (73.12)}\\
\bottomrule
\end{tabular}
\label{tab:tcga_top_genes}
\end{table}

\begin{figure}[h!]
    \centering
    \includegraphics[width=0.65\textwidth]{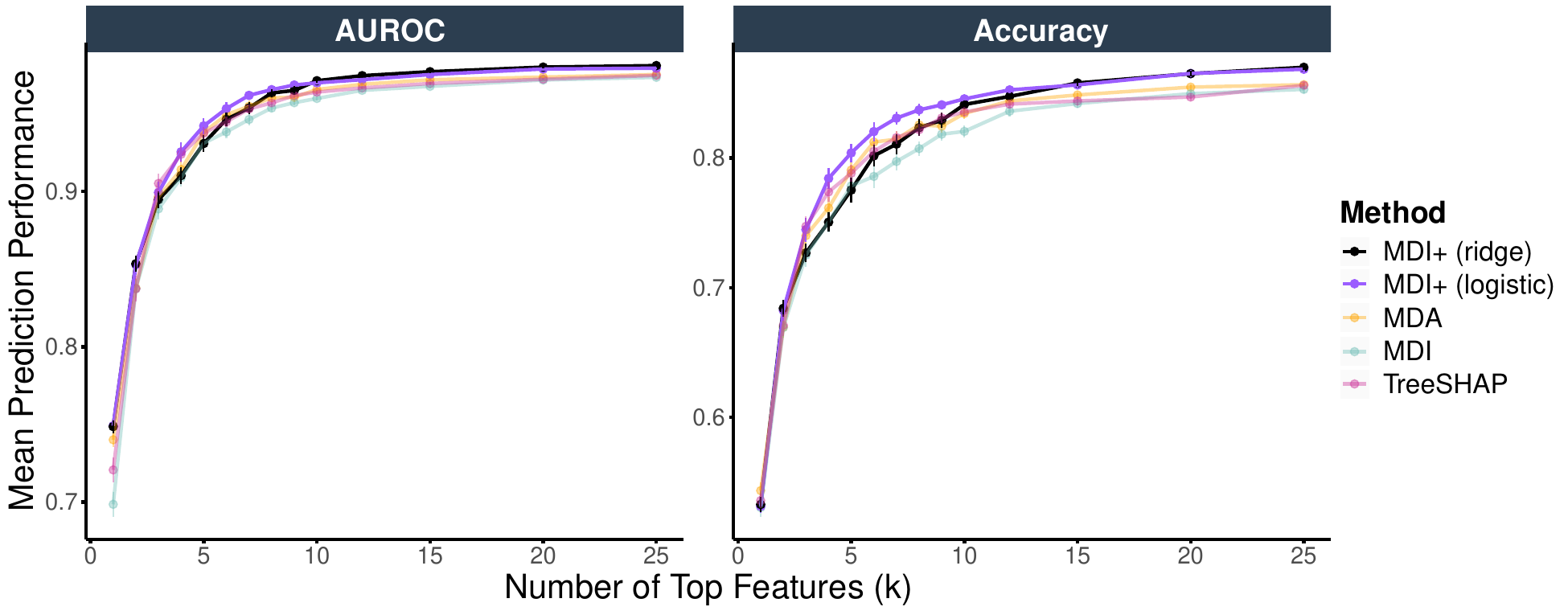}
    \caption{RF prediction performance, measured via test AUROC and classification accuracy, using the top $k$ features from each feature importance methods in the TCGA case study. Results are averaged across 32 train-test data splits.}
    \label{fig:tcga-prediction} 
\end{figure}

\newpage
\bibliography{refs}

\end{document}